\documentclass[aps,prb,reprint,10pt,groupedaddress,showpacs,amssymb,amsfonts,floatfix]{revtex4-1}
\usepackage{amsmath,amssymb,latexsym,epsfig,bm,subfigure}

\usepackage[usenames,dvipsnames]{color}
\usepackage[colorlinks,bookmarks=false,citecolor=blue,linkcolor=red,urlcolor=blue]{hyperref}
\usepackage{verbatim}

\def\beq{\begin{equation}}
\def\eeq{\end{equation}}
\def\be{\begin{equation}}
\def\ee{\end{equation}}

\def\ket#1{\vert #1 \rangle}

\newcommand{\zz}{\mathbb{Z}_2}
\newcommand{\z}{\mathbb{Z}}
\newcommand{\floor}[1]{\lfloor #1 \rfloor}

\def\cL{\mathcal{L}}
\def\cT{\mathcal{T}}

\def\cC{\mathcal{C}}

\def\r{{\bf r}}
\def\xhat{\hat{\mathbf{x}}}
\def\yhat{\hat{\mathbf{y}}}

\def\bS{\boldsymbol{S}}

\def\down{\downarrow}
\def\up{\uparrow}

\def\aem#1{\marginpar{\small AE: #1}}
\def\mhm#1{\marginpar{\small MH: #1}}

 \def\aem#1{}
 \def\mhm#1{}
\begin{document}

\title{Classifying fractionalization: symmetry classification of gapped $\zz$ spin liquids in two dimensions}

\author{Andrew M.~Essin}
\affiliation{Department of Physics, 390 UCB, University of Colorado,
Boulder CO 80309, USA}
\author{Michael Hermele}
\affiliation{Department of Physics, 390 UCB, University of Colorado,
Boulder CO 80309, USA}
\date{\today}

\begin{abstract}
We classify distinct types of quantum number fractionalization occurring in two-dimensional topologically ordered phases, focusing in particular on phases with $\zz$ topological order, that is, on gapped $\zz$ spin liquids.  We find that the \emph{fractionalization class} of each anyon is an equivalence class of projective representations of the symmetry group, corresponding to elements of the cohomology group $H^2(G, \zz)$.  This result leads us to a symmetry classification of gapped $\zz$ spin liquids, such that two phases in different symmetry classes cannot be connected without breaking symmetry or crossing a phase transition.  Symmetry classes are defined by specifying a fractionalization class for each type of anyon.  The fusion rules of anyons play a crucial role in determining the symmetry classes.  For translation and internal symmetries, braiding statistics plays no role, but can affect the classification when point group symmetries are present.  For square lattice space group, time reversal and ${\rm SO}(3)$ spin rotation symmetries, we find $2\,098\,176 \approx 2^{21}$ distinct symmetry classes.  Our symmetry classification is not complete, as we exclude, by assumption, permutation of the different types of anyons by symmetry operations.  We give an explicit construction of symmetry classes for square lattice space group symmetry in the toric code model.  Via simple examples, we illustrate how information about fractionalization classes can in principle be obtained from the spectrum and quantum numbers of excited states.  Moreover, the symmetry class can be partially determined from the quantum numbers of the four degenerate ground states on the torus.  We also extend our results to arbitrary Abelian topological orders (limited, though, to translations and internal symmetries), and compare our classification with the related projective symmetry group classification of parton mean-field theories.  Our results provide a framework for understanding and probing the sharp distinctions among symmetric $\zz$ spin liquids, and are a first step toward a full classification of symmetric topologically ordered phases.
\end{abstract}


\maketitle

\tableofcontents

\section{Introduction}
\label{sec:introduction}

One of the characteristic features of topologically ordered states of matter\cite{wen89,wen90,wen12} in two dimensions is the presence of anyons -- quasiparticle excitations with non-trivial braiding statistics.  Another important feature is quantum number fractionalization:  if some degree of symmetry is present, the anyons can carry fractional quantum numbers.  The charge $e/3$ quasiparticles of the $\nu = 1/3$ Laughlin fractional quantum Hall state\cite{laughlin83} are a celebrated example of this phenomenon.  The fractional charge of these excitations has been directly observed,\cite{kane94,depicciotto97,saminadayar97} while a direct, unambiguous measurement of their statistics remains elusive.\cite{stern08}  As in this case, it is important to recognize that fractionalization may often be easier to detect than other characteristic features of topological order.

Given the important role of fractionalization in topologically ordered states of matter, it is important to develop a better understanding of the interplay among symmetry, fractionalization, and topological order.  Many of the most basic questions along these lines are not well understood, for instance, among states having the same topological order and the same symmetry, are there distinct types of fractionalization that can be used to distinguish phases?  If so, how can we describe and classify distinct types of fractionalization? What types of fractionalization are consistent with a given type of topological order?  In this paper, we answer these questions for one of the simplest types of topological order, namely the topological order of the deconfined phase of $\zz$ gauge theory,\cite{wegner71} which we refer to as $\zz$ topological order.  
We will introduce the notion of \emph{fractionalization class} of an anyon, which describes its characteristic type of fractionalization.

To be more specific, we shall confine our attention to two dimensions, to zero temperature, and to local bosonic models (\emph{i.e.}, spin models with finite-range or exponentially decaying interactions).  For simplicity, we exclude the possibility of spontaneously broken symmetry.  In this setting, states with $\zz$ topological order are referred to as gapped\cite{gapcomment} $\zz$ spin liquids.\cite{chakraborty89,read91,wen91,balents99,senthil00,moessner01a, moessner01b,balents02,kitaev03}  Despite the restriction to two dimensions, it should be noted that nowhere will we assume a \emph{strict} two-dimensional system.  That is, our two-dimensional system may lie on the boundary of a gapped three-dimensional bulk.  This point may have interesting implications for future work, and we return to it in Sec.~\ref{sec:discussion}.

Our results can also be viewed through the lens of classification of distinct phases of matter.  More specifically, we may ask for a classification of all distinct phases of matter sharing a given fixed topological order and fixed symmetry group.  In this situation, it is known that many distinct ``symmetry enriched'' topological phases exist.\cite{wen02,wen03,fwang06,kou08,kou09,yao10,huh11,gchen12,levin12,cho12}  The distinctions among these phases disappear upon breaking of all symmetries, while the topological order is unaffected.  

We shall see that specifying fractionalization classes for each type of anyon defines a \emph{symmetry class}, so named because it determines the action of symmetry on the topological degrees of freedom.  Two states (\emph{i.e.}, states of matter) in different symmetry classes are distinct phases, and cannot be adiabatically connected without closing a gap or breaking symmetry.  This is only a partial classification of phases, though, because a given symmetry class may contain more than one distinct phase.  Nonetheless, symmetry classification is a first step toward classification of all phases sharing a fixed topological order and symmetry group.  We provide such a symmetry classification for gapped $\zz$ spin liquids, for an arbitrary symmetry group.  It should be noted that the symmetry classification we give here is not complete; for simplicity, we do not consider cases where some symmetry operations permute the different types of anyons.  This kind of action of the symmetry group is ``beyond fractionalization,'' and its study is left for future work.

Our approach is not limited to $\zz$ topological order.  Indeed, if only translation and  internal symmetries are considered, we describe the straightforward extension of our results to arbitrary Abelian topological orders in Sec.~\ref{sec:general-abelian}.  Including full space group symmetry, with further work, we believe our approach could be extended to give a symmetry classification for an arbitrary Abelian topological order.  We have not yet considered extensions to non-Abelian topological order or to topological order in three dimensions.

Classifications such as ours are useful because they provide a systematic basis for understanding sharp distinctions among phases.  Along these lines, we hope that our results will be of use in finding new ways to identify and distinguish different spin liquids in numerical simulations and in experiments.  Indeed, our focus on gapped $\zz$ spin liquids is partly motivated by the recent striking evidence that such states are present in simple, fairly realistic $S = 1/2$ Heisenberg spin models on the $J_1$-$J_2$ square lattice and the kagome lattice.\cite{yan11,lwang11,jiang12,depenbrock12,jiang12b,mezzacapo12}  We do present some results here touching on determination of fractionalization and symmetry classes in numerical simulations (Secs.~\ref{sec:fc-gen} and~\ref{sec:qndgs}), but substantial further progress is likely possible, and we hope to stimulate further work in this direction.

We begin with the familiar observation that quantum mechanics allows for symmetries to be realized projectively.  One classic example is the fact that rotation by $2\pi$ gives a phase $-1$ when acting on a wavefunction for a single half-odd integer spin. Another is the magnetic translation group of a single particle in a uniform magnetic field, where two translation operators $T_x$ and $T_y$ do not commute but instead satisfy $T_x T_y = e^{i \phi} T_y T_x$.  More generally, but somewhat loosely, we say symmetries are realized projectively when identities among group elements hold only up to a phase when acting on a quantum state.  Group representations with this property are called projective representations.

On the other hand, for any local bosonic model describing a spin system built from an even number of electrons (or, for that matter, an even number of neutral atoms), symmetry operations act linearly---as opposed to projectively---on many-body wave functions.  For instance, $T_x T_y = T_y T_x$.  If one has a system with an odd number of electrons, we can always consider a larger system with an even number, so, with this constraint on lattice size in mind, we assert that symmetries act linearly on the many-body wave functions of any physically reasonable local bosonic model.

However, it is well known that, in general, symmetries act projectively on anyons.  For instance, ${\rm SO}(3)$ spin rotation symmetry acts projectively on the $S = 1/2$ spinon quasiparticles appearing in many gapped $\zz$ spin liquids.  The crucial issue is how to describe and distinguish such projective actions to arrive at a set of fractionalization and symmetry classes.

Before proceeding, we first have to briefly mention some facts about $\zz$ topological order.  There are four particle types or classes of quasiparticle excitations, denoted $1$, $e$, $m$, and $\epsilon$.  It may be helpful to think of these in terms of $\zz$ gauge theory coupled to bosonic matter fields; the deconfined phase of such a theory is a concrete realization of $\zz$ topological order.  The $e$ particles are $\zz$ electric charges,  the $m$ particles are $\zz$ magnetic fluxes, and $\epsilon$ particles are $e$-$m$ bound states.  $1$-particles, also referred to as ``trivial'' particles, are excitations that are not part of the topological structure.  The non-trivial particles (\emph{i.e.}, anyons) have non-trivial braiding statistics:  any two distinct  anyons (\emph{e.g.}, an $e$ and an $m$) have $\theta = \pi$ mutual statistics.  The $e$- and $m$-particles are bosons, but  $\epsilon$-particles are fermions due to the mutual statistics of $e$ and $m$.  $1$-particles are bosonic, and have trivial mutual statistics with the other particle types.  The fusion of two particles gives a unique third particle type, according to the fusion rules
\begin{eqnarray}
 e \times e &=& m \times m = \epsilon \times \epsilon = 1 \label{eqn:fusion} \\ 
 1 \times 1 &=& 1 \text{ , } e \times 1 = e \text{ , } m \times 1 = m \text{ , } \epsilon \times 1 = \epsilon \text{,} \nonumber \\
  e \times m &=& \epsilon \text{ , }  e \times \epsilon = m \text{ , } m \times \epsilon = e  \text{.} \nonumber
\end{eqnarray}
It is important to note that these properties are unchanged under the relabeling $e \leftrightarrow m$.

Now, to state our results, the action of the symmetry group on each type of topological quasiparticle is given by a projective representation, which is associated with a $\zz$ central extension of the symmetry group.  (For $1$-particles, this is always the trivial extension.)  These central extensions can be grouped into equivalence classes, which we call fractionalization classes.  Fractionalization classes are in one-to-one correspondence with elements of the cohomology group $H^2(G, \zz)$.  A symmetry class is then defined by specifying the fractionalization class for each anyon.  The symmetry class is a universal property of a $\zz$ spin liquid phase; that is, two states (\emph{i.e.}, states of matter) with different symmetry classes cannot be adiabatically connected without breaking symmetry.  The fractionalization class for each anyon follows from the other two by fusion, so only two elements of $H^2(G, \zz)$ need be specified.  Equivalently, one can instead specify a single element of $H^2( G, \zz \times \zz)$.

Pairs of elements of $H^2(G, \zz)$ are not quite in one-to-one correspondence with distinct symmetry classes.  This occurs because pairs of $e$ and $m$ fractionalization classes related by relabeling $e \leftrightarrow m$ are not distinct.

If $G$ consists only of translations and internal symmetries, braiding statistics play no role in this classification.  In this case, the fractionalization class of, say, $\epsilon$ is given simply by the $H^2(G, \zz)$ group product of the classes for $e$- and $m$-particles.  However, the statistics can enter when $G$ contains more general space group operations, and in this case the $H^2(G, \zz)$ product can be ``twisted'' by statistics.

We hope that the reader is not discouraged at this stage by the appearance of perhaps unfamiliar mathematics. The necessary terminology and results are explained in a self-contained fashion in Sec.~\ref{sec:math}.  In our opinion, learning this material does not require any special mathematical sophistication.  Group cohomology is certainly more sophisticated, but only the second cohomology group appears, and that only as a convenient name for the group of equivalence classes of group extensions.

\subsection{Prior work}
\label{sec:prior}

The idea that symmetry acts projectively on topological quasiparticles also lies at the heart of X.-G.~Wen's projective symmetry group (PSG) classification of mean-field spin liquid states,\cite{wen02} which is a key inspiration for our work and can be viewed as an attempt to answer some of the same questions.  However, PSG classification, while a very useful tool, does not give a symmetry classification.  PSG classification begins with a parton construction, where, for instance, the spin operator is written as a bilinear of bosonic or fermionic spinon operators.  One then constructs a mean-field theory in terms of the partons, and such mean-field theories are classified by PSG.  Fluctuations about mean-field theory can be incorporated by coupling the partons to a dynamical gauge field, giving a true low-energy effective theory. 

The PSG classification is inherently tied to parton effective theories.  A symmetry classification should be built only on the essential, defining properties of $\zz$ topological order---namely, the types of anyons and their fusion and braiding properties.  Parton constructions provide a concrete means to realize these properties, but there is no reason to believe they do more than this.  Put another way, $\zz$ topological order does not seem to be essentially linked to parton theory, so, in our view, parton theory and PSG do not provide the right language to construct a symmetry classification.  We provide a more detailed discussion contrasting PSG classification with our symmetry classification in Sec.~\ref{sec:psg-comparison}.

We also note that some ideas having significant overlap with ours were outlined previously by A.~Kitaev, in Appendix F of Ref.~\onlinecite{kitaev06}.  In particular, taking the liberty of translating results presented there into the language of this paper, it was asserted that for a general topological order the symmetry classes are given by elements of $H^2(G, \Gamma_2)$, where $G$ is the symmetry group and $\Gamma_2$ is a finite Abelian group determined by the type of topological order.  For Abelian topological orders, $\Gamma_2$ is the group of fusion rules,\cite{kitaevpc} so $\Gamma_2 = \zz \times \zz$ for $\zz$ topological order, agreeing with our results.  In fact, for an arbitrary Abelian topological order, we show in Sec.~\ref{sec:general-abelian} that our approach also reproduces Kitaev's result if $G$ consists only of translations and internal symmetries.  This may also hold for more general space group symmetries, but there are subtleties having to do with the role of braiding statistics that we have only addressed for $\zz$ topological order.

A number of other prior works have also investigated related questions.\cite{ranunpub,wen03,paramekanti04,fwang06,kou08,kou09,yao10,huh11,gchen12,levin12,cho12}
In particular, in Ref.~\onlinecite{wen03}, the idea of using a pair of PSGs, one for $\zz$ charges and one for $\zz$ fluxes, was introduced.  This idea enters our symmetry classification via the need to specify two fractionalization classes (for instance, the $e$ and $m$ fractionalization classes).  Reference \onlinecite{wen03} also showed that distinct pairs of charge and flux PSGs can be realized in the toric code model, in close connection to our analysis of the same model in Sec.~\ref{sec:toriccode}.

\subsection{Outline}

We begin in Sec.~\ref{sec:z2gen} with a review of $\zz$ topological order in two dimensions, introducing many of the basic ideas important for our symmetry classification, as well as much of the notation used in the rest of the paper.  Of particular importance is the concept of superselection sectors.  Next, in Sec.~\ref{sec:z2toric} we briefly review the toric code model,\cite{kitaev03} the simplest concrete realization of $\zz$ topological order.

Sections~\ref{sec:fracclasses} and~\ref{sec:symclasses} present the central results of the paper.  In Sec.~\ref{sec:internal-trans}, the notion of fractionalization class of an anyon is introduced, focusing on the case of translation and internal symmetry.  The notions of symmetry localization and one-particle symmetry operators are also introduced, and play a crucial role.  (One technical detail is relegated to Appendix~\ref{app:non-singlet-gs}.)  We show that for translation symmetry alone there are two fractionalization classes, and similarly for ${\rm U}(1)$ or ${\rm SO}(3)$ symmetry.  Next, in Sec.~\ref{sec:math}, we introduce the mathematical language needed to describe fractionalization classes, followed by a general discussion of the structure of fractionalization classes in Sec.~\ref{sec:fc-gen}.  In Sec.~\ref{sec:fc-gen}, via simple examples, we also explain how fractionalization class information can manifest itself physically in the excitation spectrum and quantum numbers of excited states.  As part of this discussion, we introduce a ``coarsened'' classification by ${\rm U}_T(1)$ fractionalization classes, which reflect information that is in a sense physically simpler than that contained in the full classification.   Because point group operations can move anyons large distances, full space group symmetry requires the further considerations of Sec.~\ref{sec:fc-sg}.  Finally, in Sec.~\ref{sec:square-example}, we work out the fractionalization classes for the example of square lattice space group, time-reversal, and spin rotation symmetry, showing there are $2^{11}$ such classes (some technical details are given in Appendix~\ref{app:genset}).

In Sec.~\ref{sec:sc-results}, we describe our symmetry classification of $\zz$ spin liquids, which amounts to specifying fractionalization classes for the $e$ and $m$ anyons.  The crucial issue is to determine how the $\epsilon$ fractionalization class follows from the $e$ and $m$ classes.  Following a discussion of the counting of distinct symmetry classes, we move on to the case of translation and internal symmetry in Sec.~\ref{sec:sc-internal-trans}, where we show that the $\epsilon$ class is given simply by the $H^2(G, \zz)$ group product of the $e$ and $m$ classes.  We also describe the symmetry classes for the case of translation symmetry alone, and for ${\rm SO}(3)$ spin rotation alone.  Section~\ref{sec:sc-sg} discusses symmetry classes for space group symmetry, where the mutual statistics of $e$ and $m$ particles leads, in general, to a twisting of the group product determining the $\epsilon$ fractionalization class in terms of the $e$ and $m$ classes.  We explicitly work out this twisting for the square lattice space group.

In Sec.~\ref{sec:general-abelian}, we extend our results to general Abelian topological orders for the case of translation and internal symmetry.  In Sec.~\ref{sec:toriccode}, we explicitly construct one-particle symmetry operators for the generators of the square lattice space group in the toric code model, and show that four symmetry classes can be realized there, by tuning the signs of the two terms in the Hamiltonian.  Section~\ref{sec:qndgs} shows that some of the symmetry class information can be extracted directly from the quantum numbers of degenerate ground states, as illustrated for the case of translation symmetry alone.  We conclude with a comparison between our classification and PSG classification in Sec.~\ref{sec:psg-comparison}, and a discussion of open issues and future directions in Sec.~\ref{sec:discussion}.

\section{Review: $\zz$ topological order}
\label{sec:z2}

\subsection{General discussion}
\label{sec:z2gen}

Here, we review $\zz$ topological order.  We employ the language of topological quasiparticle types and the associated superselection sectors.\cite{kitaev06}  The notion of topological superselection sectors is particularly important for our classification.  We begin by introducing these notions abstractly, and then, in Sec.~\ref{sec:z2toric}, illustrate them using the exactly solvable Kitaev toric code model.\cite{kitaev03}  The discussion here focuses on $\zz$ topological order, but can be generalized to arbitrary Abelian topological orders.

We are concerned with local bosonic lattice models with an energy gap in two dimensions.  One way to characterize $\zz$ topological order is by properties of excitations above the ground state.  As discussed briefly in Sec.~\ref{sec:introduction}, there are four topological particle types, denoted $1$, $e$, $m$, and $\epsilon$.  The $e$, $m$, and $\epsilon$ particles are anyons, while the $1$-particles (``trivial'' particles) are not.  Under exchange, all the particle types are bosons except $\epsilon$, which is a fermion.  Excluding $1$-particles, any pair of distinct particles have $\theta = \pi$ mutual statistics.  $1$-particles have trivial mutual statistics with the other particle types.

The particles also obey the fusion rules given in Eq.~(\ref{eqn:fusion}).
For example, the $e \times e = 1$ fusion rule expresses the fact that two nearby $e$-particles can be viewed as a single $1$-particle.  Because the fusion and braiding rules are invariant under $e \leftrightarrow m$, we are always free to relabel $e \leftrightarrow m$ if we wish. Other Abelian topological orders can be described in the same way; that is, one specifies a set of particle types, fusion rules, and both exchange and mutual statistics.

At this point, it is useful to introduce some terminology.  We consider a region $R$ defined as some subset of all lattice sites.  Without worrying too much about precision, we also assume that $R$ is has no small holes or rough edges.  That is, we want to be able to define $R$ by drawing one or more sufficiently smooth boundary curves, and selecting the lattice sites in the interior.  We will almost always assume $R$ is a union of disjoint simply connected regions.
We denote the complement of $R$ by $\bar{R}$.  The full Hilbert space is the tensor product ${\cal H}(R) \otimes {\cal H}(\bar{R})$, where ${\cal H}(R)$ is the Hilbert space of region $R$.   We say an operator ${\cal O}$ is supported on $R$ if it can be written ${\cal O} = {\cal O}_R \otimes 1_{\bar{R}}$.  That is, if ${\cal O}$ is supported on $R$, it may act non-trivially on $R$, but acts as the identity operator on $\bar{R}$.  

It is a crucially important defining property that no local operator can create a single isolated anyon. 
However, for instance, a pair of $e$-particles can be created locally due to the fusion rule $e \times e = 1$, since isolated $1$-particles can be created locally.  Two $e$-particles created in this way can then be separated to obtain isolated $e$-particles.  This separation can be accomplished by acting with a string operator, which is supported on a linear region connecting the initial and final positions of a single $e$-particle.  String operators only modify locally observable properties of a state on which they act near the ends of the string.  That is, there is no way to discern that a string operator has been applied to some state by making local measurements along the length of the string, away from the ends.  String operators need not have ends, and can form closed loops.  
Noncontractible closed strings are related to the topological ground-state degeneracy, which we discuss below.

\begin{figure}
\includegraphics[width=1.0\columnwidth]{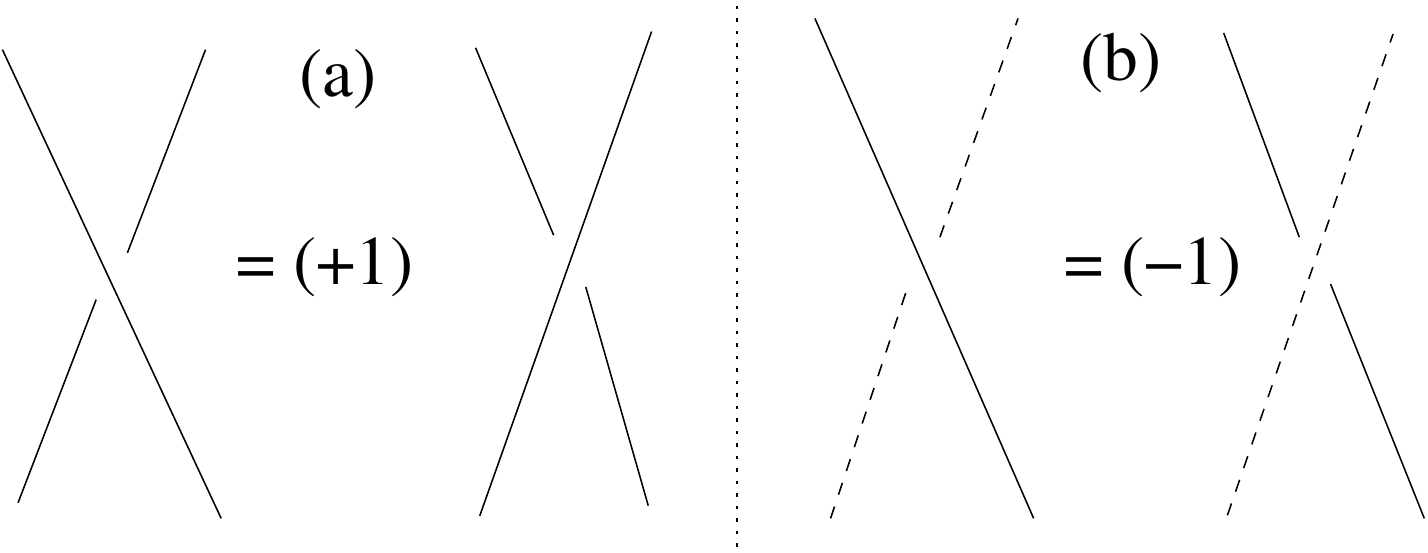}
\caption{(a) Two $e$-string operators (solid lines)  with a single crossing point commute.  (b) $e$-string (solid line) and $m$-string (dashed line) operators with a single crossing point anticommute.}
\label{fig:stringcomm}
\end{figure}

There are three distinct types of string operators, associated with the three types of anyons.  For instance, the string operators associated with $e$-particles are referred to as $e$-strings. Two string operators of the same type commute [see Fig.~\ref{fig:stringcomm}(a)]. This holds as long as their ends are well separated; if that is not the case, the commutation relations will depend on details of the ends.  On the other hand, two string operators of different types anticommute if they cross an odd number of times and commute if they cross an even number of times [see Fig.~\ref{fig:stringcomm}(b)].  This property, which follows from the mutual statistics, can be expressed by saying that we get a minus sign whenever we move a string of one type through a string of another type, at a single crossing point.  In Fig.~\ref{fig:stringcomm}, and throughout the paper, we adopt the graphical convention that strings are drawn on top when the corresponding operator lies to the left in a product of operators.  That is, the strings on the bottom act first on a state, followed by strings on top.

\begin{figure}
\includegraphics[width=1.0\columnwidth]{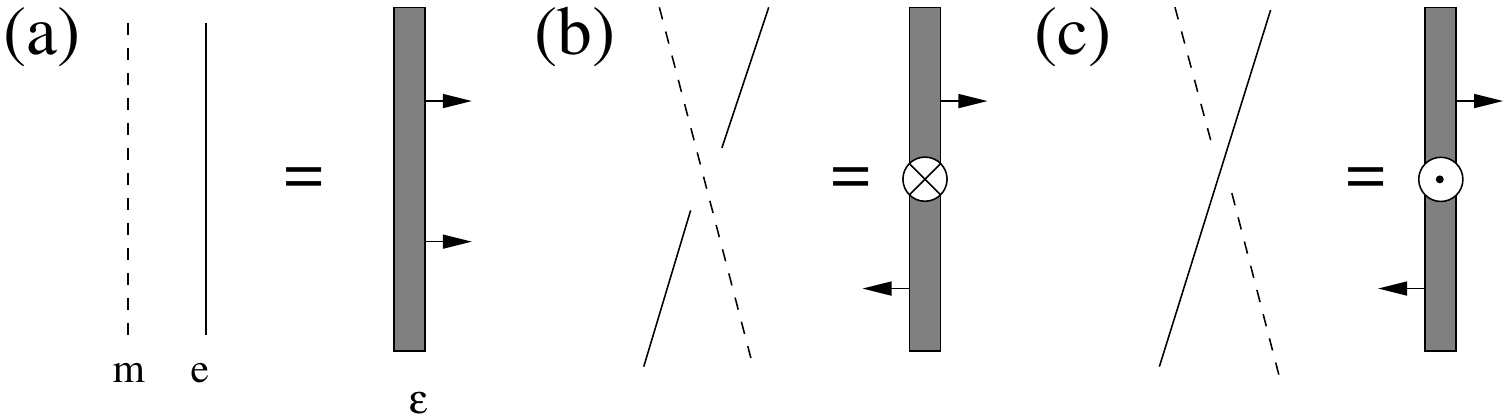}
\caption{(a) $e$-string (solid line) and $m$ (dashed line) strings can be viewed in combination as an $\epsilon$-string.  The arrow points from the $m$ string toward the $e$-string.  (b) Twisted $\epsilon$ string where the $e$ string passes underneath the $m$-string.  (c) Twisted $\epsilon$ string where the $e$ string passes over the $m$ string.  Note that the configurations in (b) and (c) are related by a minus sign.}
\label{fig:epsilon-string}
\end{figure}

An $\epsilon$ string can be viewed as a pair of nearby, parallel $e$ and $m$ strings, as depicted in Fig.~\ref{fig:epsilon-string}(a).  Since the $e$ string has to lie on one side or the other, $\epsilon$ strings thus carry an orientation [see Fig.~\ref{fig:epsilon-string}(a)].  The orientation changes at a point where one of the $e$ or $m$ constituent strings passes under the other; at such a point we say the $\epsilon$ string is twisted.  There are two kinds of twists, 
since the $e$-string can pass over or under the $m$-string [see Figs.~\ref{fig:epsilon-string}(b) and \ref{fig:epsilon-string}(c)]; these two twists are related by a minus sign.

We will primarily be interested in considering $\zz$ topologically ordered states on a torus (\emph{i.e.}, with periodic boundary conditions).  In this case, we can form non-contractible loops in both $x$- and $y$-directions.
Let $\cL_x^e$ be a closed $e$-string operator winding once around the system in the $x$-direction, with corresponding definitions for $\cL_y^e$, $\cL_x^m$, and $\cL_y^m$.  We can think of the product $\cL_x^e \cL_x^m$ as an $\epsilon$ string running in the $x$ direction, so it is not necessary to introduce more operators to represent non-contractible $\epsilon$-strings.  These string operators satisfy the following commutation and anticommutation relations:
\begin{align}
\{ \cL_x^e, \cL^m_y \} = \{ \cL_x^m, \cL^e_y \} &= 0, \notag\\
[ \cL_x^e , \cL_y^e ] = [ \cL_x^m , \cL_y^m ] 
= [ \cL_x^e , \cL_x^m ] = [ \cL_y^e , \cL_y^m ] &= 0 . \label{eqn:loop1}
\end{align}
We also assume
\begin{equation}
(\cL_x^e)^2 = (\cL_y^e)^2 = (\cL_x^m)^2 = (\cL_y^m)^2 = 1 \text{.} \label{eqn:loop2}
\end{equation}
This can be justified by noting that, for instance, moving a pair of nearby $e$-particles (equivalent to a 1-particle) around a closed loop should be equivalent to doing nothing, except perhaps accumulating a phase that can be removed by a trivial redefinition of the string operators.

We refer to Eqs.~(\ref{eqn:loop1}) and (\ref{eqn:loop2}) as the loop algebra.  This algebra has a single, four-dimensional irreducible representation, and this implies that the ground state on a torus must be fourfold degenerate.  We note that there are situations where the loop algebra and the topological ground-state degeneracy are both modified by choice of boundary conditions,\cite{you12} but we will not consider such cases.

Associated to each of the four particle types is a topological superselection sector.  To understand what this means, it is first helpful to think about a single isolated and localized $e$-particle in an infinite plane, where no other excitations are present.  Starting from such a state, we define an arbitrarily large but finite connected region $R^e$ containing the $e$-particle.  It is important that the system be ``locally in the ground state'' near the boundary of $R^e$ (and outside of $R^e$), in the sense that local measurements in these areas should give the same result as in the ground state.  Then we can obtain all states in the $e$-sector by acting with (almost)\cite{opR} arbitrary operators supported on $R^e$.
The resulting states correspond to moving the $e$-particle to different positions, ``dressing'' it in various ways, modifying any internal quantum numbers it may carry, and so on.  The $e$-sector is thus closed under action of operators supported on $R^e$.  Moreover, no operator supported on $R^e$ can act on an $e$-sector state and turn it into a state belonging to a different superselection sector.

We will also have occasion to consider regions that are not connected.  To handle this situation, we make the following definition:  an $s$ operator on $R$ is an operator that, restricted to $\bar{R}$, consists entirely of string operators (of any type) connecting disconnected components of $R$.   If an $s$-operator on $R$ has only, say, $e$ strings in $\bar{R}$, we call such an operator an $e$ operator on $R$.  Again, if the region $R^e$ contains an isolated $e$ particle and no other excitations, we can obtain all $e$-sector states (for the region $R^e$) by acting with $s$-operators on $R^e$, and the $e$-sector is closed under the action of such operators.

\begin{figure}
\includegraphics[width=1.0\columnwidth]{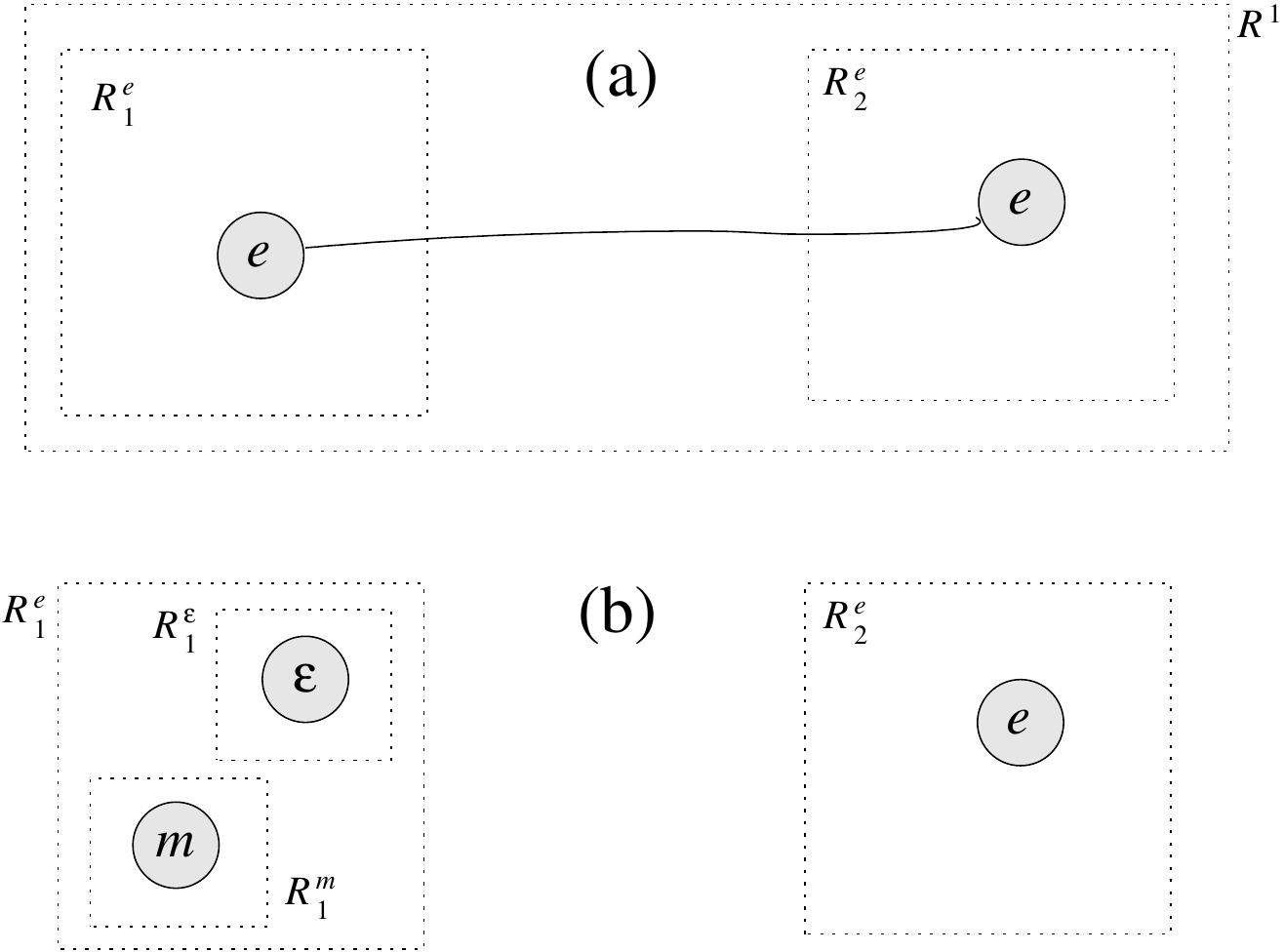}
\caption{(a)  State with two isolated $e$-particles, with $e$-sector regions $R^e_1$ and $R^e_2$.  The $e$-string connecting the particles is shown as a solid line.  The two regions can be combined together to give the $1$-sector region $R^1$.  (b)  State as in (a), but where the $e$-particle in $R^e_1$ has split into isolated $m$ and $\epsilon$ particles as allowed by the $e = m \times \epsilon$ fusion rule.  $R^e_1$ can thus be subdivided into $R^m_1$ and $R^\epsilon_1$ as shown.  Strings connecting the particles are not shown.}
\label{fig:regions}
\end{figure}

This discussion needs to be modified on a finite torus, where non-trivial particles must occur in pairs.  Indeed, in this situation, all physical states belong to the $1$-sector, because any state can be obtained from a ground state by acting with operators supported on the whole system.  To see that the superselection sectors still have meaning here, consider a state with two localized and well-separated $e$-particles, with no other excitations present [see Fig.~\ref{fig:regions}(a)].  Any operator acting on a ground state to create such a two-particle state will involve an $e$-string connecting the positions of the two particles; therefore it is useful to think of the particles as being connected by an $e$-string.

Now, as illustrated in Fig.~\ref{fig:regions}(a), we define two regions $R^e_1$ and $R^e_2$ by drawing a box around each $e$-particle.  The boundaries of these regions, and the space outside the regions, should be locally in the ground state.  From such a reference state, acting with arbitrary operators supported on, say $R^e_1$, one generates all $e$-sector states in the region $R^e_1$.

More generally, we can decompose a state into regions $R^1_i$, $R^e_i$, $R^m_i$, and $R^\epsilon_i$.  Such a decomposition is not unique and can be modified according to the fusion rules.  For instance, going back to the example of two $e$-particles,  if we draw a larger box containing both $R^e_1$ and $R^e_2$, the resulting new region $R^1$ contains a state in the $1$-sector due to the fusion rule $e \times e = 1$.  On the other hand, due to the  $e = m \times \epsilon$ fusion rule, there can be a state in $R^e_1$ consisting of a localized $m$-particle well-separated from a localized $\epsilon$-particle, with no other excitations.  In this case, we can subdivide $R^e_1$ into $R^\epsilon_1$ and $R^m_1$, as shown in Fig.~\ref{fig:regions}(b).

\subsection{Toric code model}
\label{sec:z2toric}

\begin{figure}
\includegraphics[width=0.6\columnwidth]{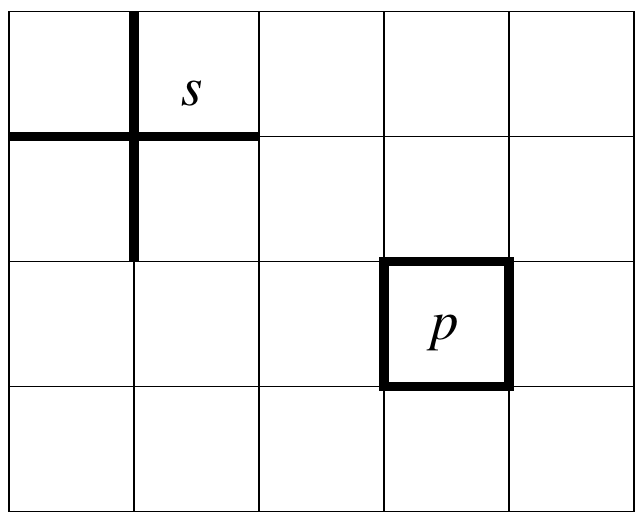}
\caption{Thick bonds depict the four edges meeting the vertex $s$ and bounding the plaquette $p$.}
\label{fig:toric}
\end{figure}

The toric code is a spin model that makes manifest all the essential features of the $\zz$ theory with a minimum of extra structure.   It is therefore a very useful testbed for our discussion, and in Sec.~\ref{sec:toriccode}, we will present explicit constructions that give substance to the general considerations of Secs.~\ref{sec:fracclasses} and \ref{sec:symclasses}.  The model can be defined on any lattice in two dimensions, but we restrict ourselves to the square lattice. 

The model consists of spin-1/2 degrees of freedom on the edges of the square lattice.  We label lattice sites by $\r$, and write Pauli matrices acting on the nearest-neighbor edge $(\r, \r')$ as $\sigma^z_{\r, \r'}$, and so on.  Four edges meet at each site to form a vertex, denoted $s$,  and four edges bound each plaquette, denoted $p$ (see Fig.~\ref{fig:toric}).  The Hamiltonian is built from the following products, associated respectively with vertices and plaquettes:
\beq
A_s = \prod_{(\r, \r')  \in s} \sigma^x_{\r, \r'}, \qquad B_p = \prod_{(\r, \r') \in p} \sigma^z_{\r, \r'}.
\eeq
These operators can be viewed as measuring $\zz$ charge and flux, respectively.  The Hamiltonian is
\beq
H_{tc} = -K_e \sum_s A_s - K_m \sum_p B_p \text{.}
\eeq 
The exact eigenstates of $H_{tc}$ are easily constructed,  because $[H_{tc} , A_s] = [H_{tc}, B_p] = [ A_s, B_p ] = 0$.

\begin{figure}
\includegraphics[width=0.7\columnwidth]{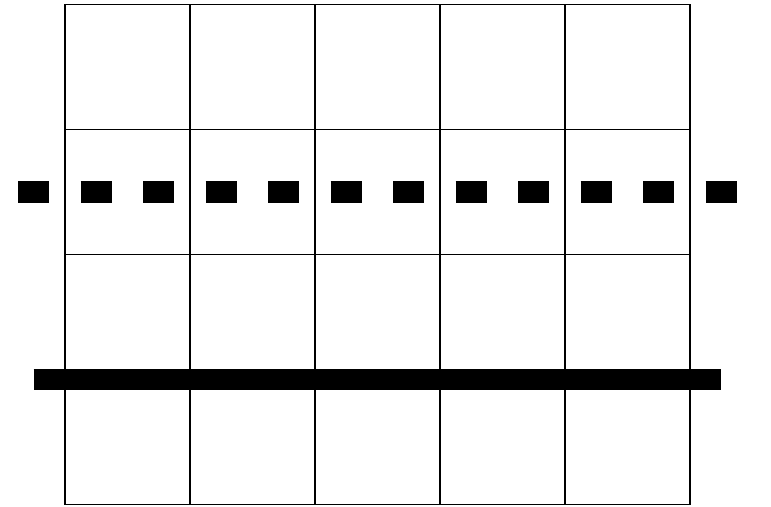}
\caption{Depiction of contours $C^e_x$ (thick solid line) and $C^m_x$ (thick dashed line) used to define the loop operators ${\cal L}^e_x$ and ${\cal L}^m_x$, respectively.}
\label{fig:loop-contours}
\end{figure}

Here we assume $K_e,K_m>0$, although in Sec.~\ref{sec:toriccode} we consider more general situations.  Any ground state satisfies $A_s = B_p = 1$.  On a finite torus there are four such ground states, which can be seen by explicitly constructing the loop operators as products of Pauli matrices,
\begin{eqnarray}
\cL^e_{x,y} &=& \prod_{(\r, \r') \in C^e_{x,y}} \sigma^z_{\r,\r'} \\
\cL^m_{x,y} &=& \prod_{(\r, \r') \perp C^m_{x,y} } \sigma^x_{\r, \r'} \text{,}
\end{eqnarray}
as illustrated in Fig.~\ref{fig:loop-contours}.  Here the contours $C^e_{x,y}$ consist of a path of lattice edges winding around the system in the $x,y$ directions, where the $C^m_{x,y}$ contours pass through perpendicular edges. It is straightforward to check that these operators satisfy the loop algebra.  More generally, $e$ ($m$) strings are products of $\sigma^z_i$ ($\sigma^x_i)$ along an appropriate contour.

We can be even more explicit by constructing the state
\begin{equation}
| \psi_0 \rangle = \Big[ \prod_s \tfrac{1}{\sqrt{2}} \big(1 + A_s \big) \Big] | \{ \sigma^z_i = 1\} \rangle \text{.}
\label{eqn:uniformstate}
\end{equation}
This is easily seen to be a ground state, and $\cL^e_{x,y} | \psi_0 \rangle = |\psi_0 \rangle$.  However, this state is not an eigenstate of $\cL^m_{x,y}$, and the other three ground states are $\cL^m_x | \psi_0 \rangle$,  $\cL^m_y | \psi_0 \rangle$ and  $\cL^m_x \cL^m_y | \psi_0 \rangle$.

Along the same lines, we can also construct contractible $e$- and $m$-strings.  Any such $e$-string can be written as a product of $B_p$ operators; a single $B_p$ operator is an elementary contractible $e$-string encircling a single plaquette.  Similarly, contractible $m$-string operators are products of $A_s$ operators.  Therefore, the ground states are eigenstates of all contractible string operators, with eigenvalue unity.

There is a gap to excited states, which have vertices where $A_s = -1$ and/or plaquettes where $B_p = -1$.  Vertices with $A_s = -1$ are $e$-particles, and plaquettes with $B_p = -1$ are $m$-particles.  (Again,  which we call $e$ and which $m$ is arbitrary.)  To create, for instance, a state with two isolated $e$-particles, one can act on a ground state with a product of $\sigma^z_i$ Pauli matrices (an $e$-string) on a contour connecting the desired particle positions.  Based on the above discussion, it is clear that we can ``slide around'' the string connecting the two particles with no effect whatsoever on the state.  This statement needs to be weakened slightly when we consider states with both $e$ and $m$-particles; in such cases moving strings around can change the state by a minus sign, either when we bring an $e$-string through an $m$-string, or when a string of one type ``slides over'' a particle of the other type.  Still, the string positions are clearly unobservable.

It is important to recognize that the toric code model is highly fine-tuned.  In particular, the quasiparticles have no dynamics (dispersion) and a vanishing correlation length.  These features are convenient for our study when we come to an explicit implementation in Sec.~\ref{sec:toriccode}.  To consider the generic properties of a phase of matter, however, one must allow all finite-range terms consistent with  symmetry  to be added to the Hamiltonian.  This will introduce one or more time scales beyond which the quasiparticles should not be considered isolated.  In the absence of such processes the fusion rules would have little physical relevance, but we will find that, in general, the fusion rules play a crucial role in determining the possible fractionalization classes.

\section{Fractionalization Classes}
\label{sec:fracclasses}

In this section, we will discuss and classify the action of symmetries on a single type of anyon in the $\zz$ theory.  Most of the discussion will focus on $e$ particles, with occasional comments on $\epsilon$ particles, due to the different nature of $\epsilon$ strings.  Everything we say also clearly holds for $m$ particles, since the topological properties do not change under relabeling $e \leftrightarrow m$.  

We assume that symmetry operations do not change one type of anyon into another.  The symmetry group may consist of internal symmetries (including anti-unitary time reversal), translation symmetry, and general space group operations.  We begin by introducing the notion of fractionalization classes for  translation  and internal symmetry (see Sec.~\ref{sec:internal-trans}). Next, introducing the mathematics of group extensions and their equivalence classes (see Sec.~\ref{sec:math}), we discuss the general structure of fractionalization classes (see Sec.~\ref{sec:fc-gen}).  We then show that the same general structure continues to hold for space group symmetry in Sec.~\ref{sec:fc-sg}.

\subsection{Translation and internal symmetries}
\label{sec:internal-trans}

To introduce the notion of fractionalization classes, we begin with translation symmetry, then argue that the same structure holds for internal symmetry (including time reversal).  We discuss $e$-particles for concreteness, but all statements apply just as well to $\epsilon$-particles, except where explicitly noted.  

\begin{figure}
\includegraphics[width=1.0\columnwidth]{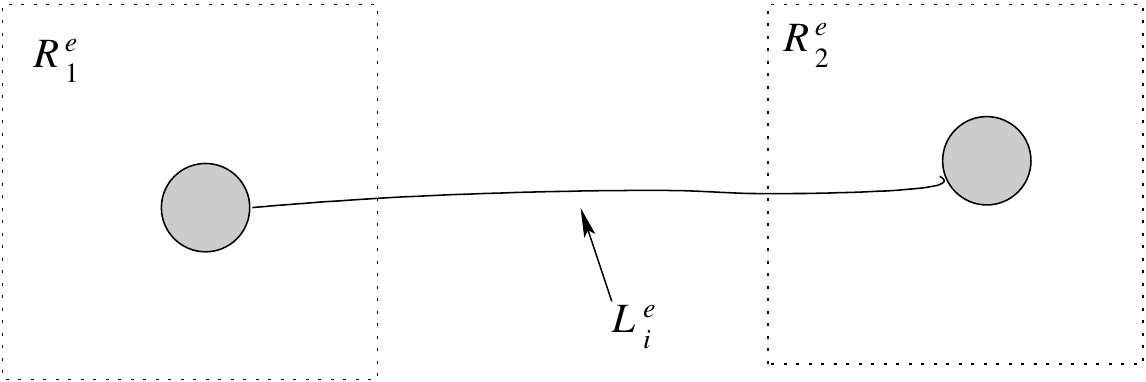}
\caption{Depiction of the state $| \psi_\alpha \rangle$ with two $e$-particles.  The operator ${\cal O}_\alpha$ creating this state is supported on the two shaded circular regions and along the solid line connecting them.  ${\cal O}_\alpha$ is an $e$-string operator along the solid line, which is referred to as $L^e_i$ outside of the regions $R^e_i$.  ${\cal O}_\alpha$ is thus an $e$-operator on the union of the $R^e_i$.}
\label{fig:2e-state}
\end{figure}

We consider translation symmetry generated by $T_x$ and $T_y$, satisfying the relation $T_x T_y T^{-1}_x T^{-1}_y = 1$, which holds acting on all physical states (see Sec.~\ref{sec:introduction}), and thus on the $1$-sector.  We wish to understand the action of translations on states with two localized, well separated $e$-particles.  More formally, we consider a family of states $ \{ | \psi_\alpha \rangle \}$ (labeled by $\alpha$),  that can be decomposed into two fixed, connected $e$-sector regions $R^e_i$ ($i = 1,2$), as described in Sec.~\ref{sec:z2gen}, and are otherwise locally in the ground state (see Fig.~\ref{fig:2e-state}).  
We also assume that $|\psi_\alpha \rangle = {\cal O}_{\alpha} |\psi_0 \rangle$, where $| \psi_0 \rangle$ is a ground state.
The operator ${\cal O}_\alpha$ is an $e$-operator on the union of $R^e_{1}$ and $R^e_2$, with an $e$-string running between the two components. For simplicity, we assume that $| \psi_0 \rangle$ satisfies $T_x | \psi_0 \rangle = T_y | \psi_0 \rangle = | \psi_0 \rangle$.  This assumption is not necessary and we describe how it can be relaxed in Appendix~\ref{app:non-singlet-gs}.  Any desired combination of $e$-sector states in the two regions $R^e_i$ can be produced by the above construction.

We proceed by making the crucial assumption that symmetry operations can be localized to regions surrounding the excitations in the states $|\psi_\alpha \rangle$.  We refer to this property as \emph{symmetry localization}.  The basic idea is that a symmetry operation (such as translation) changes the $e$-particle state in one of the regions to another $e$-particle state in the same region, and that it should be possible to accomplish such a change locally.
Formally,
\begin{equation}
T_x | \psi_\alpha \rangle = T^e_x(1) T^e_x(2) | \psi_\alpha \rangle \text{,} \label{eqn:product}
\end{equation}
where $T^e_x(i)$ is supported on $R^e_i$.  The operators $T^e_x(i)$ are independent of $\alpha$.  The corresponding statements are also assumed for $T_y$. The operator $T^e_x(1)$, for instance, can be interpreted as a ``one-particle'' translation operator, that translates the $e$-particle in region $R^e_1$ against the translation-invariant medium of the ground state $|\psi_0\rangle$.  It is straightforward to generalize this discussion to a state with multiple $e$-sector regions.  We further examine and justify the assumption of symmetry localization at the end of this section.

We note that symmetry localization fails when a symmetry operation changes one type of anyon into another type, because this cannot be accomplished by any local operator.  However, we exclude this situation by assumption.  

Now we have
\begin{eqnarray}
| \psi_\alpha \rangle &=& T_x T_y T^{-1}_x T^{-1}_y | \psi_\alpha \rangle \\
&=&  \prod_{i = 1,2} \big( T^e_x(i) T^e_y(i) [T^e_x(i)]^{-1} [T^e_y(i)]^{-1} \big)  | \psi_\alpha \rangle \text{.} 
\nonumber
\end{eqnarray}
For this to hold for all $\alpha$, we must have 
\begin{equation}
T^e_x(i) T^e_y(i) [T^e_x(i)]^{-1} [T^e_y(i)]^{-1} 
= e^{i \phi_i} \text{.}
\end{equation}
If this were not true, there would be a state $| \psi_\alpha \rangle$ on which the identity operator acts nontrivially.  We have $e^{i \phi_1} e^{i \phi_2} = 1$, which is a consequence of the fusion rule $e \times e = 1$.  Since there is no  difference between the two regions, we also expect $e^{i \phi_1} = e^{i \phi_2}$.  To show this, consider instead a state with four $e$-sector regions $R^e_i$ ($i = 1,\dots,4$), with $e^{i \phi_i}$ defined as above.  Then for any pair $i \neq j$, we can fuse $R^e_i$ and $R^e_j$ to obtain a $1$-sector region, acting on which we must have $T_x T_y T^{-1}_x T^{-1}_y = 1$, implying $e^{i \phi_i} e^{i \phi_j} = 1$.  This implies the $e^{i \phi_i}$ are all equal and  $e^{i \phi_i} = \pm 1$. Therefore, dropping the label distinguishing the two regions, we write
\begin{equation}
T^e_x T^e_y (T^e_x)^{-1} (T^e_y)^{-1} \equiv \sigma^e_{txty} = \pm 1 \text{.}
\end{equation}

The parameter $\sigma^e_{txty}$ defines the fractionalization class of the $e$-sector.  Evidently, there are two fractionalization classes in the case of translation symmetry alone.  It is important to emphasize, as follows from the discussion above, that $\sigma^e_{txty}$ is constant on the $e$-sector.  Putting essentially the same argument in  different terms, suppose there is one type of $e$-particle with $\sigma^e_{txty} = 1$ and another with $\sigma^e_{txty} = -1$.  We could then fuse these to obtain a $1$-particle acting on which $T_x T_y T^{-1}_x T^{-1}_y = -1$, a contradiction.  Since  $\sigma^e_{txty}$ is discrete and is constant on the $e$-sector, it cannot change within a $\zz$ spin liquid phase, so long as translation symmetry is preserved.  Therefore, $\sigma^e_{txty}$ is a universal property of $\zz$ spin liquids with translation symmetry.

There is some arbitrariness in the definition of $T^e_x$.  Looking at Eq.~(\ref{eqn:product}), clearly we can redefine $T^e_x(i) \to - T^e_x(i)$ with no physical effect.   More generally, the redefinition $T^e_x(i) \to e^{i \phi} T^e_x(i)$ sends $T_x \to e^{i n \phi} T_x$, acting on a state with $n$ $e$-particles.  This transformation should leave $T_x$ unchanged (apart from possible overall multiplication by a phase, independent of $n$), which only occurs when $\phi = 0, \pi$, in which case $T_x \to T_x$ since $n$ is even.  Therefore we are allowed to redefine $T^e_x(i) \to - T^e_x(i)$, and similarly $T^e_y(i) \to - T^e_y(i)$.  Note that such redefinitions do not affect $\sigma^e_{txty}$.

Now we generalize the above discussion to the case of unitary internal symmetry, perhaps also combined with translation symmetry.  By internal symmetry, we roughly mean any symmetry operation that does not move the lattice.  This includes on-site symmetries such as spin rotation, but we need not limit ourselves to strictly on-site symmetry.  If $S$ is a symmetry operation, we again assume symmetry localization, that is,
\begin{equation}
S | \psi_\alpha \rangle = S^e(1) S^e(2) | \psi_\alpha \rangle \text{,}
\end{equation}
where $S^e(i)$ is supported on $R^e_i$.
The logic is identical to the case of translation: it should be possible to accomplish the operation $S$ by making local modifications near the two quasiparticle excitations.  Again, we are free to redefine $S^e \to - S^e$ with no effect on the physics.  Suppose we have a relation among symmetry operations of the form $S_1 S_2 \cdots S_k = 1$.  Then, following the arguments above,
\begin{equation}
S^e_1 S^e_2 \cdots S^e_k = \pm 1 \text{.}
\end{equation}
Specifying such $\zz$-valued parameters for all group relations among symmetry operations specifies the fractionalization class of the $e$-sector.  At the present stage of the discussion, it may not be clear how to make this last statement precise.  This can be accomplished in a straightforward fashion after the discussion of the following section, where the mathematics of group extensions and their equivalence classes is introduced.

It is worth explicitly discussing the particularly simple and familiar cases of ${\rm U}(1)$ and ${\rm SO}(3)$ internal symmetry.  In the case $G = {\rm U}(1)$,  $1$-sector states (\emph{i.e.}, physical states) carry integer ${\rm U}(1)$ charge, or more generally, they are superpositions of states with different integer ${\rm U}(1)$ charges.    Alternatively, denoting with $R(\theta)$ a ${\rm U}(1)$ rotation by angle $\theta$, we have $R(2\pi) = 1$.  On the $e$-sector, however, we may have $R^e(2\pi) = \pm 1$, corresponding to integer ($+1$) and half-odd integer ($-1$) ${\rm U}(1)$ charges.  These are the only two fractionalization classes.  For instance, $e$-particles cannot have other charges (\emph{e.g.}, $1/3$ charge), since combining two $e$-particles must always give an integer charge due to the $e \times e = 1$ fusion rule.  Moreover, $e$-particles with charge $1/2$ and charge $3/2$ are not distinct classes;  starting with a charge-$1/2$ $e$-particle, one can fuse it with a charge-$1$ $1$-particle to obtain a charge-$3/2$ $e$-particle.  Therefore, charge-$1/2$ and charge-$3/2$ $e$-particles always appear together in the spectrum.  This example points out that fractionalization classes are \emph{not} simply distinct irreducible representations of the symmetry group, and are instead a coarser type of classification.  The situation for $G = {\rm SO}(3)$ spin rotation is similar.  Denoting with $R_s(\theta \hat{n})$ a spin rotation by $\theta$ about the $\hat{n}$-axis, on the $e$-sector we have $R^e_s( 2\pi \hat{n} ) = \pm 1$, corresponding to the two fractionalization classes of integer spin ($+1$) and half-odd-integer spin ($-1$).

Finally, we consider the case of anti-unitary time reversal, which can be written ${\cal T} = U_T K$, where $K$ is complex conjugation and $U_T$ is a unitary operator.  Complex conjugation is a global operation on a wave function and cannot sensibly be localized to a region, so in this case symmetry localization takes the form
\begin{equation}
{\cal T} | \psi_\alpha \rangle = U^e_T(1) U^e_T(2) K | \psi_\alpha \rangle \text{,}
\end{equation}
where $U_T(i)$ is supported on $R^e_i$.
The relation ${\cal T}^2 = 1$ implies
\begin{equation}
U^e_T (U^e_T)^* \equiv ( {\cal T}^e)^2 =  \pm 1 \text{.}
\end{equation}
Here, in the interest of concise notation, we have made a formal definition of $( {\cal T}^e )^2$.

We can also consider relations involving time-reversal and unitary symmetry operations.  For instance, suppose ${\cal T} S {\cal T}^{-1} S^{-1} = 1$ for some symmetry operation $S$.  This implies
\begin{equation}
U^e_T (S^e)^* (U^e_T)^{-1} (S^e)^{-1} \equiv {\cal T}^e S^e  ({\cal T}^e)^{-1} (S^e)^{-1} = \pm 1 \text{,}
\end{equation}
where again the expression involving ${\cal T}^e$ is a formal definition.

\begin{figure}
\includegraphics[width=1.0\columnwidth]{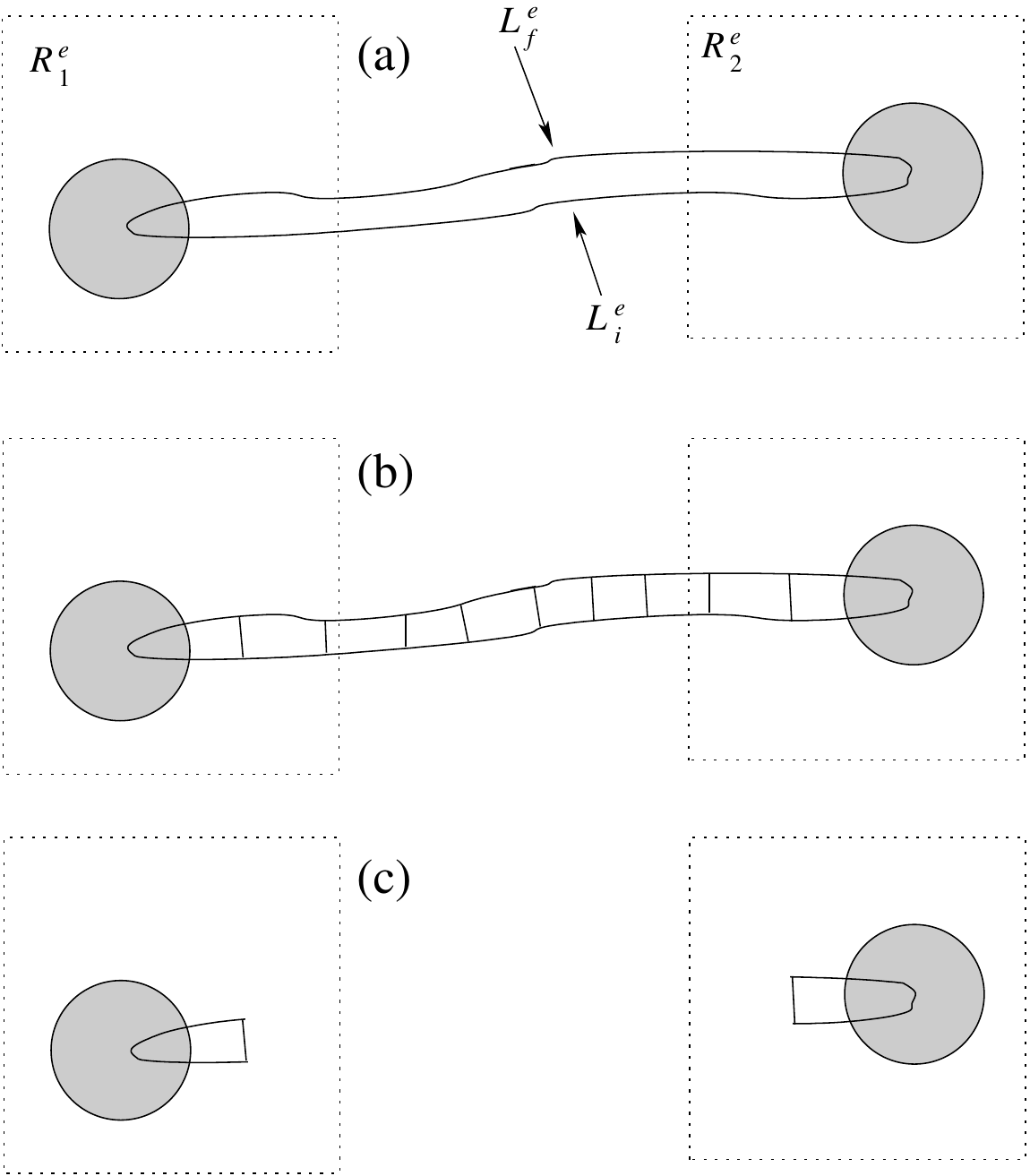}
\caption{Depiction of the action of a symmetry operation $S$ on a two $e$-particle state, where $S$ is either translation or an internal symmetry operation.  The initial state on which $S$ acts is shown in Fig.~\ref{fig:2e-state}.  $S$ acts nontrivially in the shaded circular regions, and also as the closed $e$-string loop (solid line), as shown in (a).  Outside the regions $R^e_i$, $L^e_i$ is the $e$-string of the initial state $ | \psi_\alpha \rangle$, and $L^e_f$ is the $e$-string of the transformed state $S | \psi_\alpha \rangle$.  In (b), the $e$-string loop is divided into a product of smaller loops, chosen so that each gives unity acting on the ground state, except possibly the two loops at the ends.  Therefore, the loops in the center can be eliminated (c), leaving only two loops contained entirely within the regions $R^e_i$, which can be absorbed into the definition of $S^e(i)$.}
\label{fig:2e-transform}
\end{figure}

We now discuss the assumption of symmetry localization in more detail.  Let $R$ be the union of the $R^e_i$ regions, and $\bar{R}$ the complement of $R$. In addition to acting on the $e$-particles in $R$ with some symmetry operation $S$, we also have to act on the $e$-string operator in $\bar{R}$.  Now, moving or otherwise modifying the string can only result in an overall phase, but the question is then whether this phase always factors into a product of two phases associated with each region.  In general, we expect
\begin{equation}
S | \psi_\alpha \rangle = S^e(1) S^e(2) L^e(1,2) | \psi_\alpha \rangle \text{,}
\end{equation}
where $L^e(1,2)$ is a closed $e$-string loop that accomplishes the necessary transformation of the $e$-string in $\bar{R}$ [see Fig.~\ref{fig:2e-transform}(a)].  Restricted to $\bar{R}$, $L^e(1,2) = L^e_f L^e_i$, where $L^e_i$ is identical to ${\cal O}_{\alpha}$ restricted to $\bar{R}$; that is, $L^e_i$ is the $e$-string of the ``initial'' state $| \psi_\alpha \rangle$.  $L^e_f$ is the desired ``final'' state $e$-string.  Since $L^e(1,2)$ is a contractible loop operator, it can be written as a product of smaller contractible loops of $e$-string as shown in Fig.~\ref{fig:2e-transform}(b).  We assume that $e$-string operators square to unity, so the elementary contractible loop operators have eigenvalues $\pm 1$.  Assuming some degree of spatial homogeneity, by combining and splitting the elementary loops as needed, it should be possible to choose all the elementary loops to have eigenvalue unity, except possibly those at the ends. Therefore, $L^e(1,2)$ can be broken in the middle and deformed as shown in Fig.~\ref{fig:2e-transform}(c), without accumulating any phase factors.  The remaining loops on the two ends can then be absorbed into $S^e(1)$ and $S^e(2)$, and symmetry localization holds.

The discussion of transforming the string also goes over to the case of $\epsilon$ particles, where the string is oriented.  It is still possible to construct the necessary closed $\epsilon$-string loop operator $L^\epsilon(1,2)$ as above for $e$-particles.  The only difference from the case of $e$-particles is that, depending on the relative orientations of the $L^\epsilon_i$ and $L^\epsilon_f$ strings, it may be necessary for the $L^\epsilon(1,2)$ to be twisted near the ends, inside the two $\epsilon$-regions.  Otherwise, the discussion proceeds exactly in the $e$-particle case.

\subsection{Group extensions and their equivalence classes}
\label{sec:math}

Here, we give an account of group extensions and their equivalence classes. For the most part, we find it clearer to separate the mathematics from the physics, so the discussion here focuses on the mathematics. In learning this mathematics, we found it useful to consult Refs.~\onlinecite{hamermeshbook,murthy66, maclanebook, vandenbroek76, chen11a,chen11b}.
 Application of the mathematics to the physics of fractionalization follows in Sec.~\ref{sec:fc-gen}.  
 
Consider a group $G$ with elements $g \in G$.  We consider a projective  representation, where the group element $g$ is represented by a unitary matrix $\Gamma(g)$.  (We include anti-unitary group operations below.)  When multiplying two $\Gamma$'s we have
\begin{equation}
\Gamma(g_1) \Gamma(g_2) = \omega(g_1, g_2) \Gamma(g_1 g_2) \text{,} \label{eqn:fs-defn}
\end{equation}
where $\omega(g_1, g_2) \in {\rm U}(1)$ is a phase.  The presence of these ${\rm U}(1)$ phases is what it means for the representation to be projective.  An ``ordinary'' representation where $\omega(g_1, g_2) = 1$ for all $g_1, g_2$ is referred to as a linear representation.  We allow for the possibility that the projective representation $\Gamma$ may be a linear representation; that is, any linear representation is a projective representation, but not vice-versa.  To connect with the discussion of Sec.~\ref{sec:internal-trans}, $\Gamma$ arises physically as the action of the symmetry group on one of the superselection sectors.

We  restrict $\omega(g_1, g_2) \in A$, where $A$ is a subgroup of ${\rm U}(1)$.  
In the physical applications of this paper, we will be interested in the case $A = \zz$.  The function $\omega(g_1, g_2)$ satisfies an associativity constraint, because
\begin{eqnarray}
\Gamma(g_1) \Gamma(g_2) \Gamma(g_3) &=& \omega(g_1, g_2) \omega(g_1 g_2, g_3) \Gamma(g_1 g_2 g_3)  \\
&=& \omega(g_1, g_2 g_3) \omega(g_2, g_3)  \Gamma(g_1 g_2 g_3) \text{,}  \nonumber
\end{eqnarray}
where the two results are obtained by the two different ways of using associativity to evaluate the product of three $\Gamma$'s.  The associativity constraint is then
\begin{equation}
 \omega(g_1, g_2) \omega(g_1 g_2, g_3)  = \omega(g_1, g_2 g_3) \omega(g_2, g_3) \text{.} \label{eqn:assoc}
\end{equation}
Any function $\omega(g_1, g_2) \in A$ satisfying the associativity constraint is called a \emph{factor set}, or sometimes an $A$-factor set.

If $G$ includes anti-unitary operations, \emph{and} if $A \neq \zz$, the above discussion needs to be modified.  If $g$ is anti-unitary, then so is $\Gamma(g)$, which acts nontrivially on elements of $A$ by complex conjugation.  For example,
\begin{equation}
\Gamma(g_1) \Big[ \omega(g_2, g_3) \Gamma(g_2 g_3) \Big] = \omega^{-1}(g_2, g_3) \Gamma(g_1)  \Gamma(g_2 g_3) 
\end{equation}
for $g_1$ anti-unitary.  This modifies the associativity constraint.  We define $s(g) = 1$ for $g$ unitary and $s(g) = -1$ for $g$ anti-unitary, and then
\begin{equation}
 \omega(g_1, g_2) \omega(g_1 g_2, g_3)  = \omega(g_1, g_2 g_3) \omega^{s(g_1)}(g_2, g_3) \text{.}
 \label{eqn:assoc-modified}
\end{equation}
To remind ourselves of the nontrivial action of anti-unitary operations, we will call such factor sets $A_T$-factor sets.  Since we are mostly interested in $A = \zz$, this complication is not relevant to much of our discussion.  However, we will also have occasion to consider ${\rm U}_T(1)$-factor sets, so we will include the possibility of nontrivial action of anti-unitary operations in the discussion below.

If $\omega_a(g_1, g_2)$ and $\omega_b(g_1, g_2)$ are factor sets, then $\omega_{ab}(g_1, g_2) = \omega_a(g_1, g_2) \omega_b (g_1, g_2)$ is also a factor set.  It is simple to check that this product defines an Abelian group structure on the collection of all factor sets.  The product of factor sets is associated with tensor products of representations:   if $\omega_a, \omega_b$ are the factor sets of the representations $\Gamma_a , \Gamma_b$, respectively, then the product $\omega_{ab} = \omega_a \omega_b$ is the factor set of the tensor product representation $\Gamma_a \otimes \Gamma_b$.

Suppose that we allow a redefinition of the $\Gamma$'s by
\begin{equation}
\Gamma'(g) = \lambda(g) \Gamma(g) \text{,} \label{eqn:gamma-redef}
\end{equation}
where $\lambda(g) \in A$.  This induces the following transformation of the factor set:
\begin{equation}
\omega'(g_1, g_2) = \lambda(g_1) \lambda^{s(g_1)} (g_2) \lambda(g_1 g_2)^{-1} \omega(g_1, g_2) \text{.} \label{eqn:fs-equivalence}
\end{equation}
$\omega'$ is also a factor set (\emph{i.e.}, satisfies the associativity constraint).  Two factor sets $\omega$ and $\omega'$ are said to be \emph{equivalent} if they are related by Eq.~(\ref{eqn:fs-equivalence}) for some $\lambda(g)$, and in this case we write $\omega \sim \omega'$.  This notion of equivalence is reflexive ($\omega \sim \omega$), symmetric ($\omega' \sim \omega$ if $\omega \sim \omega'$) and transitive (if $\omega \sim \omega'$ and $\omega' \sim \omega''$, then $\omega \sim \omega''$), therefore $\sim$ defines an equivalence relation that partitions the set of factor sets into equivalence classes.  We denote the equivalence class of $\omega$ by $c(\omega)$.  Note that $c(\omega) = c(\omega')$ if and only if $\omega \sim \omega'$.  Given a class $c(\omega)$, we say $\omega$ is a representative of the class.

The equivalence classes themselves form an Abelian group, with product defined by
\begin{equation}
c(\omega_1) c(\omega_2) = c(\omega_1 \omega_2) \text{.}
\end{equation}
This product is well-defined, in the sense that it does not depend on the representatives we choose for each class.  The Abelian group of factor set equivalence classes is isomorphic to the cohomology group $H^2(G, A_T)$.  If anti-unitary operations are not present, or if they act trivially on $A$ as when $A = \zz$, we leave off the $T$ subscript and write $H^2(G, A)$.  We shall not bother to give a definition of $H^2(G, A_T)$ in terms of group cohomology, because, for our purposes, it is sufficient to view the group of factor set equivalence classes as the \emph{definition} of $H^2(G, A_T)$ (see footnote 17 of Ref.~\onlinecite{mermin92}).  We shall often refer to factor set equivalence classes as cohomology classes [\emph{i.e.}, elements of $H^2(G, A_T)$].

With this discussion behind us, we note that the projective representation $\Gamma$ (with anti-unitary operations) is associated with an $A_T$ extension of the group $G$.  Roughly, such an extension is a new group $E$ in which it makes sense to multiply elements of $A$ and elements of $G$.
The advantage of a defining an extension is that it is characterized entirely by $G$, $A$ and $\omega$, so we can equivalently speak of classifying factor sets or classifying group extensions.  
Formally, an $A_T$ extension is a group $E$ such that: (1) $A$ is a normal subgroup of $E$.  (2) $G = E / A$.  Elements of $G$ can be viewed as cosets $A u(g)$ in $E$, where $u(g) \in E$ is a representative of $g$.  The choice of representative is arbitrary, that is, we are free to redefine $u'(g) = a(g) u(g)$ where $a(g) \in A$.  A general element of $E$ is of the form $a u(g)$, where $a \in A$.  This leads us to the third and final condition defining an $A_T$-extension:  (3) $u(g) a = a^{s(g)} u(g)$.

We have $u(g_1) u(g_2) = \omega(g_1, g_2) u(g_1 g_2)$, where $\omega(g_1, g_2) \in A$ is an $A_T$-factor set again satisfying the associativity constraint Eq.~(\ref{eqn:assoc-modified}).  At this point, it is clear that all the structure of factor sets and their equivalence classes is identical to the discussion given above, and that we can also view these classes as equivalence classes of  $A_T$ extensions.

If $G$ contains no anti-unitary operations, or, more importantly for the purposes of this paper, if $A = \zz$, then condition (3) above reduces to the statement that $A$ lies in the center of $E$.  Such $A_T$-extensions are called $A$-central extensions.

For the most part, it is not necessary to use the terminology of group extensions; we can just as well talk about projective representations and factor sets.  One advantage of the above more abstract discussion is that there  is no requirement that $A$ be a subgroup of ${\rm U}(1)$; it can be any Abelian group.

Coming back to projective representations, we see that any projective representation belongs to a cohomology class.  For each class, there are one or more unitarily inequivalent irreducible representations.  Therefore, classifying projective representations by cohomology class is coarser than classification by unitary equivalence.

We now consider a few simple examples to get a feeling for the general structure we have been describing. In these examples, we choose $\Gamma(1) = 1$; this can always be done and implies  $\omega(1,1) = \omega(g,1) = \omega(1,g) = 1$.  When doing practical calculations for discrete groups it is often convenient to specify the group in terms of generators and relations.  For instance, $G = \zz$ is generated by $a$, subject to the relation $a^2 = 1$.  If we consider $A = \zz$ and a projective representation $\Gamma$, the single relation becomes $[\Gamma(a)]^2 = \sigma = \pm 1$.  Specifying the relation in this way defines a factor set $\omega(1,1)=\omega(1,a)=\omega(a,1) = 1$, and $\omega(a,a) = \sigma$.  There are two cohomology classes labeled by $\sigma$, and $H^2(\zz, \zz) = \zz$.  Each class has two one-dimensional irreducible representations:  $\Gamma_{\sigma = 1}(a) = \pm 1$, and $\Gamma_{\sigma = -1}(a) = \pm i$.

For any discrete group, we can follow this procedure of writing down generators and relations.  We can write the relations so that the right-hand side of each is unity.  Then, passing to a projective representation $\Gamma$, the right-hand side of each relation is replaced by an element of $A$.  Let us work out an example to illustrate the procedure.  Suppose $G = \zz \times \zz$ and $A = {\rm U}(1)$.  We choose generators $a$ and $b$, satisfying the relations
\begin{eqnarray}
a^2 &=& 1 \label{eqn:z2z2-r1} \\
b^2 &=& 1 \\
a b a^{-1} b^{-1} &=& 1 \text{.} \label{eqn:z2z2-r3}
\end{eqnarray}
Passing to a projective representation $\Gamma$, we have
\begin{eqnarray}
\Gamma(a)^2 &=& \sigma_a \\
\Gamma(b)^2 &=& \sigma_b \\
\Gamma(a) \Gamma(b) \Gamma(a)^{-1} \Gamma(b)^{-1} &=& \sigma_{ab} \text{,} \label{eqn:pz2z2-r3}
\end{eqnarray}
where $\sigma_a, \sigma_b, \sigma_{ab} \in {\rm U}(1)$.  

At this point a couple of issues arise.  First, the $\sigma$'s are not, in general, in one-to-one correspondence with cohomology classes.  That is, there is some redundancy that has to be eliminated.  Second, some choices of the $\sigma$'s may be inconsistent and not give a legitimate factor set.  Both these issues arise in this example.  We can set $\sigma_a, \sigma_b \to 1$ by redefining the phase of $\Gamma(a)$ and $\Gamma(b)$.  Therefore we can write Eq.~(\ref{eqn:pz2z2-r3}) as
\begin{equation}
[ \Gamma(a) \Gamma(b) ]^2 = \sigma_{a b} \text{.}
\end{equation}
Multiplying this on the left and right by $\Gamma(a)$, we also obtain
\begin{equation}
[ \Gamma(b) \Gamma(a) ]^2 = \sigma_{a b} \text{.}
\end{equation}
Since $[\Gamma(b) \Gamma(a) ] = [ \Gamma(a) \Gamma(b) ]^{-1}$, these two equations are only consistent if $\sigma_{ab} = \pm 1$.  Therefore we find $H^2(\zz \times \zz, {\rm U}(1) ) = \zz$.

Suppose we consider again $G = \zz \times \zz$, but now $A = \zz$.  In this case, we proceed as above, but $\sigma_a, \sigma_b, \sigma_{ab} \in \zz$.  We can no longer eliminate $\sigma_a$ and $\sigma_b$, and all choices of the $\sigma$'s are consistent, so we have $H^2(\zz \times \zz, \zz) = \zz \times \zz \times \zz$.  Notice that the number of classes increased upon changing $A$ from ${\rm U}(1)$ to $\zz$.  Indeed, since every $\zz$-factor set is also a ${\rm U}(1)$-factor set, we can group the $\zz$ classes together into ${\rm U}(1)$ classes:  the four $\zz$ classes with $\sigma_{ab} = 1$ belong to the same ${\rm U}(1)$ class, and similarly for the four classes with $\sigma_{ab} = -1$.  

We can always ``coarsen'' the $\zz$ classification in this way, grouping $\zz$ classes together into ${\rm U}_T(1)$ classes.  Note the appearance of the $T$ subscript, which is important if $G$ contains anti-unitary operations. We denote the resulting group of ${\rm U}_T(1)$ classes by $\bar{H}^2(G, \zz)$.  In general, $\bar{H}^2(G, \zz)$ is the subgroup of $H^2(G, {\rm U}_T(1))$ generated by all elements of order 2.  In the above example with $G = \zz \times \zz$, $\bar{H}^2(G, \zz) = H^2(G, {\rm U}_T(1))$, but this is not true in general.  We will see that this coarsening has an important physical interpretation.

If $G$ is a continuous group, it is natural that there should be some kind of continuity condition on $\omega(g_1, g_2)$.  The naive choice of requiring $\omega(g_1, g_2)$ to be a continuous function on the group is not adequate;\cite{chen11b} for instance, it is easily seen that $\omega$ is discontinuous for the $S = 1/2$ representation of ${\rm SO}(3)$.  Instead, we believe the correct prescription, following Ref.~\onlinecite{chen11b},
 is to require that $\omega(g_1, g_2)$ be a measurable function on $G$ (that is, to classify extensions by Borel cohomology).  For practical purposes, when dealing with continuous groups it is often possible to work out the fractionalization classes by simple elementary arguments, as illustrated by the discussion of Sec.~\ref{sec:internal-trans} for $G = {\rm U}(1)$ and $G = {\rm SO}(3)$.

\subsection{General structure, and physical manifestations in excited states}
\label{sec:fc-gen}

We are now in a position to state the result that fractionalization classes for each superselection sector are given by elements of $H^2(G, \zz)$, where $G$ is the symmetry group.  We focus on the $e$-sector only to simplify the notation; all statements also hold  for $m$ and $\epsilon$ sectors.
To connect with the discussion of the previous two sections, we can say that the action of symmetry operations on the $e$-sector states in a region $R^e$ is given by the projective representation $\Gamma^e$, satisfying
\begin{equation}
\Gamma^e(g_1) \Gamma^e(g_2) = \omega_e(g_1, g_2) \Gamma^e(g_1 g_2) \text{,}
\end{equation}
where $\omega_e(g_1, g_2) \in \zz$ is a $\zz$-factor set.

If a state $| \psi \rangle$ decomposes into $e$-sector regions $R^e_1, \dots, R^e_k$, then symmetry localization holds,
\begin{equation}
U(g) | \psi \rangle = \Gamma^e(g, 1) \cdots \Gamma^e(g, k) | \psi \rangle \text{,}
\end{equation}
where $U(g)$ is the unitary operator representing $g$, and $\Gamma^e(g, i)$ is an $e$-operator on $R^e_i$. (For a discussion of anti-unitary time reversal, see Sec.~\ref{sec:internal-trans}.) The notion of $e$-operator was introduced in Sec.~\ref{sec:z2gen}, and is important when the $R^e_i$ are not connected, which will be the case for point group operations as discussed in Sec.~\ref{sec:fc-sg}.
Physical properties are invariant under
\begin{equation}
\Gamma^e(g) \to \lambda(g) \Gamma^e(g) \text{,} \label{eqn:z2-invariance}
\end{equation}
 where $\lambda(g) \in \zz$.  This invariance, and the fact that $\omega_e(g_1, g_2) \in \zz$, is a consequence of the fusion rule $e \times e = 1$, which also implies $k$ must be even.  Due to this invariance, the fractionalization class is given by the $\zz$ cohomology class of the factor set $\omega_e$.  

The fractionalization class is a universal property of a $\zz$ spin liquid phase, so long as symmetry is preserved.  To see this,  suppose that somehow two $e$-particles have different factor sets $\omega_{e1}$ and $\omega_{e2}$, in different classes.  Then we can fuse them to obtain a $1$-particle with factor set $\omega_{e1} \omega_{e2}$.  But since $\omega_{e1}$ and $\omega_{e2}$ are assumed to be in different classes, $c (\omega_{e1} \omega_{e2} ) \neq  c(1)$; that is, we have found a $1$-particle that does not transform in the class of linear representations.  This is a contradiction, so all $e$-particles must have the same cohomology class.  Since cohomology classes are discrete, they are then a robust property of a phase, so long as symmetry is preserved.

At this point it is important to ask what type of physical information is encoded the fractionalization class, and how this information can be extracted.  First, mathematically, the classification by $H^2(G, \zz)$ can be coarsened to classification by $\bar{H}^2(G, \zz)$, if we allow for ${\rm U}(1)$ transformations of $\Gamma^e(g)$ [\emph{i.e.}, $\lambda(g) \in {\rm U}(1)$].  We say that  elements of  $\bar{H}^2(G, \zz)$ specify the ${\rm U}_T(1)$ fractionalization class, to distinguish it from the $\zz$ class specified by elements of $H^2(G, \zz)$.  Physically, ${\rm U}(1)$ transformations leave measurable properties invariant in a process during which $e$-particles do not fuse (and are not created in pairs), \emph{except} for overall phases (and hence eigenvalues) of symmetry operators.  This can occur, for instance,  if we have several $e$-particles that are very far apart and remain far apart on some timescale of interest.  During such a process the number $\hat{n}_e$ of $e$-particles is a well defined \emph{integer} conserved quantity.  The transformation $\Gamma^e(g) \to e^{i \phi} \Gamma^e(g)$ modifies $U(g) \to e^{i \phi \hat{n}_e } U(g)$. (The same holds for anti-unitary time reversal.)  In general, the transformed $U(g)$ is not a symmetry operation, but it is during the process of interest (by assumption).  Therefore, we can think of the ${\rm U}_T(1)$ fractionalization class as capturing some properties of individual anyons, while the additional information in the $\zz$ class can only obtained when we consider fusion of anyons or eigenvalues of symmetry operators.

The attentive reader may notice an apparent conflict between the roles of symmetry localization and fusion processes in our classification.  Indeed, symmetry localization requires a set of $e$-particles to be well-separated on the scale of the correlation length (\emph{i.e.}, the characteristic size of an $e$-particle).  On the other hand, fusion of two $e$-particles requires them to come close together.  Is it possible for two well-separated $e$-particles, to which symmetry localization can be applied, to fuse?  The answer is yes, and this is important for the validity of the $\zz$ (as compared to ${\rm U}_T(1)$) classification.  To see this, consider two $e$ particles, well separated on the scale of the correlation length.  Now suppose an infinitesimal finite-range perturbation is added to the Hamiltonian, which has a non-zero matrix element fusing the two $e$-particles into the vacuum (or into a local excitation in the $1$-sector).  It is certainly possible to find such a perturbation, which changes the total number of $e$-particles by two and thus transforms nontrivially under general ${\rm U}(1)$ transformations of $\Gamma^e(g)$.  Therefore, only $\zz$ transformations leave all physical properties invariant.

We now illustrate the relationship between $\zz$ and ${\rm U}_T(1)$ classes with the example of $G = \zz \times \zz$ unitary internal symmetry, which also gives some sense of how fractionalization class information may be extracted physically.  We take generators $a$ and $b$ for $\zz \times \zz$, satisfying the relations given in Eqs.~(\ref{eqn:z2z2-r1})--(\ref{eqn:z2z2-r3}).  As discussed in Sec.~\ref{sec:math}, there are two ${\rm U}_T(1)$ classes, depending on whether $a$ and $b$ commute ($\sigma_{ab} = 1$) or anticommute ($\sigma_{ab} = -1$).  Suppose we consider energy eigenstates with two localized $e$-particles that are held fixed in space, far enough apart so they do not interact with one another.  Also suppose that we consider such states on the sphere, so we do not have to worry about the global topological degeneracy.  When $\sigma_{ab} = 1$, there are four one-dimensional projective irreducible representations.  Because the irreducible representations are one-dimensional, in the absence of other symmetries, the states we consider will be nondegenerate.  However, for $\sigma_{ab} = -1$, there is a single two-dimensional irreducible representation.  This implies that the states we consider are \emph{fourfold degenerate}, because each $e$-particle has internal degrees of freedom described by a two-dimensional Hilbert space.

Using degeneracy of levels works to distinguish the two ${\rm U}_T(1)$ classes, but it does not distinguish the $\zz$ classes within a given ${\rm U}_T(1)$ class.  This makes sense in light of the physical interpretation we gave of ${\rm U}_T(1)$ versus $\zz$ classes; degeneracy of levels has to do with the projective irreducible representations associated with individual $e$-particles, but does not involve fusion or eigenvalues of symmetry operators. Moreover, given one of the four $\zz$ classes within a given ${\rm U}_T(1)$ class, the other three can be realized by making transformations $\Gamma^e(a) \to i \Gamma^e(a)$ and $\Gamma^e(b) \to i \Gamma^e(b)$; this does not affect the dimensions and multiplicities of irreducible representations. 

Extracting the additional $\zz$ class information is more subtle.  Suppose we consider the two classes with $\sigma_b = \sigma_{ab} = 1$, and suppose we make the assumption that the two $e$-particles are identical.  In the class $\sigma_a = 1$, in any irreducible representation $\Gamma(a) = \pm 1$, so acting on a state with two identical particles we have $U(a) = 1$.  On the other hand, in the class with $\sigma_a = -1$, $\Gamma(a) = \pm i$ in any irreducible representation, so $U(a) = -1$ on a state of two identical $e$-particles.  Subtle information of this kind is not captured in the ${\rm U}_T(1)$ class, as  the eigenvalues of symmetry operators are involved.  Somewhat less obviously, fusion is also involved via the implicit assumption that the ground state (no $e$-particles) satisfies $U(a) = 1$; in fact, what we are doing is comparing eigenvalues of $U(a)$ for the ground state and a state of two identical $e$-particles.  This comparison is not well defined if fusion processes are suppressed, because in that case, the phase of $U(a)$ can be adjusted separately for the two states being compared.

\subsection{Space group symmetry}
\label{sec:fc-sg}

Much of the discussion above carries over for general space group symmetry, but the notion of symmetry localization needs to be modified.  This is so because space group operations such as reflection and rotation move some points by large distances, and can thus move an anyon out of the region in which it is localized.  We again focus on $e$-particles for concreteness, but all statements also hold for $m$ and $\epsilon$ particles.

\begin{figure}
\includegraphics[width=0.9\columnwidth]{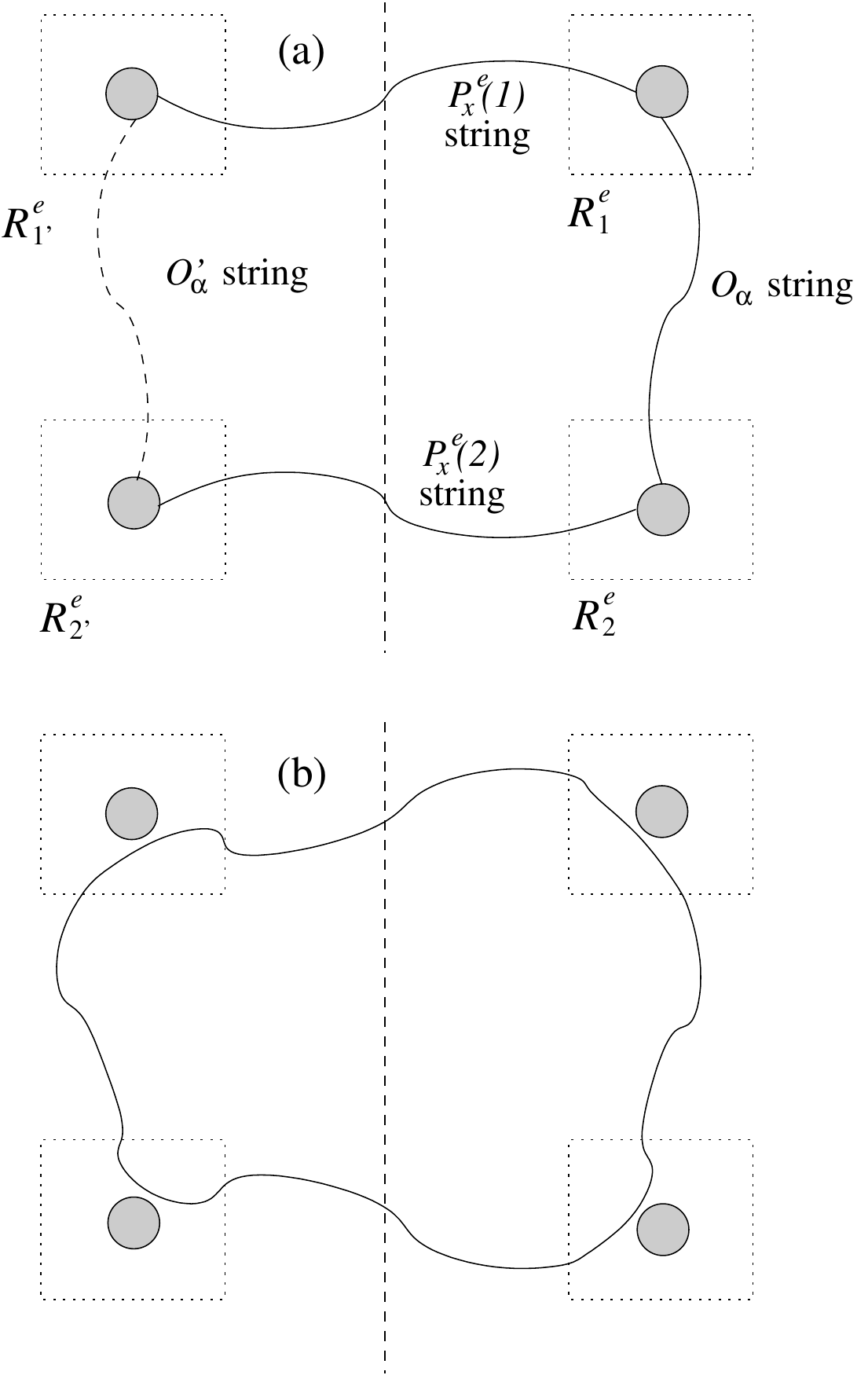}
\caption{(a) Illustration of the initial state $|\psi_\alpha\rangle$, and the action of $P_x$ on $|\psi_\alpha\rangle$.  The vertical dashed line is the reflection axis, and the regions $R^e_i$ and $R^e_{i'}$ are defined in the main text.  The $e$-strings connecting regions for operators ${\cal O}_{\alpha}$ and $P^e_x(i)$ are shown.  The dashed-line ${\cal O}'_{\alpha}$ string is the image of the ${\cal O}_{\alpha}$ string under $P_x$.  (b) By acting with a closed $e$-string loop as shown, the ${\cal O}_{\alpha}$ and $P^e_x(i)$ strings can be eliminated in favor of the ${\cal O}'_\alpha$ string.  Following the argument depicted in Figs.~\ref{fig:2e-transform}b and~\ref{fig:2e-transform}c, the closed loop can be decomposed into smaller  regions with unit eigenvalue acting on the ground state, and the effect of transforming the string can be absorbed into the definition of $P^e_x(i)$.  In the $\epsilon$-particle case, depending on the orientations of the strings in (a), the loop in (b) may need to be twisted in some of the regions, to ensure the correct orientation of the ${\cal O}'_\alpha$ string.}
\label{fig:px-example}
\end{figure}

It is simplest to illustrate the differences from the case of translation and internal symmetries by focusing on a concrete example.  We consider the reflection symmetry $P_x$ sending $x \to -x$, $y \to y$, and satisfying the relation $P_x^2 = 1$.  As in Sec.~\ref{sec:internal-trans}, we consider states $| \psi_\alpha \rangle$ decomposed into two $e$-sector regions $R^e_i$, $i = 1,2$ (see Fig.~\ref{fig:px-example}).  Reflection maps these regions to image regions, $P_x : R^e_i \to R^e_{i'}$.  Symmetry localization can again be expressed by writing
\begin{equation}
P_x | \psi_{\alpha} \rangle = P^e_x(1) P^e_x(2) | \psi_{\alpha} \rangle \text{,}
\end{equation}
but now $P^e_x(i)$ is an $e$-operator on $\tilde{\boldsymbol{R}}^e_i$, defined as the union of $R^e_i$ and $R^e_{i'}$.  $P^e_x(i)$ has a single $e$-string connecting $R^e_i$ and $R^e_{i'}$ (see Fig.~\ref{fig:px-example}).
Again, the physical interpretation is that $P^e_x(i)$ is a ``one-particle'' symmetry operator.  This operation can no longer be accomplished entirely locally, because the $e$-particle must be moved from $R^e_i$ to $R^e_{i'}$, hence the presence of the string operator.  However, once the $e$-particle is moved, any remaining operations can be accomplished locally in $R^e_i$ and $R^e_{i'}$.

Just as for the case of translation and internal symmetries, we should also consider the effect of any phase factor obtained by transforming the string connecting the two $e$-particles.  Here, one can follow essentially the same argument, illustrated in Fig.~\ref{fig:px-example}(b), to show that this phase factors into a product of phases associated with the individual particles.

Now consider the relation $P_x^2 = 1$.  Arguing as before, we have $P^e_x(1') P^e_x(1) = P^e_x(2') P^e_x(2) = \pm 1$, where $P^e_x(i')$ is the operator giving the action of $P_x$ on the transformed $e$-particle in region $R^e_{i'}$.  $P^e_x(i') P^e_x(i)$ is an $e$-operator on $\tilde{\boldsymbol{R}}^e_i$.

More generally, suppose we consider a group relation $S_1 \cdots S_k = 1$.  For the $e$-particle in $R^e_1$ in $| \psi_{\alpha} \rangle$, we then have $S^e_1 \cdots S^e_k = \pm 1$, where for simplicity we have suppressed region labels for the $S^e_i$ operators.  Each of the $S^e_i$, and thus the product  $S^e_1 \cdots S^e_k$, is an $e$-operator on $\tilde{\boldsymbol{R}}^e_1$, defined as the union of $R^e_1$, $S_k (R^e_1)$, $S_{k-1} S_k (R^e_1)$, and so on.  Note that we are assuming that the symmetry operators for one particle commute with those for the other.  This will be the case if $\tilde{\boldsymbol{R}}^e_1$ and $\tilde{\boldsymbol{R}}^e_2$ do not overlap, which we assume.  This amounts to assuming that the different particles occupy generic, \emph{i.e.}, not symmetry related, positions.  It would be interesting to consider the implications of relaxing this assumption, but we leave this for future work.

Note that $\tilde{\boldsymbol{R}}^e_i$ depends on the group relation considered.  This is unappealing, because these regions serve to define the $e$-sector states associated with each particle, in which the $S^e$ symmetry operators act.  A solution to this is instead to define $\boldsymbol{R}^e_i$ to be the union of all regions that can be obtained as images of $R^e_i$ under all point group operations with some arbitrary fixed center of symmetry.  We can choose the generators of the symmetry group to leave the chosen center of symmetry fixed (or nearly fixed).  Then, for any relation involving a small number of generators, $\boldsymbol{R}^e_i$ includes  $\tilde{\boldsymbol{R}}^e_i$ as a subset.

With these modifications, the general structure described in Sec.~\ref{sec:fc-gen} continues to hold.  In particular, fractionalization classes are again given by elements of $H^2(G, \zz)$, where $G$ is the full symmetry group including space group operations.

\subsection{Example: square lattice space group, time reversal and spin rotation}
\label{sec:square-example}

Because it is important for discussing the toric code model, and also to make contact with projective symmetry group classification, we discuss the example of square lattice space group symmetry, combined also with time reversal and ${\rm SO}(3)$ spin rotation.  For clarity of notation, we focus on the $e$-sector.

The symmetry group $G$ is generated by: (1) $P_x$, reflection $x \to -x$, $y \to y$; (2) $P_{xy}$, reflection $x \leftrightarrow y$; (3) $T_x$, translation by one lattice constant along the $x$-axis; (4) time reversal $\cT$; and (5) $R_s(\theta \hat{\bf{n}})$, spin rotation about axis $\hat{\bf{n}}$ through angle $\theta$.  It is convenient to use $T_y = P_{xy} T_x P^{-1}_{xy}$ in some of the relations, which is a translation by one lattice constant along the $y$-axis.  In the $e$-sector, including factors of $\pm 1$ to specify a non-trivial factor set, the relations are
\begin{subequations}
\begin{eqnarray}
(P^e_x)^2 &=& \sigma^e_{px},  \label{eqn:px-relation}  \\
(P^e_{xy})^2 &=& \sigma^e_{pxy},  \\
(P^e_x P^e_{xy})^4 &=& \sigma^e_{pxpxy}, \\
T^e_x T^e_y (T^e_x)^{-1} (T^e_y)^{-1} &=& \sigma^e_{txty}, \\
T^e_x P^e_x T^e_x (P^e_x)^{-1} &=& \sigma^e_{txpx}, \\
T^e_y P^e_x (T^e_y)^{-1} (P^e_x)^{-1} &=& \sigma^e_{typx}, \\
({\cT}^e)^2 &=& \sigma^e_T, \\
\cT^e T^e_x {\cT^e}^{-1} (T^e_x)^{-1} &=& \sigma^e_{Ttx}, \\
 \cT^e P^e_x (\cT^e)^{-1} P^e_x &=& \sigma^e_{Tpx}, \\
 \cT^e P^e_{xy} (\cT^e)^{-1} P^e_{xy} &=& \sigma^e_{Tpxy}, \\
R^e_s(2\pi) &=& \sigma^e_R \label{eqn:spin-relation} \text{,}
\end{eqnarray}
\end{subequations}
where the $\sigma$'s are $\zz$-valued parameters.  We also have
\begin{equation}
R^e_s(\theta \hat{\bf{n}}) {\cal G}^e = {\cal G}^e R^e_s(\theta \hat{\bf{n}}) \text{,}
\end{equation}
where we can substitute ${\cal G}^e = P^e_x, P^e_{xy}, T^e_x,  {\cal T}^e$.
In the second set of relations, one might worry that the right-hand side can be multiplied by a measurable (but not continuous) $\pm 1$-valued function $f(\theta)$, where we must have $f(0) = 1$, since $R^e_s(0) = 1$.  However, it can be shown that $f(\theta) = 1$ for all $\theta$ by assuming $R^e_s(\theta \hat{n} ) = e^{i \theta X_{\hat{n}} }$, and solving for
\begin{equation}
f(\theta) = e^{i \theta X_{\hat{n}} }  {\cal G}^e e^{- i \theta X_{\hat{n}} }  {\cal G}^{e -1} \text{.}
\end{equation}
This is manifestly continuous in $\theta$, and therefore $f(\theta) = 1$.

The relation (\ref{eqn:spin-relation}) simply tells us whether $e$-particles carry integer ($\sigma^e_R = 1$) or half-odd-integer ($\sigma^e_R = -1$) spin.  The other relations are all clearly invariant under $\zz$-valued redefinitions of any of the $e$-sector generators, as in Eq.~(\ref{eqn:z2-invariance}).  Moreover, it is shown in Appendix~\ref{app:genset} that all choices of the $\sigma$'s are consistent.  Therefore, we have shown that $H^2(G, \zz) = \zz^{11}$, and there are $2^{11}$ fractionalization classes.  If we remove spin rotation symmetry, then $H^2(G, \zz) = \zz^{10}$, and there are $2^{10}$ fractionalization classes.

It is also interesting to work out the ${\rm U}_T(1)$ fractionalization classes, that is, to compute $\bar{H}^2(G, \zz)$.  Allowing ${\rm U}(1)$ phase redefinitions of the symmetry generators, we can choose the phase of $P^e_x$, $P^e_{xy}$ and $T^e_x$ so that $\sigma^e_{px}, \sigma^e_{pxy}, \sigma^e_{txpx} \to 1$.  Upon fixing these parameters, the residual phase freedom does not affect any of the other relations.  The anti-unitary nature of ${\cal T}$ implies that adjusting the phase of ${\cal T}$ does not affect any of the relations.  Finally,  $\sigma^e_R$ is clearly unaffected.  Therefore we have $\bar{H}^2(G, \zz) = \zz^{8}$, or, without spin rotation symmetry, $\bar{H}^2(G, \zz) = \zz^{7}$.  The latter result can be extracted from Ref.~\onlinecite{bradley73}, which is a  check on the validity of the above calculations.

\section{Symmetry classes}
\label{sec:symclasses}

\subsection{General results}
\label{sec:sc-results}

Due to the fusion rule $\epsilon = e \times m$,  fractionalization classes for the three non-trivial anyons cannot be specified independently.  Instead, knowledge of $e$ and $m$ fractionalization classes  determines the $\epsilon$ class.  Therefore, specifying $e$ and $m$ fractionalization classes specifies a \emph{symmetry class} for a $\zz$ spin liquid phase.  The crucial issue, addressed in this section, is to understand how the $\epsilon$ fractionalization class is determined by the $e$ and $m$ classes. 

We now state our results, which we establish in Secs.~\ref{sec:sc-internal-trans} and ~\ref{sec:sc-sg}, where we also provide examples.  Let the $\zz$ factor sets associated with the $e$, $m$, and $\epsilon$ fractionalization classes be $\omega_e$, $\omega_m$, and $\omega_\epsilon$, respectively.  With only translation and internal symmetries, $\omega_\epsilon = \omega_e \omega_m$.  That is, $\omega_\epsilon$ is given in terms of $\omega_e$ and  $\omega_m$ by the $H^2(G, \zz)$ group product.  In the general case where $G$ includes point group operations, then $\omega_\epsilon = \omega_t \omega_e \omega_m$, where $\omega_t$ is another $\zz$ factor set depending on the group in a manner specified below.  We refer to the presence of $\omega_t$ as a ``twisting'' of the $H^2(G, \zz)$ group product.  Physically, this twisting is a consequence of the nontrivial braiding statistics of $e$ and $m$, and occurs because products of point group operations can braid an $e$ and $m$ bound together to form an $\epsilon$, in contrast to translation and internal symmetries.

Whether or not point group symmetry is present, the symmetry class can be specified by two elements of $H^2(G, \zz)$, one for the $e$-sector and one for the $m$-sector.  Equivalently, we can specify a single element of $H^2(G, \zz \times \zz)$.  We now discuss the number of distinct symmetry classes.  The number of fractionalization classes is $\Omega_f = | H^2(G, \zz) |$.  Naively we might say that the number of symmetry classes is simply $\Omega_f^2$, but this is not correct, because two classes related by relabeling $e \leftrightarrow m$ are not in fact distinct.  This means that elements of $H^2(G, \zz \times \zz)$ are not actually in one-to-one correspondence with symmetry classes.  Taking this into account, the number of symmetry classes is
\begin{equation}
\Omega_c = \frac{1}{2} ( \Omega_f^2 - \Omega_f) + \Omega_f \text{.}
\end{equation}

We apply this result to the case of square lattice space group, time-reversal, and spin rotation symmetryies where $\Omega_f = 2^{11}$, and so $\Omega_c = 2\, 098\, 176 \approx 2^{21}$, or, removing spin rotation, $\Omega_f = 2^{10}$, so $\Omega_c = 524\, 800 \approx 2^{19}$.

\subsection{Translation and internal symmetries}
\label{sec:sc-internal-trans}

To relate the $\epsilon$ fractionalization class to the $e$ and $m$ classes, we consider $\epsilon$-particles formed as $e$-$m$ bound states, and compute the fractionalization class of the bound state in terms of the classes of its constituents.  We do this first for the simpler case of translation and internal symmetries.

We consider states $| \psi_{\alpha} \rangle$ that can be decomposed into two fixed $\epsilon$-sector regions $R^\epsilon_i$ ($i = 1,2$).  Each of these regions is further decomposed into an $e$-sector ($R^e_i$) and an $m$-sector ($R^m_i$) region.  We assume that 
\begin{equation}
| \psi_\alpha \rangle = {\cal O}^e_\alpha {\cal O}^m_{\alpha} | \psi_0 \rangle \text{,} \label{eqn:psi-oeom}
\end{equation}
where $| \psi_0 \rangle$ is a ground state.  ${\cal O}^e_\alpha$ is an $e$-operator on the union of $R^e_1$ and $R^e_2$, with an $e$-string connecting the two regions.  Similarly, ${\cal O}^m_{\alpha}$ is an $m$-operator on the union of $R^m_1$ and $R^m_2$.  
Again, for simplicity but not by necessity, we assume $|\psi_0\rangle$ is a singlet under all symmetry operations. 

Now let $S_a$ ($a = 1,2,3$) be translations or unitary internal symmetry operations, satisfying $S_1 S_2 = S_3$.  Symmetry localization for the $\epsilon$-sector regions is expressed by writing
\begin{equation}
S_a | \psi_\alpha \rangle = S^\epsilon_a(1) S^\epsilon_a(2) | \psi_\alpha \rangle \text{,}
\end{equation}
where $S^\epsilon_a(i)$ is supported on $R^\epsilon_i$.  The symmetry can be further localized to the $e$ and $m$ subregions, that is
\begin{equation}
S^\epsilon_a(i) = S^e_a(i) S^m_a(i) \text{,} \label{eqn:epsilon-to-e-m}
\end{equation}
where $S^e_a(i)$ and $S^m_a(i)$ are supported respectively on $R^e_i$ and $R^m_i$.  The $S^e_a(i)$ are the same operators appearing in the localization of $S_a$ if the $m$-particles are not present, and correspondingly for the $S^m_a(i)$.

Now, suppose
\begin{eqnarray}
S^e_1(i) S^e_2(i) &=& \omega_e(1,2) S^e_3(i) \\
S^m_1(i) S^m_2(i) &=& \omega_m(1,2) S^m_3(i) \text{.}
\end{eqnarray}
Then it follows immediately that
\begin{equation}
S^\epsilon_1(i) S^\epsilon_2(i) =  \omega_e(1,2) \omega_m(1,2) S^\epsilon_3(i) = \omega_\epsilon(1,2) S^\epsilon_3(i)   \text{,}
\end{equation}
and therefore
\begin{equation}
\omega_\epsilon = \omega_e \omega_m \text{,} \label{eqn:omega-epsilon-product}
\end{equation}
which is the desired result.  The same statement holds when we consider anti-unitary time-reversal symmetry; this is easily seen following the discussion of Sec.~\ref{sec:internal-trans}.

We now apply these results to the simple case of translation as the sole symmetry. In Sec.~\ref{sec:internal-trans}, we found that there are two fractionalization classes in this case, parametrized for the $e$-sector by 
$T^e_x T^e_y (T^e_x)^{-1} (T^e_y)^{-1} = \sigma^e_{txty} = \pm 1$, with corresponding relations for $m$ and $\epsilon$ sectors.  Equation~(\ref{eqn:omega-epsilon-product}) implies $\sigma^{\epsilon}_{txty} = \sigma^e_{txty} \sigma^m_{txty}$.  At this point, we might naively conclude there are four symmetry classes labeled by ordered pairs $(\sigma^e_{txty}, \sigma^m_{txty})$.  However, the $(+1, -1)$ and $(-1, +1)$ classes are related by relabeling $e \leftrightarrow m$, and thus are not actually distinct.  Therefore, there are three symmetry classes in the case of translation symmetry alone.

\begin{figure}
\includegraphics[width=0.6\columnwidth]{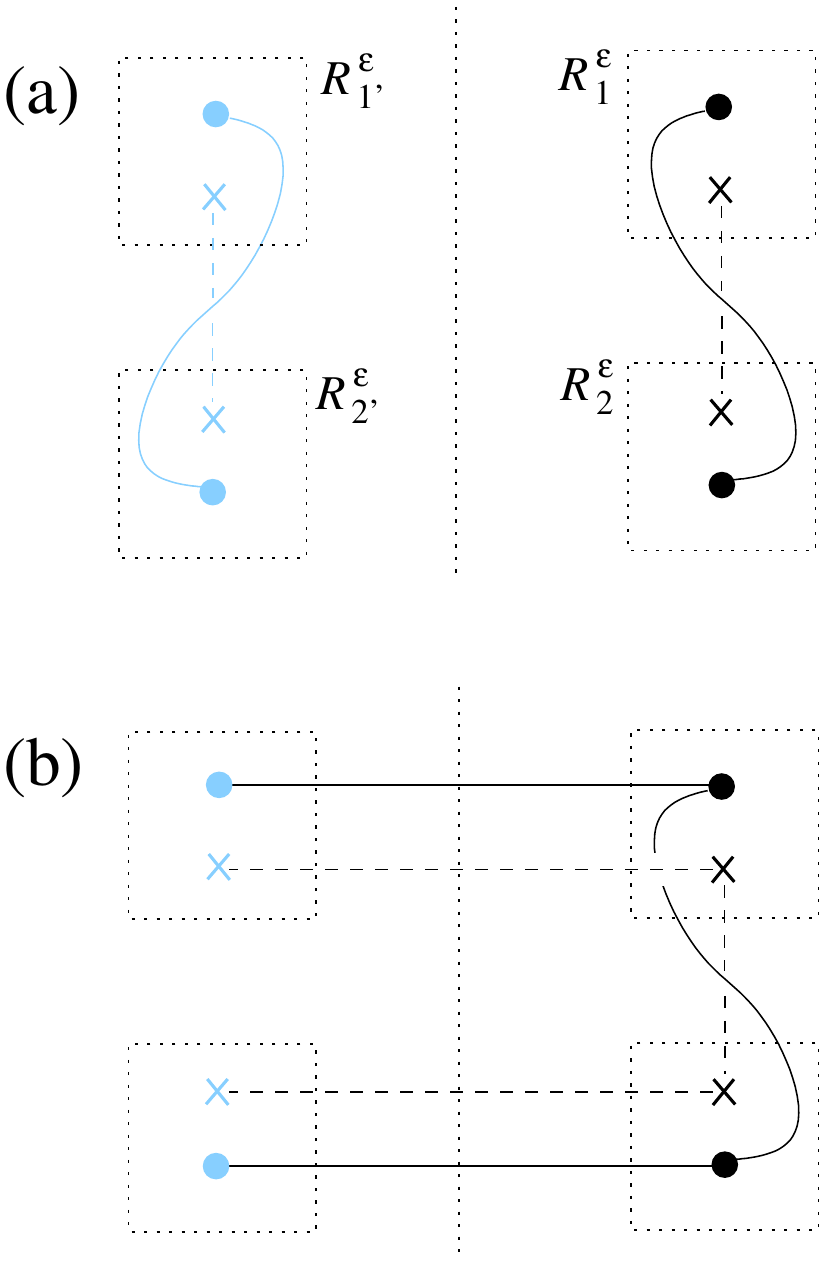}
\caption{(Color online) Computation of $W(i)$ phase factor for $P_x$ reflection symmetry.  (a) The initial state $|\psi_\alpha\rangle$ is depicted on the right-hand side, while the vertical dotted line is the reflection axis, and the final state $P_x|\psi_\alpha\rangle$ lies to the left of the axis.  Regions $R^\epsilon_i$ and their images under $P_x$ are indicated with dotted lines.  All particles and strings are shown in black in the initial states, and in gray (blue online) in the final states.  $e$-particles are filled circles, $m$-particles are crosses, $e$-strings are solid lines, and $m$-strings are dashed lines. (b) Depiction of the state $P_x^e(1) P_x^m(1) P_x^e(2) P_x^m(2) | \psi_\alpha \rangle$.  To compute $W(i)$, we simply bring the strings into the final state configuration shown in the left-hand side of (a).  We find $W(1) = -1$, where the minus sign arises from crossing the $m$-string beneath the $e$-string in $R^\epsilon_1$.  We also find $W(2) = -1$, where in this case the sign arises from sliding the $e$-string over the final-state $m$ particle in $R^\epsilon_{2'}$.}
\label{fig:px-Wa-example}
\end{figure}

We note that there are similarly three symmetry classes in the case of ${\rm SO}(3)$ spin rotation symmetry alone.  Upon substituting $R^e_s(2\pi \hat{{\bf n}}) = \sigma^e_R$ for the translation symmetry relation, and similarly for $m$ and $\epsilon$ sectors, the discussion above holds unchanged.

\subsection{Space group symmetry}
\label{sec:sc-sg}

When the symmetry group includes point group operations, the result in  Eq.~(\ref{eqn:omega-epsilon-product}) is modified due to the mutual statistics of $e$ and $m$ particles.  This leads to a twisting of the group product giving the $\epsilon$ fractionalization class in terms of the $e$ and $m$ classes; in particular, $\omega_\epsilon = \omega_t \omega_e \omega_m$, where $\omega_t$ is another $\zz$ factor set encoding a twisting of the $H^2(G, \zz)$ group product.  We are interested in determining the cohomology class of $\omega_t$, and thus the $\epsilon$ fractionalization class.

We proceed by considering symmetry operations $S_a$ ($a = 1,\dots, k$), satisfying the group relation $S_1 \cdots S_k = 1$.  We suppose that $S^e_1 \cdots S^e_k = \sigma_e$ and $S^m_1 \cdots S^m_k = \sigma_m$, and would like to compute $S^\epsilon_1 \cdots S^\epsilon_k = \sigma_\epsilon$.  We consider states $|\psi_\alpha\rangle$ as described above in Sec.~\ref{sec:sc-internal-trans}.  Following Sec.~\ref{sec:fc-sg}, we let $\boldsymbol{R}^e_i$ be the union of images of $R^e_i$ under arbitrary point group operations (with fixed center of symmetry), and similarly for $\boldsymbol{R}^m_i$ and $R^m_i$.  We will see that $\sigma_\epsilon = \sigma_t \sigma_e \sigma_m$, where $\sigma_t$ enters as a product over three different types of statistical phase factors.  Knowledge of $\sigma_t$ for enough group relations determines the factor set $\omega_t$.

First, we consider a single operation $S_a$ and examine the statement of symmetry localization.  We again have
\begin{equation}
S_a | \psi_{\alpha} \rangle = S^\epsilon_a(1) S^\epsilon_a(2) | \psi_\alpha \rangle \text{.}
\end{equation}
However, Eq.~(\ref{eqn:epsilon-to-e-m}) no longer holds, and is instead modified to
\begin{equation}
S^\epsilon_a(i) = W_a(i) S^e_a(i) S^m_a(i) \text{,}
\end{equation}
where $W_a(i) = \pm 1$ is a statistical phase factor originating from anticommutation of $e$ and $m$ strings.  The factor $W_a(i)$ can be computed, simply by comparing $S_a | \psi_\alpha \rangle$ with $S^e_a(1) S^m_a(1) S^e_a(2) S^m_a(2) | \psi_\alpha \rangle$.  We give an example of such a computation in Fig.~\ref{fig:px-Wa-example}.

So we have
\begin{widetext}
\begin{eqnarray}
\sigma_\epsilon | \psi_\alpha \rangle &=& S^\epsilon_1(i) \cdots S^\epsilon_k(i) | \psi_\alpha \rangle
=  [ \prod_{a = 1, \dots, k} W_a(i) ] S^e_1(i) S^m_1(i) \cdots S^e_k(i) S^m_k(i) | \psi_\alpha \rangle \label{eqn:piece-by-piece} \\
&=& \ell(i)  [ \prod_{a = 1, \dots, k} W_a(i) ] [ S^e_1(i) \cdots S^e_k(i) ] [ S^m_1(i) \cdots S^m_k(i) ]  | \psi_\alpha \rangle \text{.} \label{eqn:products-to-compute}
\end{eqnarray}
\end{widetext}
Here, $\ell(i) = \pm 1$ arises from the fact that some of the $S_a^e(i)$ and $S_a^m(i)$ may anti-commute due to crossings of strings.  In more detail, we observe that $\prod_a S_a^e(i)$  and $\prod_a S_a^m(j)$ define closed loops of $e$ and $m$ strings, respectively, with pieces of string running among the components of $\boldsymbol{R}^e_i$ and $\boldsymbol{R}^m_i$.  These two loops are assembled ``piece-by-piece,'' reading from right-to-left in the product of symmetry operations in Eq.~(\ref{eqn:piece-by-piece}); graphically, strings further to the left in the product can be drawn on top of strings further to the right.  The factor $\ell (i)$ simply measures the $\zz$ linking number of these two loops; that is $\ell(i) = (-1)^{n_c}$, where $n_c$ is the number of times $m$-strings need to be crossed below $e$-strings so that the $m$-string loop lies entirely underneath the $e$-string loop.

To compute $\sigma_\epsilon$, now we need only act on $| \psi_\alpha \rangle$ with the products of symmetry operations in Eq.~(\ref{eqn:products-to-compute}).  We have
\begin{eqnarray}
S^e_1(i) \cdots S^e_k(i) | \psi_\alpha \rangle &=& Z^{em}(i) \sigma_e  \\
S^m_1(i) \cdots S^m_k(i) | \psi_\alpha \rangle &=& Z^{me}(i) \sigma_m \text{,}
\end{eqnarray}
where $Z^{em}(i) = +1$ ($-1$) if an even (odd) number of $m$-particles in $|\psi_\alpha \rangle$ are enclosed in the $e$-string loop defined by $S^e_1(i) \cdots S^e_k(i)$, with corresponding definition for $Z^{me}(i)$, reversing the roles of $e$ and $m$.  Therefore we have found
\begin{equation}
\sigma_t = Z^{em}(i) Z^{me}(i) \ell(i)  \Big[ \prod_{a = 1, \dots, k} W_a(i) \Big] \text{.}
\label{eqn:sigmat}
\end{equation}

\begin{figure}
\includegraphics[width=0.7\columnwidth]{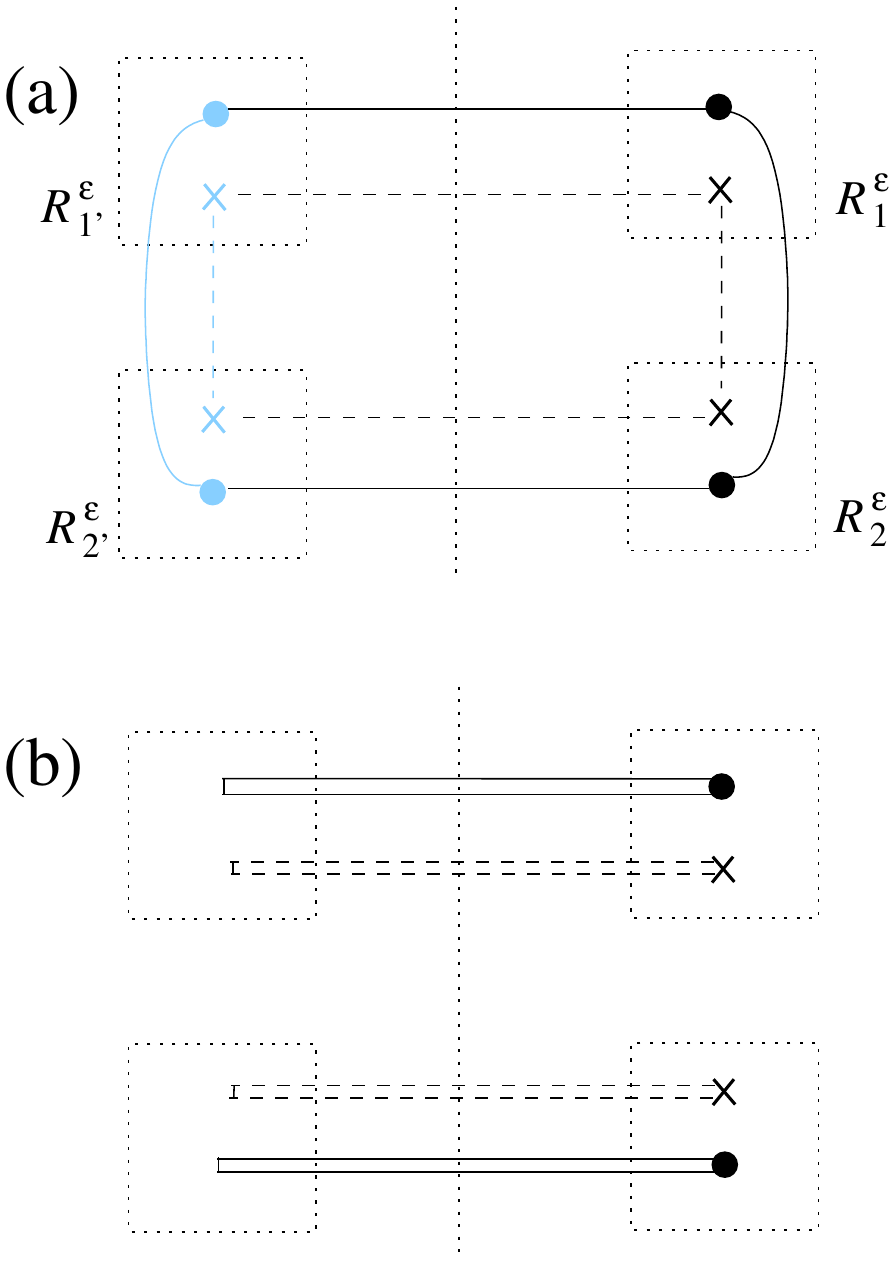}
\caption{(Color online) Computation of $\sigma^t_{px}$, for the group relation $P_x^2 = 1$.  The graphical notation used here is introduced in Fig.~\ref{fig:px-Wa-example}.  (a)  The reflection axis is the vertical dotted line.  The state $|\psi_\alpha\rangle$ is depicted to the right of the axis, with $e$ and $m$ particles and (vertical) strings drawn in black.  This is almost the same state considered in Fig.~\ref{fig:px-Wa-example}, but with a simpler arrangement of strings.  The  state $P_x |\psi_\alpha \rangle$ is shown to the left of the axis in gray (blue online).  The horizontal strings depict the operators $P^e_x(i)$ and $P^m_x(i)$.  From this figure, it can be seen that $W(i) = -1$ in the symmetry localization of $P_x | \psi_\alpha \rangle$, and the same is easily seen to hold in the symmetry localization of $P_x ( P_x | \psi_\alpha \rangle )$.  (b)  Closed loops of string obtained from $P^e_x(i') P^e_x(i)$ and $P^m_x(i') P^m_x(i)$.  Positions of $e$ and $m$ particles in $|\psi_\alpha\rangle$ are shown.  Since these loops  do not link, and, for instance, each $e$-loop encloses no $m$ particles, we have $\ell(i) = Z^{em}(i) = Z^{me}(i) = 1$.}
\label{fig:px-relation}
\end{figure}

To illustrate this discussion, we compute $\omega_t$ for the case of square lattice space group symmetry (plus time reversal).  First, we consider the relation $P_x^2 = 1$.  In Fig.~\ref{fig:px-relation}, by considering a convenient state $|\psi_\alpha\rangle$,  we show that 
\begin{equation}
(P^\epsilon_x)^2 = \sigma^\epsilon_{px} =  \sigma^e_{px} \sigma^m_{px} \text{;}
\end{equation}
that is, $\sigma^t_{px} = 1$.  In more detail, Fig.~\ref{fig:px-relation}(a) illustrates the symmetry localization of $P_x | \psi_\alpha \rangle$, showing that $W(i) = -1$ for $i = 1,2$.  The same is easily seen to be true for the symmetry localization of $P_x (P_x | \psi_\alpha \rangle)$, so these factors cancel in the computation of $\sigma^t_{px}$.  Figure~\ref{fig:px-relation}(b) illustrates the closed $e$ and $m$ string loops obtained when computing $P_x^2 = 1$ in terms of the one-particle operators $P^e_x(i)$, $P^e_x(i')$, and so on.  These loops do not link, so $\ell(i) = 1$.  Moreover, the $e$-loops do not enclose any $m$-particles, and vice versa, so $Z^{me}(i) = Z^{em}(i) = 1$.  Therefore, by Eq.~(\ref{eqn:sigmat}), $\sigma^t_{px} = 1$.  It is important to emphasize that we are free to choose the $e$ and $m$ strings of the one-particle operators $P^e_x(i)$, $P^m_x(i)$, and so on, to run horizontally as shown in Fig.~\ref{fig:px-relation}.  The result should not be affected by this choice; we have not proved this in general, but have experimented with other conventions and always find $\sigma^t_{px}$ to be unaffected.

\begin{figure}
\includegraphics[width=0.8\columnwidth]{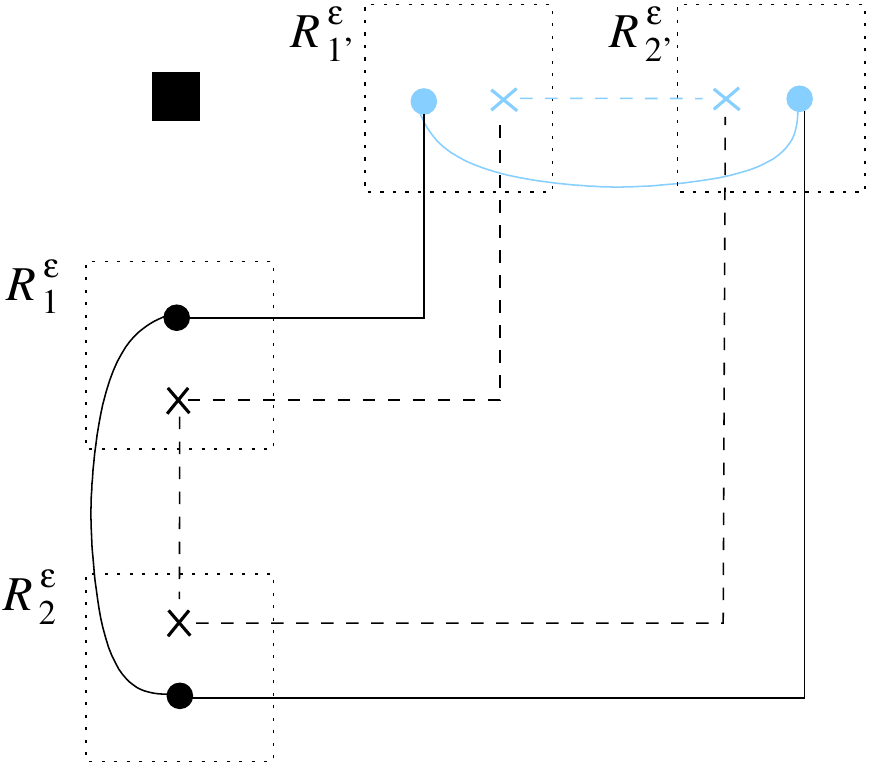}
\caption{(Color online) Symmetry localization of $\pi/2$-rotation $R_{\pi/2}$ on a state $| \psi_\alpha \rangle$.  The center of rotation symmetry is the solid square.  The initial state $| \psi_\alpha \rangle$ has $e$ and $m$ particles arranged along a line extending below the center of symmetry, and $\epsilon$-sector regions $R^\epsilon_i$ as shown.  The particles and strings of the final state are shown in gray (blue online).  The angled $e$ and $m$ strings are the strings of the $R^e_{\pi/2}(i)$ and $R^m_{\pi/2}(i)$ operators.  Inspection of this figure shows that $W(i) = 1$ for $i = 1,2$.}
\label{fig:rot-relation1}
\end{figure}

\begin{figure}
\includegraphics[width=0.9\columnwidth]{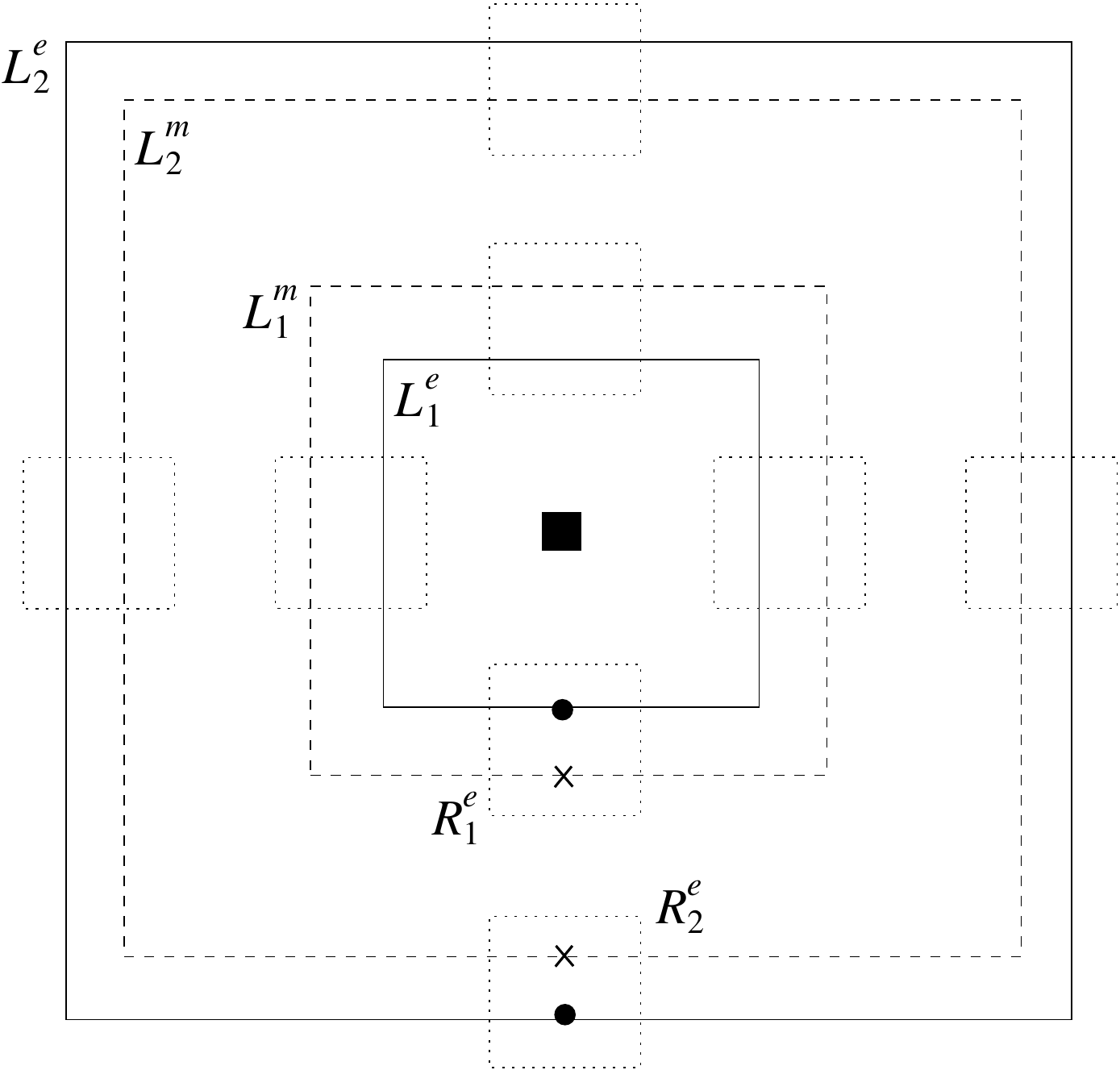}
\caption{Closed $e$ and $m$ string operators obtained upon expressing $R^4_{\pi/2} | \psi_\alpha \rangle$ in terms of products of one-particle symmetry operators acting on $|\psi_\alpha\rangle$.  The locations of $e$ and $m$ particles in $|\psi_\alpha\rangle$ are shown, and the closed strings are labeled as shown.}
\label{fig:rot-relation2}
\end{figure}

Proceeding along the same lines, we find $\sigma_t = 1$ for all other relations, \emph{except} the relation $(P_x P_{xy})^4 = 1$, which we now consider.  Noting that $R_{\pi/2} = P_x P_{xy}$ is a $\pi/2$ rotation, for simplicity we instead consider the equivalent relation $R_{\pi/2}^4 = 1$.  (We find the same result without making this simplification.)  Figure~\ref{fig:rot-relation1} illustrates the state $|\psi_\alpha\rangle$, and the $e$ and $m$ strings in $R^e_{\pi/2}(i)$ and $R^m_{\pi/2}(i)$ are chosen to run as shown.  For subsequent applications of $R_{\pi/2}$, the corresponding strings are obtained simply by rotation of Fig.~\ref{fig:rot-relation1}.  Again, we emphasize that we are always free to choose the one-particle symmetry operator strings to run in this fashion.  We find $W(i) = 1$ for all four rotation operators in the group relation.  

Evaluating $R^4_{\pi/2} | \psi_\alpha \rangle$, we find the four closed $e$ and $m$ string loops shown and labeled in Fig.~\ref{fig:rot-relation2}, obtained as products of the one-particle symmetry operators.  These loops do not link, so $\ell(i) = 1$.  We define closed $\epsilon$ loops, $L^\epsilon_i = L^e_i L^m_i$.  $L^\epsilon_2$ encloses a single $\epsilon$ particle, but this does not affect evaluation of $L^\epsilon_2 | \psi_\alpha \rangle$.  Therefore, in determining the $Z^{em}(i)$ and $Z^{me}(i)$ phase factors arising from loops enclosing particles, we can consider the $i=1$ $e$ and $m$ particles \emph{separately} from the $i=2$ particles.  For $i=1$, $L^e_1$ does not enclose any particles, so $Z^{em}(1) = 1$.  On the other hand, $L^m_1$ encloses the $i=1$ $e$-particle, so $Z^{me}(1) = -1$.  Similarly, $Z^{me}(2) = 1$, because $L^m_2$ does not enclose the $i=2$ $e$-particle.  Finally, we find $Z^{em}(2) = -1$.  Combining all the statistical phase factors, we find $\sigma^t_{pxpxy} = -1$, that is
\begin{equation}
(P^\epsilon_x P^\epsilon_{xy} )^4 = \sigma^\epsilon_{pxpxy} = - \sigma^e_{pxpxy} \sigma^m_{pxpxy} \text{.}
\end{equation}
We have thus found $\omega_t$, and shown it is a non-trivial factor set.

It should not be surprising that the $\omega_t$ twisting appears in the $R^4_{\pi/2} = 1$ group relation, because during the course of this relation the constituent $e$ and $m$ particles in each $\epsilon$ particle are braided around one another.  It is interesting, and perhaps surprising, that this twisting seems to be unavoidable for a \emph{discrete} rotation symmetry. It could thus be valuable to obtain a deeper understanding of the $\omega_t$ twisting.  Finally, we note that the lack of twisting in the other group relations is to be expected. For these relations, given an appropriate choice of $|\psi_\alpha\rangle$, there is no relative motion of the constituent $e$ and $m$ particles, and thus no way for braiding statistics to enter.

\section{General Abelian topological orders}
\label{sec:general-abelian}

Here, we briefly discuss the extension of our symmetry classification to general Abelian topological orders, with the restriction that the symmetry group $G$ consists only of translations and internal symmetries.  For simplicity, we restrict to \emph{unitary} internal symmetries, and discuss inclusion of anti-unitary time reversal at the end of this section.

We find that the symmetry classes are labeled by elements of $H^2(G, \Gamma_2)$, where $\Gamma_2$ is the group of fusion rules.  This agrees with a result asserted by Kitaev.\cite{kitaev06,kitaevpc}   Not all elements of  $H^2(G, \Gamma_2)$ describe distinct symmetry classes; elements related to others by a relabeling of anyons leaving the topological structure invariant  (\emph{e.g.}, $e \leftrightarrow m$ for $\zz$ topological order)  correspond to the same symmetry class.

The fusion group $\Gamma_2$ is a finite Abelian group, and is thus isomorphic to a product of cyclic groups. Suppose there are $p$ cyclic factors and $\Gamma_2 = \z_{k_1} \times \z_{k_2} \times \cdots \times \z_{k_p}$.  Let $e_i$ ($i = 1,\dots,p$) be the generators for these factors.  Physically, the $e_i$ form an elementary set of anyons, from which any other type of anyon can be obtained by fusion.  (In the case of $\zz$ topological order, $k_1 = k_2 = 2$,  $e_1 = e$, and $e_2 = m$.)  Therefore, the symmetry class should be determined by specifying the fractionalization class for each of the $e_i$ anyons.

Since $(e_i)^{k_i} = 1$, in physical states, anyons of $e_i$ type appear in multiples of $k_i$.  For translation and internal symmetries, the property of symmetry localization is expected to hold as above.  Therefore, in a straightforward extension of the discussion given above for $\zz$ topological order, the action of symmetry on $e_i$ anyons is a $\z_{k_i}$-central extension of $G$, and the fractionalization classes are given by elements of $H^2(G, \z_{k_i})$.    Then, because 
\begin{equation}
H^2(G, \Gamma_2) = H^2(G, \z_{k_1}) \times \cdots \times H^2(G, \z_{k_p}) \text{,}
\end{equation}
symmetry classes are labeled by elements of $H^2(G, \Gamma_2)$.  

In the future, it would be interesting to generalize this result to the case of full space group symmetry.  The simplest possibility is that the only modification needed for space group symmetry is a twisting of the $H^2$ group product giving the fractionalization classes of arbitrary anyons in terms of the $e_i$ classes.  However, unlike for $\zz$ topological order, in general $e_i$-strings do not commute with themselves at crossing points, and this may lead to new features in the classification.

Finally, we discuss inclusion of anti-unitary time reversal.  First, if all $e_i$ anyons have $k_i = 2$, then no modification of the above discussion is needed to incorporate time reversal, because anti-unitary complex conjugation acts trivially on elements of $\zz \in {\rm U}(1)$.  On the other hand, if some $k_i > 2$, we might imagine that we need to account for non-trivial action of complex conjugation on elements of $\z_{k_i} \in {\rm U}(1)$.  This is true, but is not sufficient; all such cases are beyond the scope of our classification because the assumption that time reversal does not permute different types of anyons is actually inconsistent with the topological order.  Observe that, if $k_i > 2$, the $e_i^2$ anyon must have either non-trivial self statistics, or non-trivial mutual statistics with some other anyon.  Otherwise, $e_i^2 = 1$, a contradiction. Letting the $\theta_{s2}$ be the self-statistics angle of $e_i^2$, and $\theta_{s1}$ the same for $e_i$, we have, employing the $K$-matrix Chern-Simons approach, $\theta_{s2} = 4 \theta_{s1}$.\cite{wenbook}  Similarly, letting $\theta_{m2}$ be the mutual statistics angle of $e_i^2$ and some other fixed type of anyon, and $\theta_{m1}$ the same for $e_i$ and the same other fixed anyon, we have $\theta_{m2} = 2 \theta_{m1}$.\cite{wenbook}  Therefore, in order for $e_i^2$ to have non-trivial statistics, we must have $\theta_{m1} \neq 0, \pi$, or $\theta_{s1} \neq 0, \pi/2, \pi, 3 \pi/2$.  Suppose $\theta_{s1}  \neq 0, \pi/2, \pi, 3 \pi/2$.  Then acting with time reversal on a pair of $e_i$ anyons gives a pair of anyons with new self-statistics angle $\theta'_s = - \theta_{s1} \neq \theta_{s1} \operatorname{mod} 2\pi$, which is only consistent if time reversal transforms $e_i$ into a different type of anyon.  If instead $\theta_{m1} \neq 0,\pi$, the same argument  shows that time reversal must transform $e_i$ or the other fixed anyon into a different type of anyon.  Therefore, if time reversal symmetry is present and some $k_i > 2$, time reversal must permute the different types of anyons.  
This is an additional motivation to develop a full symmetry classification for Abelian topological orders in future work, including symmetry classes ``beyond fractionalization'' where anyons are permuted by symmetry.

\section{Explicit realization: toric code}
\label{sec:toriccode}

In this section we show how these ideas work out explicitly in the example of the toric code.\cite{kitaev03}  Some related prior results were obtained in Ref.~\onlinecite{wen03}. In particular, we will work out fractionalization and symmetry classes for the case of square lattice space group symmetry alone.  It is straightforward to include time reversal in the same discussion, but we omit this for brevity.  We can understand the possible fractionalization and symmetry classes this case simply by omitting the relations of Sec.~\ref{sec:square-example} containing time reversal or spin rotation, and keeping the remaining six relations.  The $\Omega_f = 2^6$ fractionalization classes are elements of $H^2(G, \zz) = \zz^6$, and there are $\Omega_c = 2080 \approx 2^{11}$ symmetry classes.  Of these, the toric code model realizes three, using two tunable parameters. 

Guided by the general discussion of Sec.~\ref{sec:symclasses}, it is also possible to explicitly work out the $\omega_t$ twisting involved in relating the $\epsilon$ fractionalization class to the $e$ and $m$ classes.  We have done this, but do not present the results here; this essentially amounts to a more cumbersome repetition of the general constructions of Sec.~\ref{sec:sc-sg}.

The toric code model was introduced in Sec.~\ref{sec:z2toric}.  We assume that $|K_e| \neq |K_m|$, to avoid extra symmetry present in that case.
When it is necessary to know the size of the system, we take an $L\times L$ torus with $L$ divisible by 4.  For explicit calculations, we will take the coordinates of vertices $s$ of the form $\r = (x,y) = (m,n)$ with $m,n$ integers, and the coordinates of faces $p$ will take values $(x,y) = (m+1/2,n+1/2)$.  Throughout this section, using language from the $\zz$ gauge theory description of $\zz$ topological order, we shall often refer to $e$-particles as charges and $m$-particles as fluxes.

When $K_e>0$, the ground state has $A_s = 1$, and a vertex $s$ with $A_s = -1$ is an $e$-particle (a charge excitation).  When $K_e<0$, the ground state has $A_s = -1$.  Viewed in terms of the $K_e > 0$ ground state, this is a background charge of $e$-particles.  This situation often arises in theories of $\zz$ spin liquids and, when it appears for a gauge theory Hamiltonian, is referred to as \emph{odd} $\zz$ gauge theory.\cite{moessner01b}  With $K_e < 0$, the excited $e$-particles now correspond to $A_s = +1$ vertices.  Identical considerations relate $m$-particles (flux excitations) to the value of $B_p$.  The four different symmetry classes that this Hamiltonian accesses are realized by the four choices of signs of $K_e$ and $K_m$.

\subsection{Wave functions}

\begin{figure}
\includegraphics[width=0.3\columnwidth]{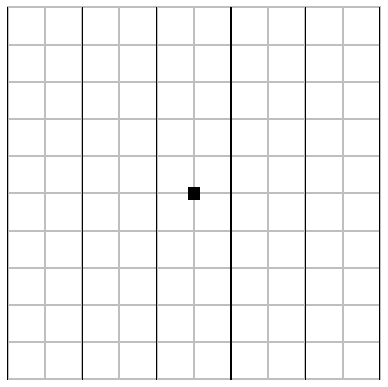}
\caption{\label{fig:refstate}
The state $\ket{\text{ref}(-1)}$.  Dark links carry $\sigma^z = -1$, others have $\sigma^z=1$.  The square at the center of all figures identifies the origin of coordinates and of point group operations.}
\end{figure}

It will be helpful to have explicit forms for the wave functions.  We build the ground state off of a reference state that minimizes the flux term of the Hamiltonian.  Let 
$s_e = \mathrm{sign}\, K_e$ and 
$s_m = \mathrm{sign}\, K_m$.  For $s_m = 1$, the reference state 
$\ket{\text{ref}(1)}$ will have 
$\sigma^z_{\r,\r'} = 1$ on all bonds, as discussed earlier 
[see Eq.~\eqref{eqn:uniformstate}].  For $s_m = -1$ the reference 
$\ket{\text{ref}(-1)}$ has $\sigma^z_{\r,\r'} = 1$ on horizontal links and 
$\sigma^z_{\r,\r'} = \pm 1$ on alternating columns of links; that is, 
$\sigma^z_{\r,\r'} = -1$ for $x=x'=\text{odd}$.  This puts one link with 
$\sigma^z_{\r,\r'} = -1$ on each plaquette $p$ (see Fig.~\ref{fig:refstate}).  
The full ground state is then 
\beq
\ket{\psi_0(s_e,s_m)} = \prod_{s} \tfrac{1}{\sqrt{2}} (1 + s_e A_s) \ket{\text{ref}\,(s_m)},  \label{eqn:psi0-se-sm}
\eeq
with
\beq
A_s{\psi_0} = s_e \ket{\psi_0}, \quad B_p{\psi_0} = s_m \ket{\psi_0}
\eeq
for all $s, p$.
We require these states to be invariant under space group symmetry.  This is manifest for $s_m=1$ since the reference state has full symmetry.  For $s_m = -1$, we can use the identity
\beq
\prod_{s} (1 + s_e A_s) = \left[ \prod_{s} (1 + s_e A_s) \right] \prod_{s\in \mathcal{G}} (s_e A_s) ,
\eeq
for any set of vertices $\mathcal{G}$.  The second product will implement a space group operation (via spin flips) on the reference state for an appropriate choice of $\mathcal{G}$.  We require $\mathcal{G}$ to contain an even number of vertices so that the sign factors $s_e$ cancel; in particular, this forces us to take $L$ divisible by 4, because the appropriate $\mathcal{G}$ for the reflection $P_{xy}$ contains $(L/2)^2$ vertices, which can be chosen as the vertices  $\r = (\text{odd},\text{odd})$.

\begin{figure}
\subfigure[]{\includegraphics[width=0.3\columnwidth]{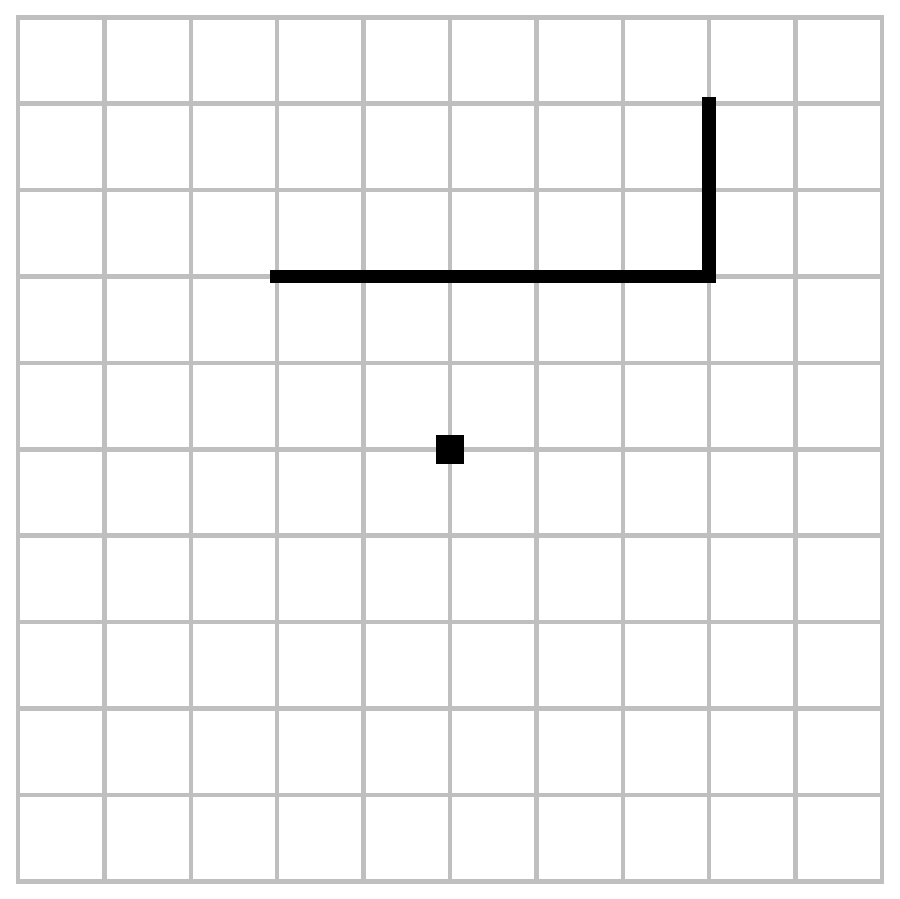} \label{fig:estring} }
\subfigure[]{\includegraphics[width=0.3\columnwidth]{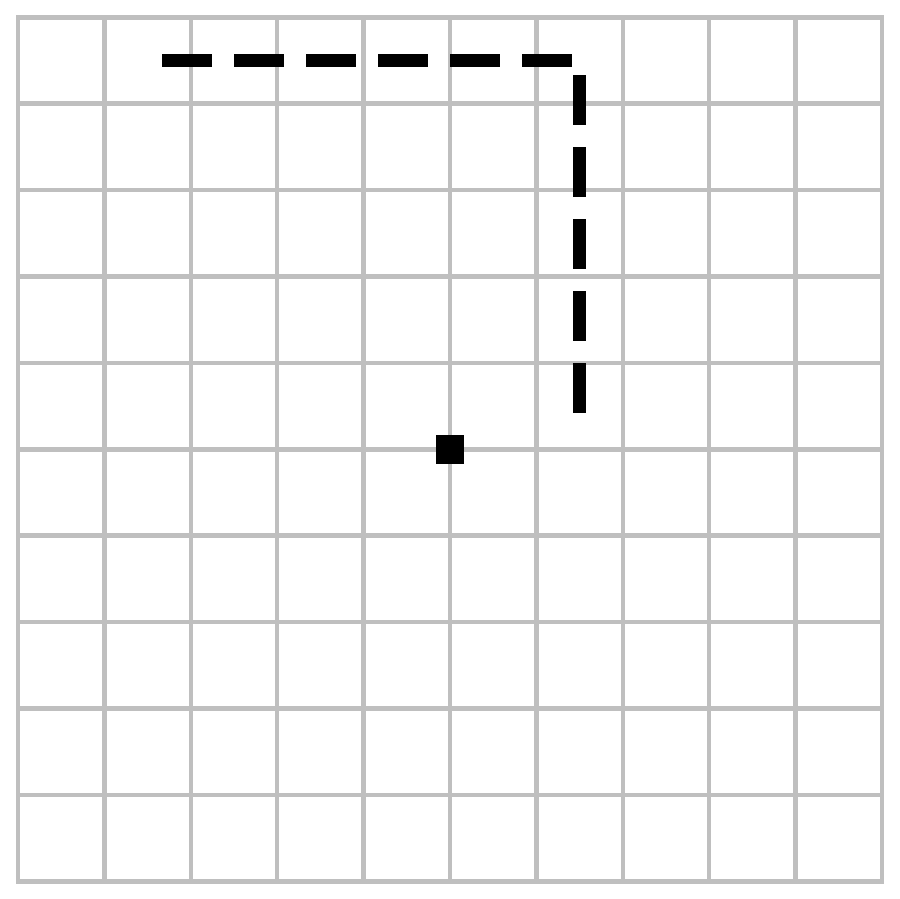} \label{fig:mstring} }
\caption{(a) Electric string.  (b) Magnetic string.}
\end{figure}

With the ground states in hand, we can work out the excited states.  Excitations come in pairs, connected by strings: $m$-strings, connecting two fluxes, consist of spin flips $\sigma^x$; $e$-strings, connecting charges, consist of $\sigma^z$.  We choose conventional contours for strings in initial states (\emph{i.e.}, those states on which we will act with some symmetry operation).  For simplicity, we take each contour $\cC(\r,\r')$ to consist of two straight segments at most; starting from the leftmost particle, the contour first goes right, then up or down as needed.  The initial-state string operators are
\begin{align}
I^e(\r_1^e,\r_2^e) = \prod_{\cC(\r,\r')} \sigma^z_{\r\r'}, \quad
I^m(\r_1^m,\r_2^m) = \prod_{\cC(\r,\r')} \sigma^x_{\r\r'} ,
\end{align}
which define corresponding two-particle states
\begin{align}
\ket{\r_1^e,\r_2^e} = I^e(\r_1^e,\r_2^e) \ket{\psi_0}, \quad 
\ket{\r_1^m,\r_2^m} = I^m(\r_1^m,\r_2^m) \ket{\psi_0} ,
\end{align}
depicted in Figs.~\ref{fig:estring} and \ref{fig:mstring}.
  
\subsection{Single-particle symmetry operators}

\begin{figure}
\subfigure[]{\includegraphics[width=0.3\columnwidth]{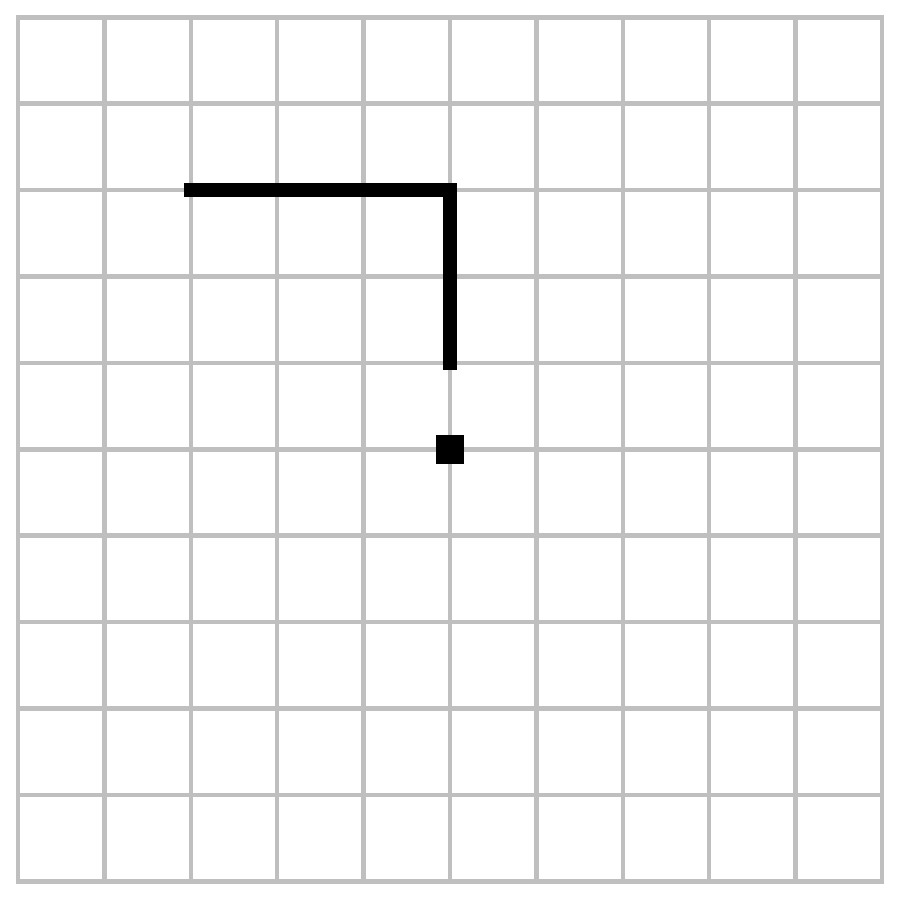} \label{fig:tx2e-a}}
\subfigure[]{\includegraphics[width=0.3\columnwidth]{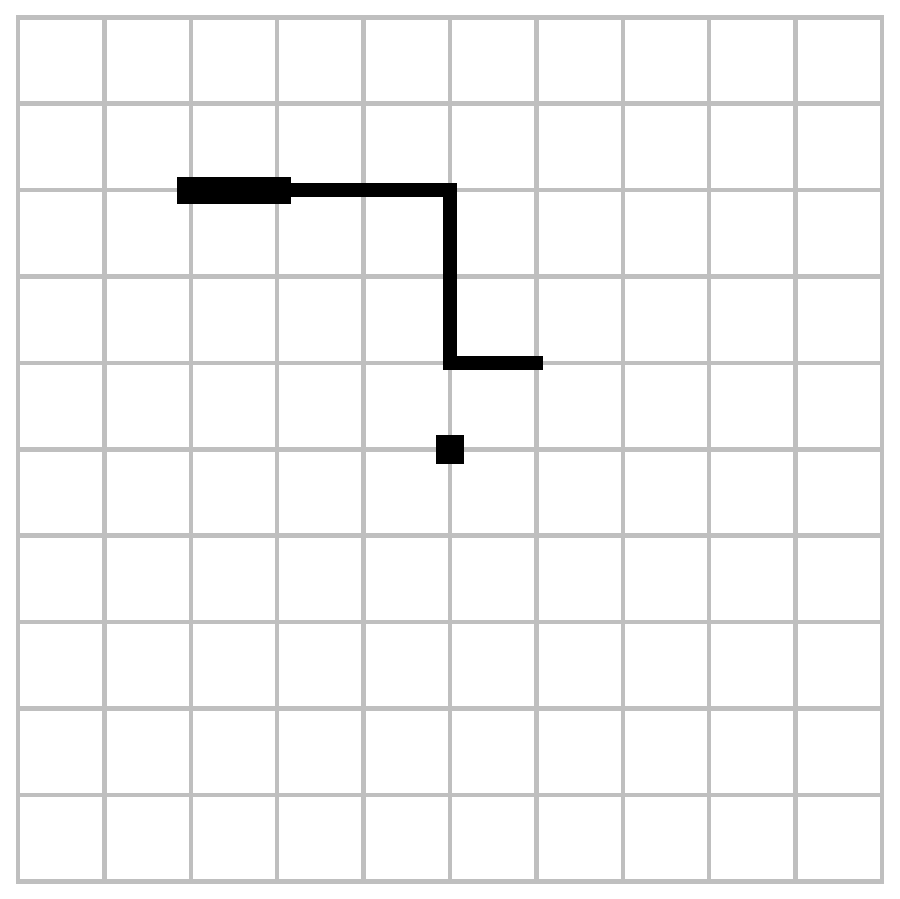} \label{fig:tx2e-b}}
\subfigure[]{\includegraphics[width=0.3\columnwidth]{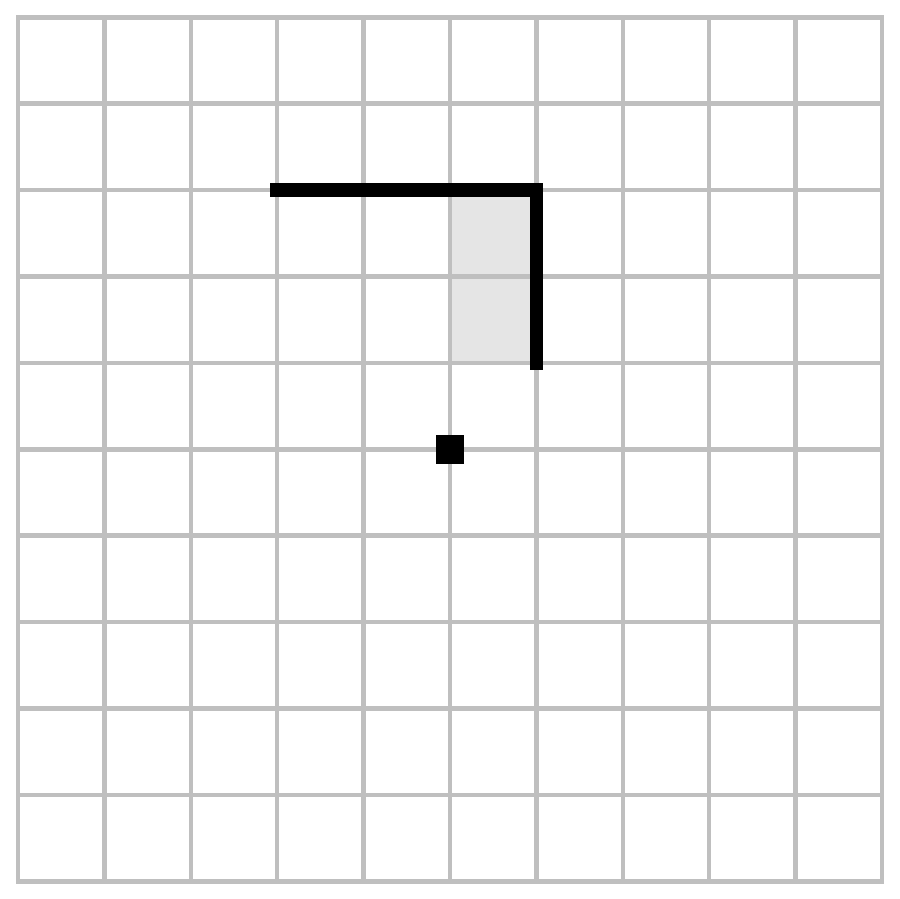} \label{fig:tx2e-c}}
\caption{\label{fig:tx2e}
The action of translation $T_x$ by one lattice spacing along $\xhat$ on the two-charge state in (a).  (b) Result of moving the quasiparticles with single spin flips, and the shaded area in (c) is swept out by sliding the strings to their final position.  In (b), the thicker string is doubled; this convention will be used in subsequent figures, except where noted.}
\end{figure}

Now consider how translation acts on a pair of charges, say.  We have 
\begin{align}
T_x \ket{\r_1^e,\r_2^e} &= \ket{\r_1^e+\xhat,\r_2^e+\xhat} \notag\\
&= s_m^{y_2-y_1} \sigma^z_{\r_1,\r_1+\xhat} \sigma^z_{\r_2,\r_2+\xhat} 
\ket{\r_1^e,\r_2^e} \notag\\
&= \left( s_m^{y_1} \sigma^z_{\r_1,\r_1+\xhat} \right)
\left( s_m^{y_2} \sigma^z_{\r_2,\r_2+\xhat} \right) \ket{\r_1^e,\r_2^e},
\end{align}
where we have dropped the superscript $e$ in many places to ease the notation.  The crucial second equality comes from sliding the vertical segment of the string $L^e$ over the possible background flux (see Fig.~\ref{fig:tx2e}).
The result factors into single-particle operators, as we argued it should on  general grounds.  Parallel arguments apply for fluxes, and for $T_y$, so we have identified single-particle operators that act on single sites $\r$,
\begin{alignat}{2}
T^e_x(\r) &= s_m^{y} \sigma^z_{\r,\r+\xhat} &\qquad
T^m_x(\r) &= s_e^{\floor{y}} \sigma^x_{\r,\r+\xhat} \notag\\
T^e_y(\r) &= s_m^{x} \sigma^z_{\r,\r+\yhat} &\qquad
T^m_y(\r) &= s_e^{\floor{x}} \sigma^x_{\r,\r+\yhat} .
\end{alignat}
Here we use the floor function $\floor{\cdot}$ so that all the phases are real.  As shorthand we write
\beq
T^e_x(\r) :\; s_m^{y},  \;\,
T^m_x(\r) :\; s_e^{\floor{y}},  \;\,
T^e_y(\r) :\; s_m^{x},  \;\,
T^m_y(\r) :\; s_e^{\floor{x}}  ,
\eeq
when we only need the single-particle phases, since the necessary factors of $\sigma^x$ and $\sigma^z$ just lie on the contours we choose to represent the operators.  

In fact, since we prefer to work with $T_x$, $P_x$, and $P_{xy}$ as generators, we will want to verify that these expressions for $T_y$ agree with the relation $T_y = P_{xy} T_x P^{-1}_{xy}$, which is done below.  Note that for fluxes, $\r$ takes values in the dual lattice, \textit{i.e.}~at the centers of faces, and the links indicated by the subscripts on $\sigma^x$ in these formulae should be thought of as links of the dual lattice.  
Also note that the toric code is special in that a quasiparticle is localized to a single vertex or plaquette.  To connect to the general formalism developed earlier, we should write quasiparticle symmetry operators that act on regions.  For the purposes of the present discussion, though, we shall just take the region $R$ to consist of a single vertex or plaquette.

\begin{figure}
\subfigure[]{\includegraphics[width=0.45\columnwidth]{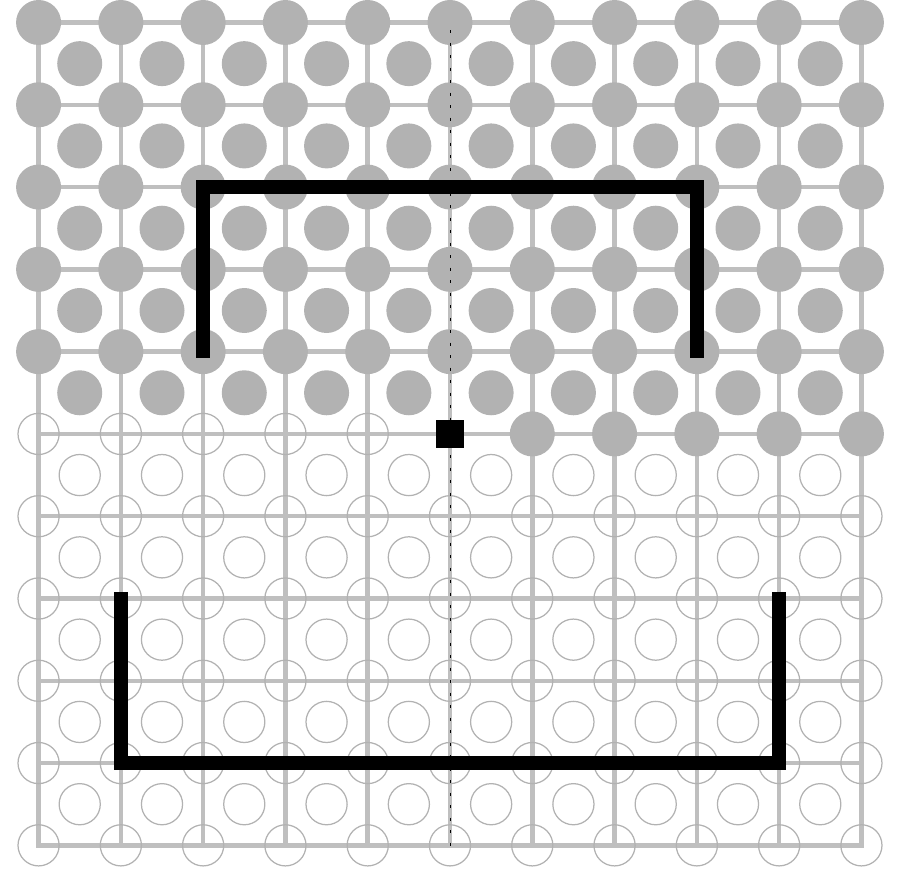} \label{fig:pxdef}}
\subfigure[]{\includegraphics[width=0.45\columnwidth]{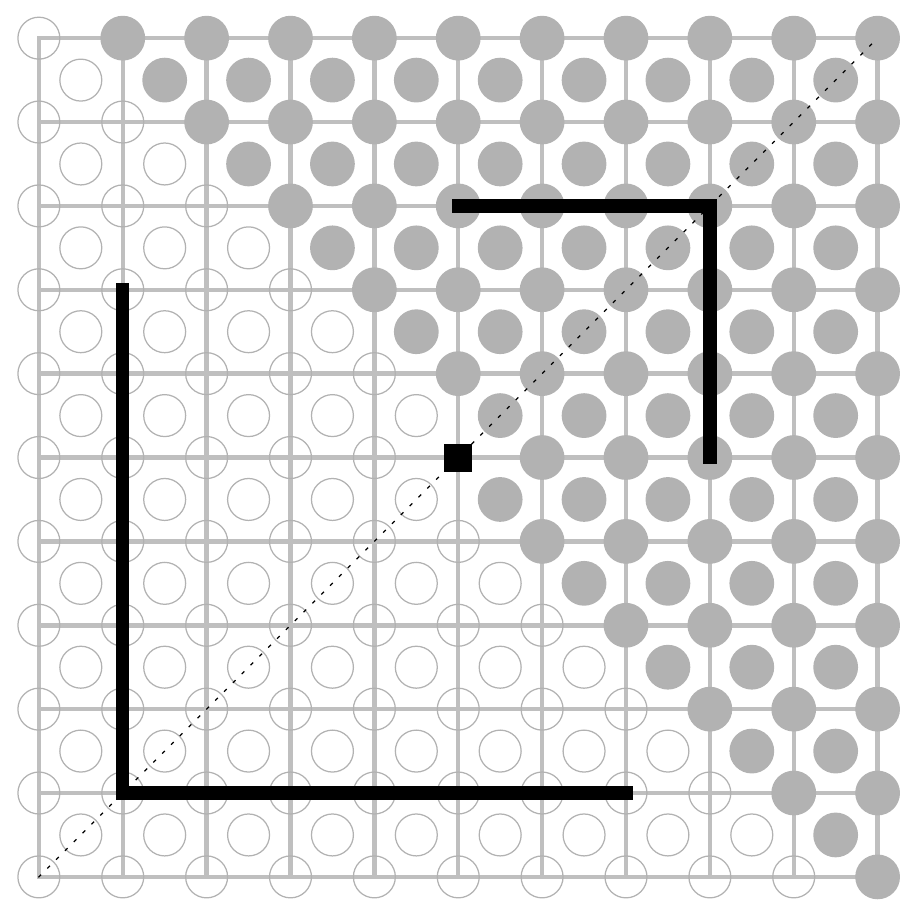} \label{fig:pxydef}}
\caption{\label{fig:pdefs}
(a) Under $P_x$, charges and fluxes in the $\bullet$ region follow contours above the origin, those in the $\circ$ region go below.  (b) The conventions for $P_{xy}$ are analogous.}
\end{figure}

The point group operations $P_x$ and $P_{xy}$ move the quasiparticles over greater distances, and so we need to choose conventional contours $\cC^e_{P_x}(\r, P_x \r)$ and so on.  We depict our choices in Fig.~\ref{fig:pdefs}.
The strings always run along the boundary of a square centered at the origin; the direction is set by the initial position of the particle, which separates into two regions, so that the string never covers more than $180^\circ$ of angle.  This choice of contour is less than obvious for $P_x$---we have chosen it to simplify computations of the product $(P_x P_{xy})^4$, which is a product of four $\pi/2$ rotations.

We define the action of a symmetry on states to transform the string operators in the natural way, by transforming the coordinates of the spin operators in the string.  Note that under point group operations, this will not always carry a conventional string to a conventional string.

Now we can work out the phases that accompany our choices of strings for point group operations.  The guiding principle is that a point group operation moves a quasiparticle along the boundary of a square centered at the origin.  The relevant distance is captured by the function
\beq
d(\r) = \max(|x|,|y|)
\eeq
in terms of the coordinates $x,y$ of the particle.  The strategy is just as for translations: (1) act on a two-particle state with the one-particle strings, and (2) identify the region over which the strings need to slide in order to arrive at the transformed state.  The calculation can be done graphically, although we show a more analytic approach as well.
The distance $d(\r)$ takes half-integer values for fluxes; suitable integer-valued functions are
\begin{align}
d_x(\r) &= \floor{d(\r)} + \theta^x_{\circ}(\r), \;\;
d_{xy}(\r) = \floor{d(\r)} + \theta^{xy}_{\circ}(\r) ,
\end{align}
where $\theta^x_{\circ}(\r)$ [$\theta^{xy}_{\circ}(\r)$] takes the value 1 on the region marked $\circ$ in Fig.~\ref{fig:pxdef} [Fig.~\ref{fig:pxydef}] and 0 otherwise.

\begin{figure}
\subfigure[]{\includegraphics[width=0.3\columnwidth]{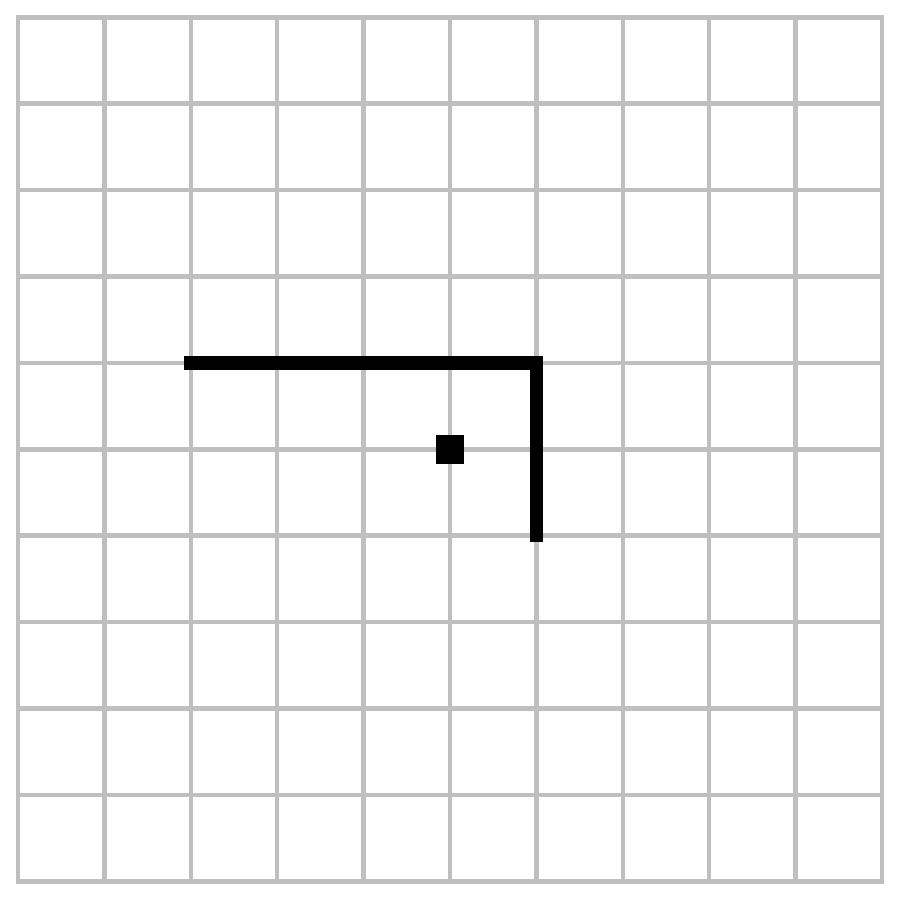} \label{fig:px2e-a}}
\subfigure[]{\includegraphics[width=0.3\columnwidth]{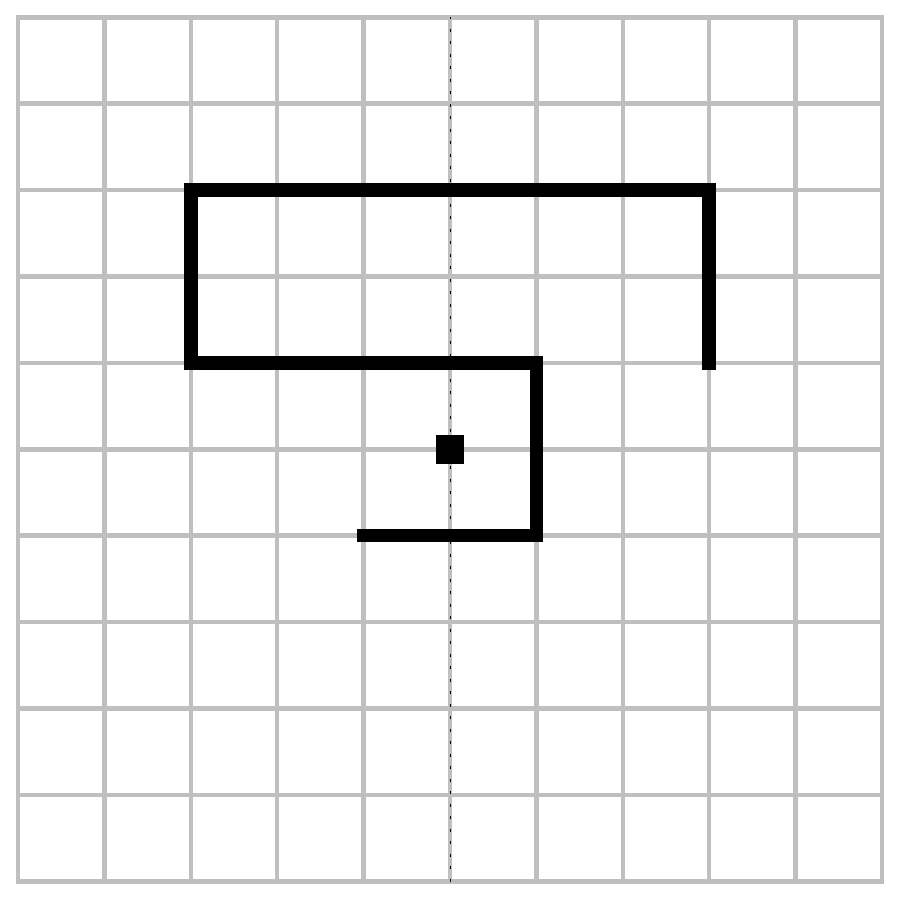} \label{fig:px2e-b}}
\subfigure[]{\includegraphics[width=0.3\columnwidth]{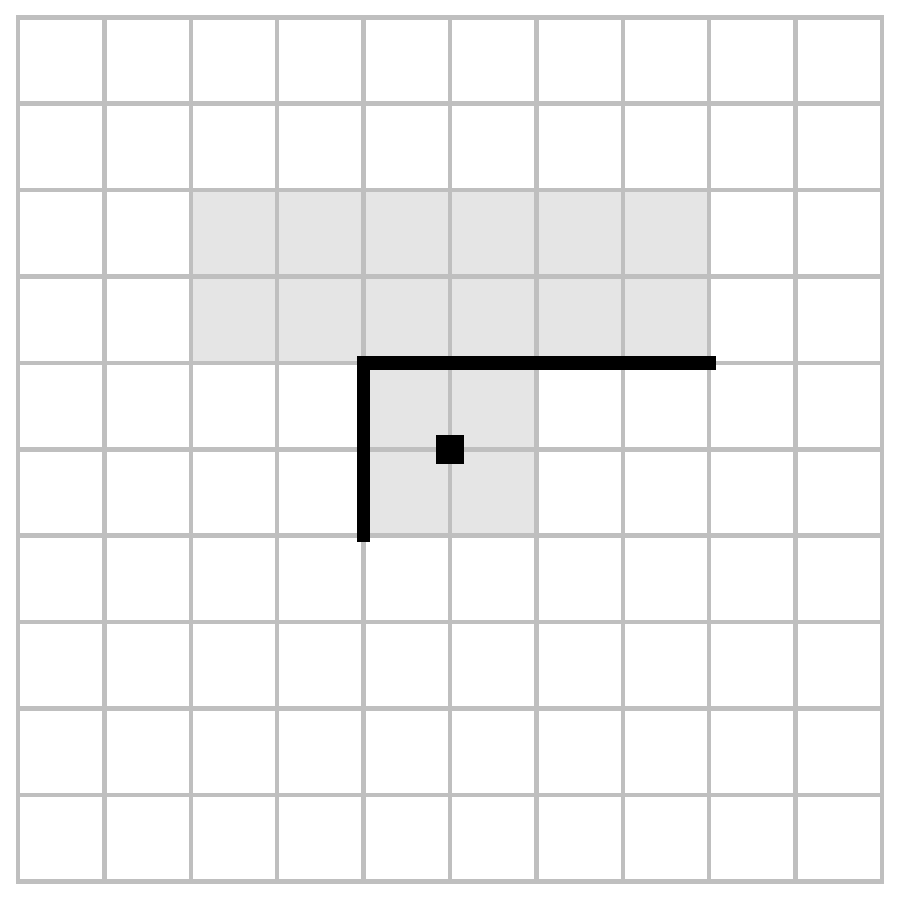} \label{fig:px2e-c}}
\caption{\label{fig:px2e}
The action of $P_x$ on a two-$e$ state.}
\end{figure}

Consider first the action of $P_x$ on a state with two charges, connected by a conventional string, as in Fig.~\ref{fig:px2e-a}.
Adding the strings of spin flips to move the quasiparticles to their reflected positions produces Fig.~\ref{fig:px2e-b}.  Then sliding all the strings to their final positions sweeps out an area, shown in Fig.~\ref{fig:px2e-c}.  If the area were odd, the wave function would pick up a sign.  However,  the relevant area will always contain an even number of background fluxes, since it will be symmetric about the axis $x=0$.  Therefore, $P_x^e$ needs no extra phase beyond the string of spin flips:
\beq
P_x^e(\r):\; 1.
\eeq

\begin{figure}
\subfigure[]{\includegraphics[width=0.3\columnwidth]{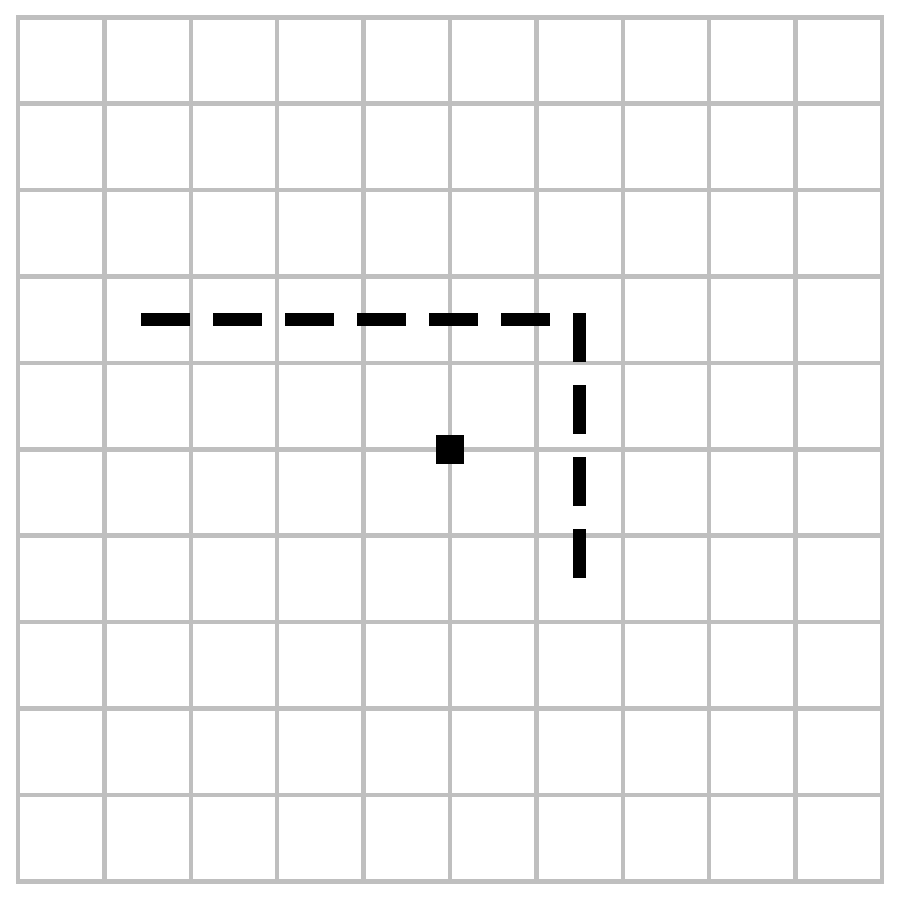} \label{fig:px2m-a}}
\subfigure[]{\includegraphics[width=0.3\columnwidth]{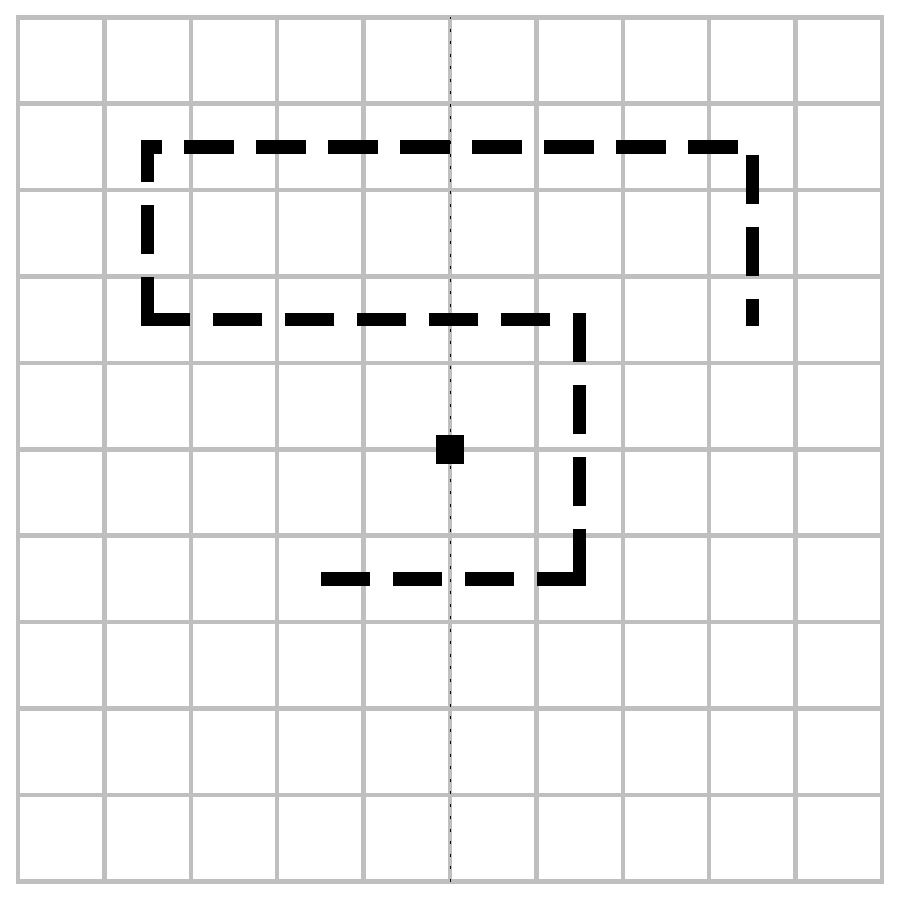} \label{fig:px2m-b}}
\subfigure[]{\includegraphics[width=0.3\columnwidth]{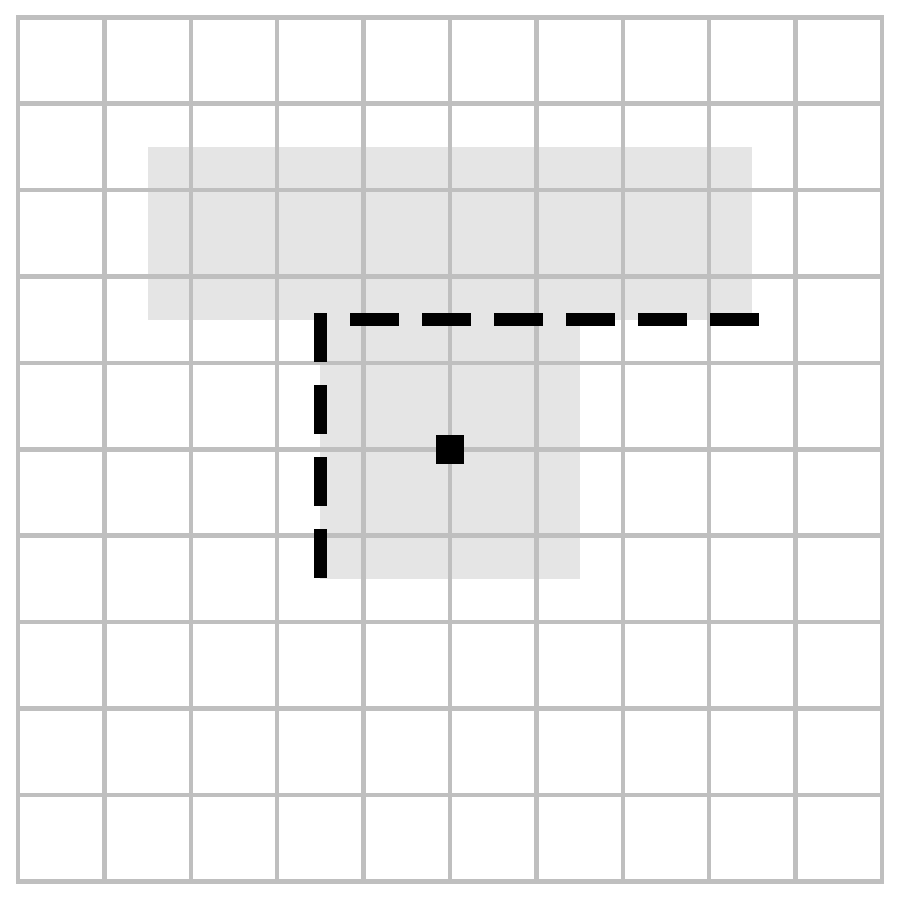} \label{fig:px2m-c}}
\caption{\label{fig:px2m}
The action of $P_x$ on a two-$m$ state.}
\end{figure}

Next, consider $P_x$ acting on a pair of fluxes (see Fig.~\ref{fig:px2m}).
Again the swept-out area has reflection symmetry, but may contain an odd number of background charges, leading to a phase factor.  All background charges contained in the swept-out area are doubled (\emph{i.e.}, come in pairs), \emph{except} those lying on the axis of reflection.  Therefore, one merely needs to count the number of charges along the reflection axis inside the swept-out area.  This is $|d_x(\r_1) - d_x(\r_2)|$ if both fluxes are in the same half plane $y\gtrless0$, and $d_x(\r_1) + d_x(\r_2)$ if one particle is at $y>0$ and one is at $y<0$, as in Fig.~\ref{fig:px2m}.  The resulting sign therefore factors as
\beq
P_x^m(\r):\; s_e^{d_x(\r)} .
\eeq

\begin{figure}
\subfigure[]{\includegraphics[width=0.3\columnwidth]{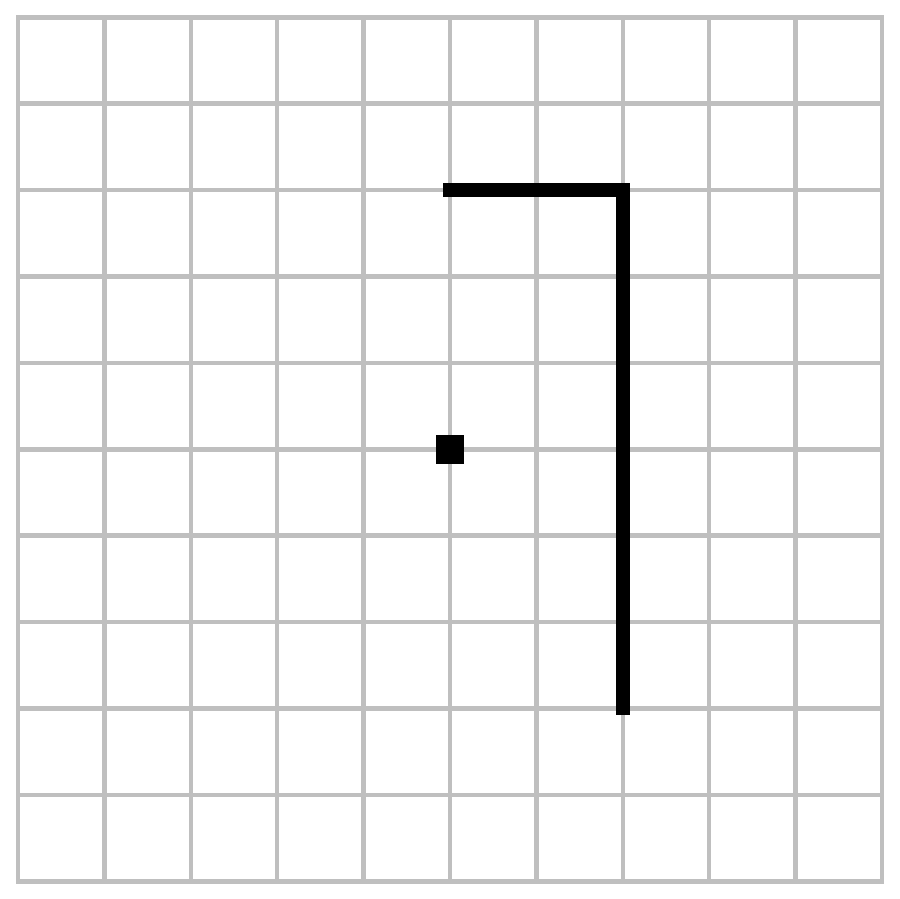} \label{fig:pxy2e-a}}
\subfigure[]{\includegraphics[width=0.3\columnwidth]{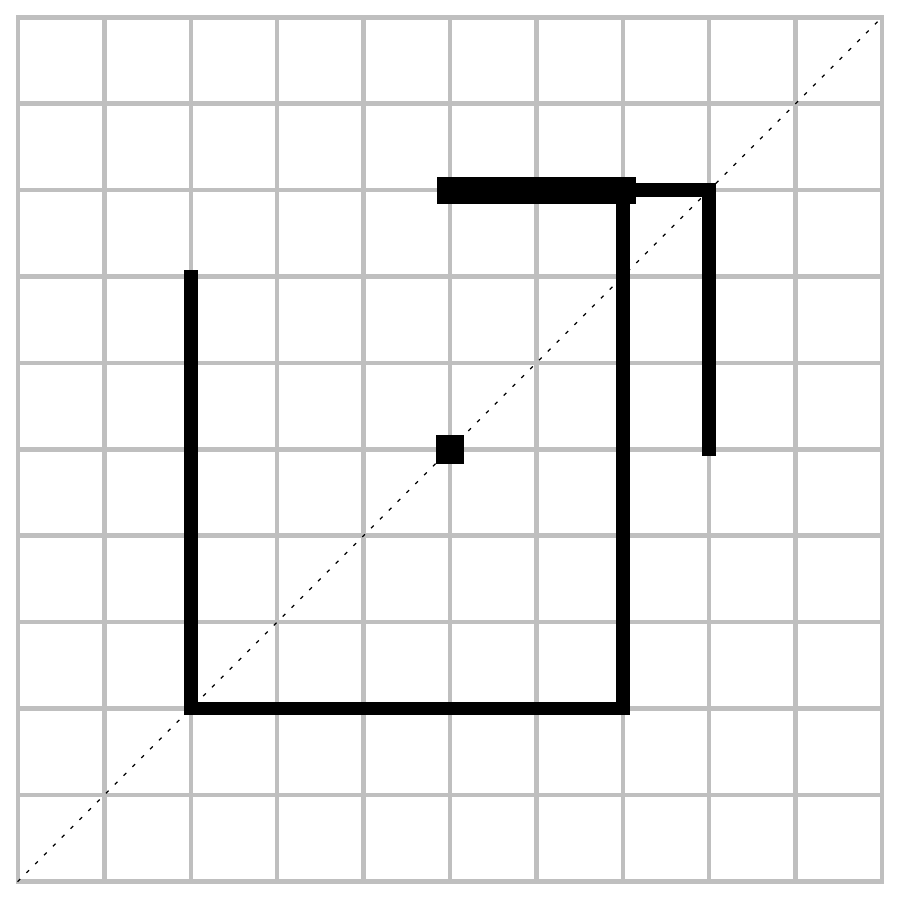} \label{fig:pxy2e-b}}
\subfigure[]{\includegraphics[width=0.3\columnwidth]{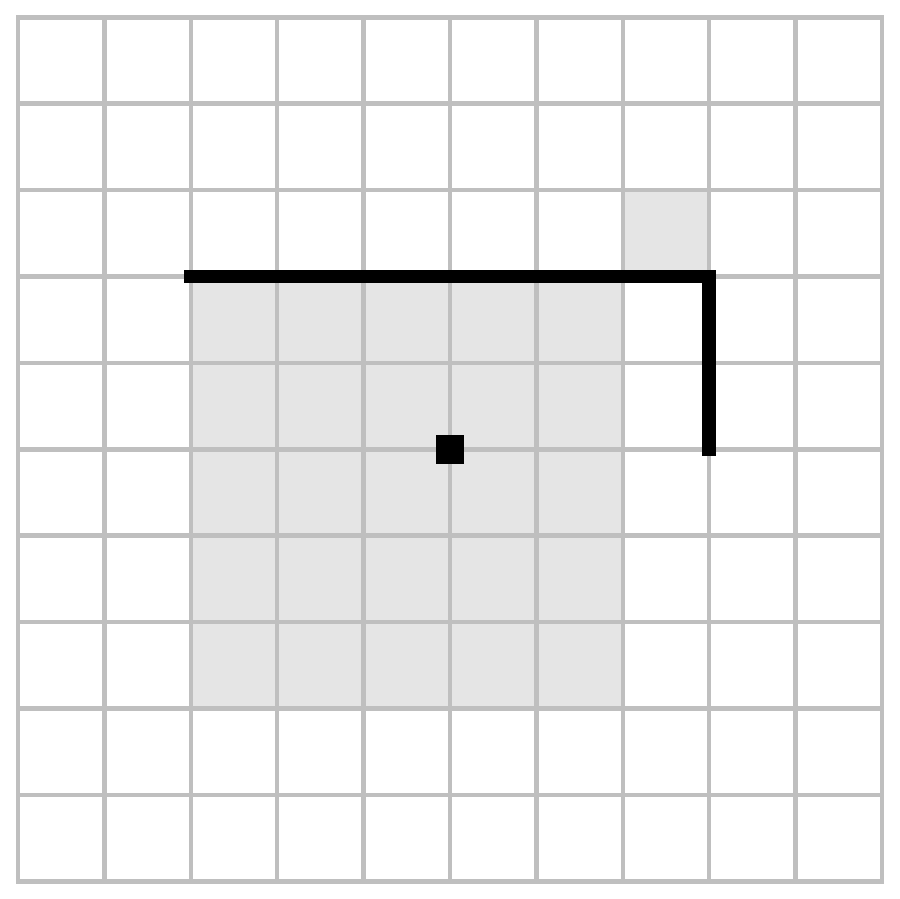} \label{fig:pxy2e-c}}
\caption{\label{fig:pxy2e}
The action of $P_{xy}$ on a two-$e$ state.}
\end{figure}

\begin{figure}
\subfigure[]{\includegraphics[width=0.3\columnwidth]{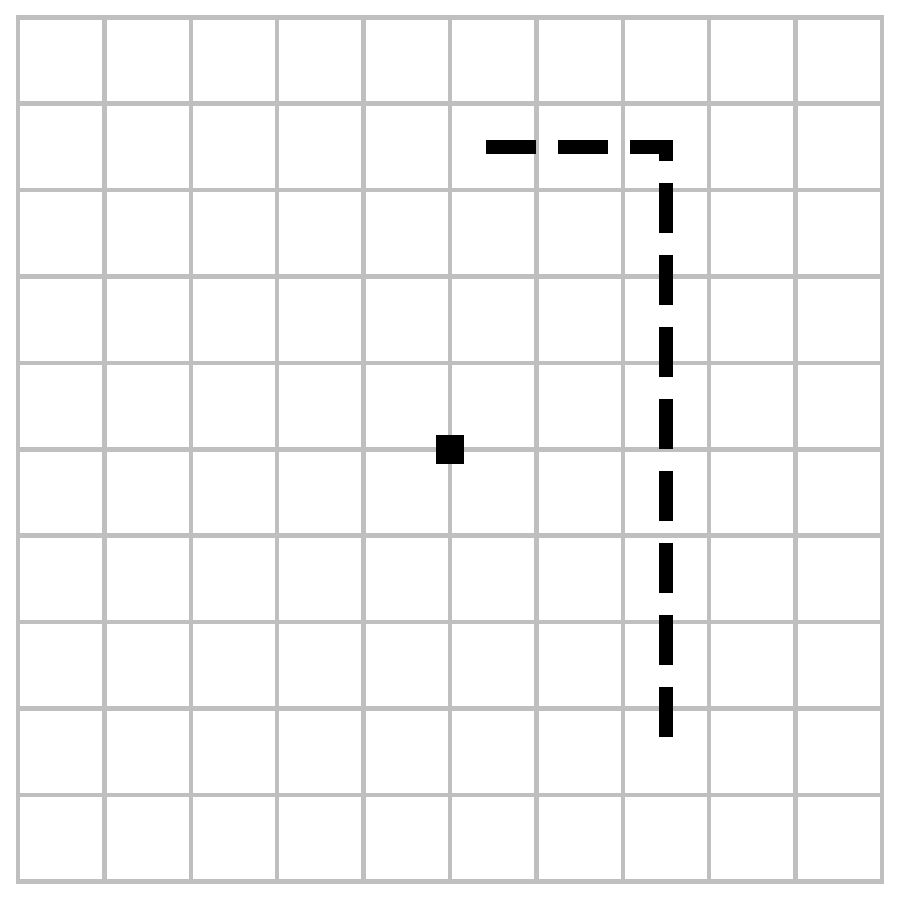} \label{fig:pxy2m-a}}
\subfigure[]{\includegraphics[width=0.3\columnwidth]{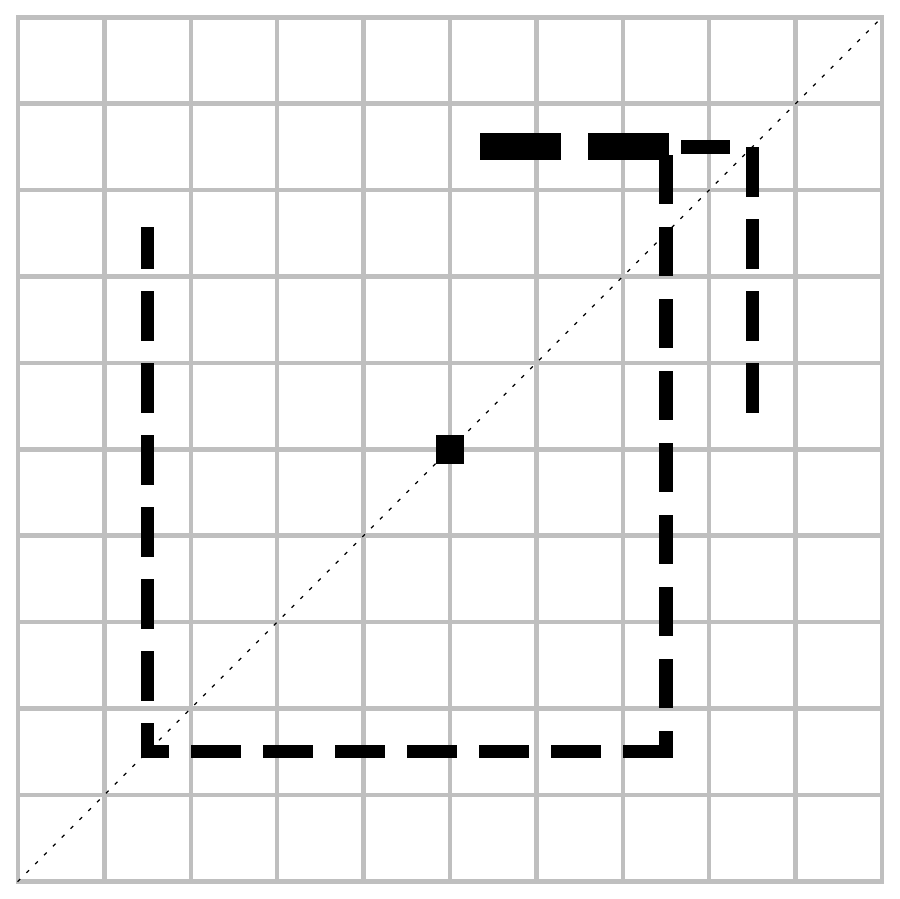} \label{fig:pxy2m-b}}
\subfigure[]{\includegraphics[width=0.3\columnwidth]{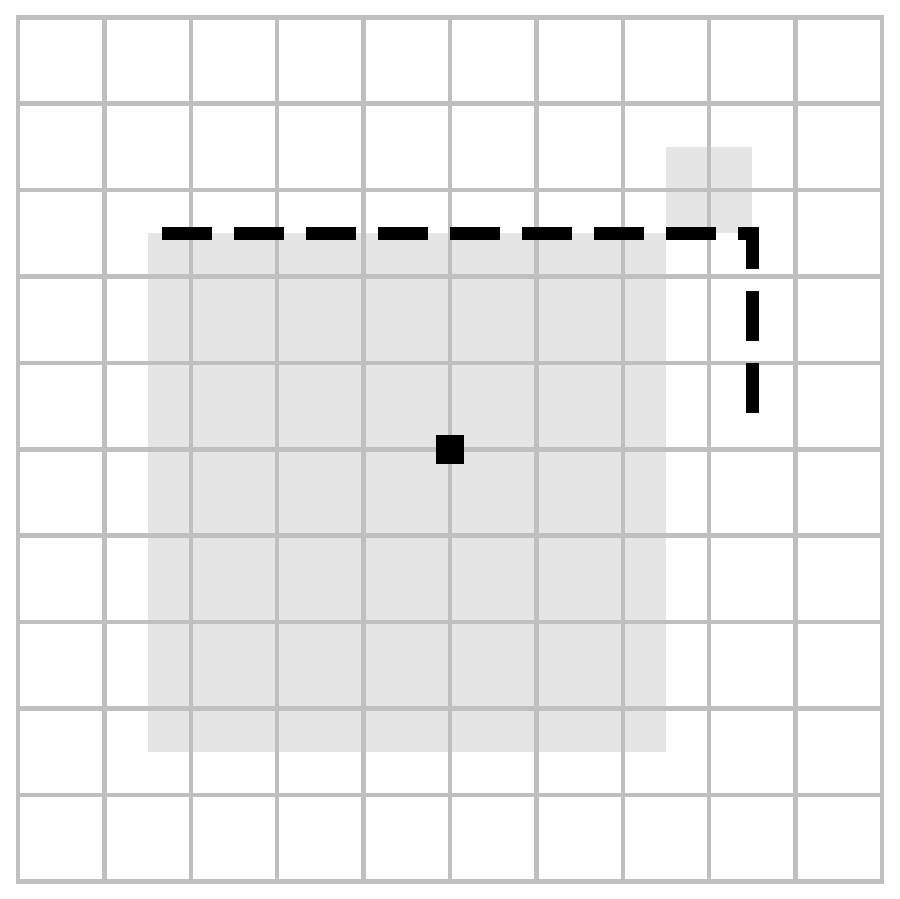} \label{fig:pxy2m-c}}
\caption{\label{fig:pxy2m}
The action of $P_{xy}$ on a two-$m$ state.}
\end{figure}

The calculation of signs for $P_{xy}$ is analogous.  In Fig.~\ref{fig:pxy2e}, we see that we need only count the number of background fluxes within the swept-out area along the line $x=y$; all the other fluxes are doubled and do not contribute a sign.
The sign is therefore set by $d(\r_1) + d(\r_2) \mod 2$, and factors as
\beq
P_{xy}^e(\r) :\; s_m^{d(\r)}.
\eeq
For fluxes, the result is similar, see Fig.~\ref{fig:pxy2m}, and
we find
\beq
P_{xy}^m(\r) :\; s_e^{d_{xy}(\r)}.
\eeq

Recall that we only want to deal with three generators, so we demand that $T_y = P_{xy} T_x P^{-1}_{xy}$ on each sector.
\begin{figure}
\subfigure[]{\includegraphics[width=0.3\columnwidth]{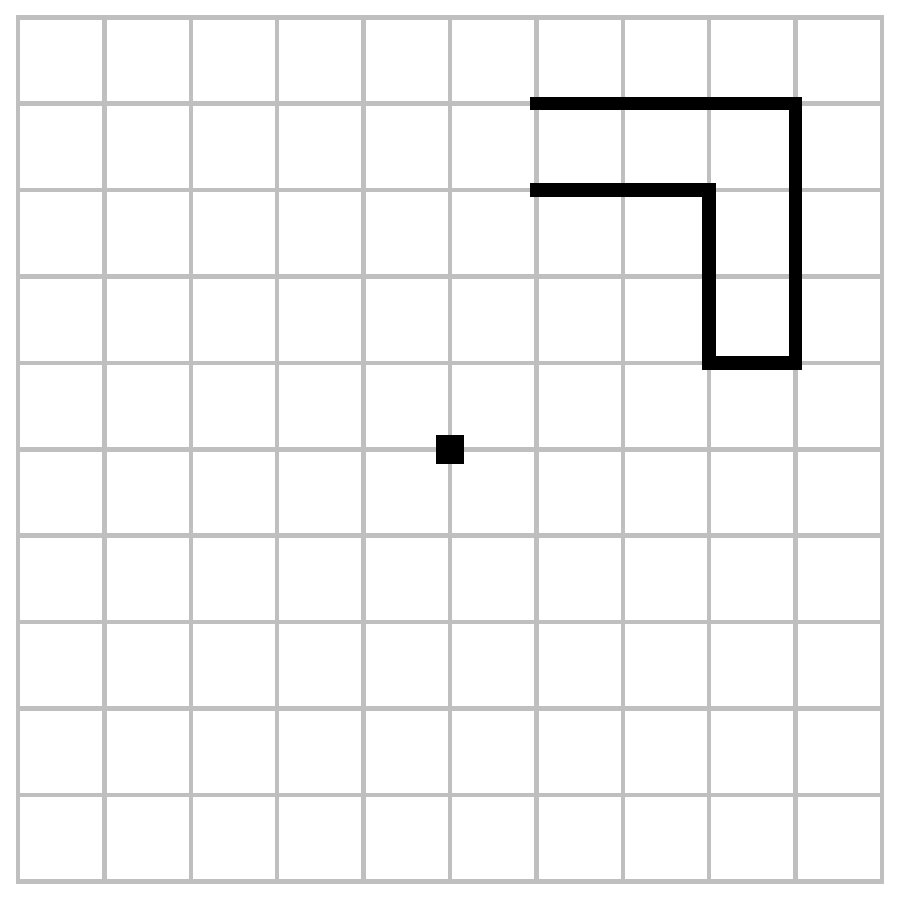} \label{fig:tye-a}}
\subfigure[]{\includegraphics[width=0.3\columnwidth]{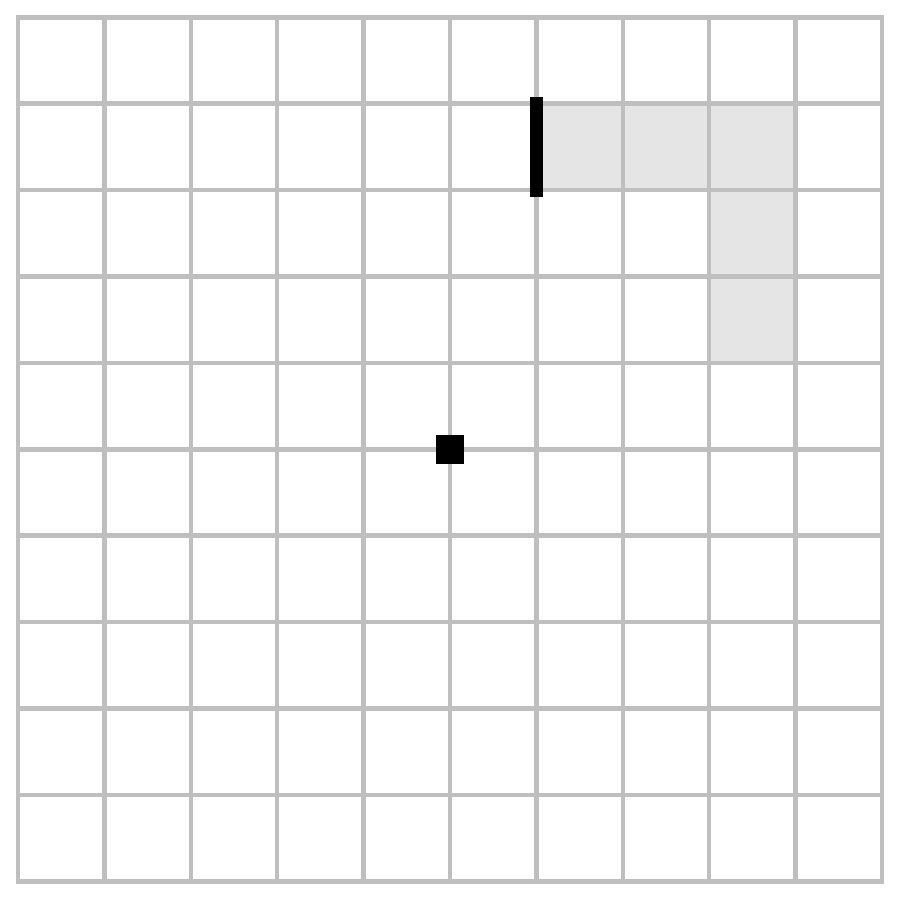} \label{fig:tye-b}}
\caption{\label{fig:tye}
(a) The action of $P_{xy} T_x P^{-1}_{xy}$ on a single charge $e$.  (b) Sliding the string of (a) to obtain a contribution to the single-particle phase of $T_y$.}
\end{figure}
Consider this sequence of operations, for example Fig.~\ref{fig:tye}.  Because $(P^e_{xy})^2 = (P^m_{xy})^2 = 1$, we can use $P_{xy}$ rather than $P^{-1}_{xy}$ for the first operation.  The phases from the single-particle operators are
\begin{multline}
P_{xy}^e(y+1,x) T_x^e(y,x) P_{xy}^e(x,y): \\
s_m^{d(y+1,x)} s_m^{x} s_m^{d(x,y)} = 
\begin{cases}
s_m^{x+1} & |y| > |x| \\ 
s_m^x & |y|<|x|
\end{cases} .
\end{multline}
We can see from Fig.~\ref{fig:tye-b} that there is an extra factor of $s_m$ from sliding the string when $|y|>|x|$, so that the appropriate phase is
\beq
T_y^e :\; s_m^x,
\eeq
as expected.  The marginal case $|y|=|x|$ gives the same result.  The same calculation gives the same result on the  $m$-sector.

\subsection{Symmetry group relations and symmetry classes}

\subsubsection{Direct computations}

\begin{figure}
\subfigure[]{\includegraphics[width=0.3\columnwidth]{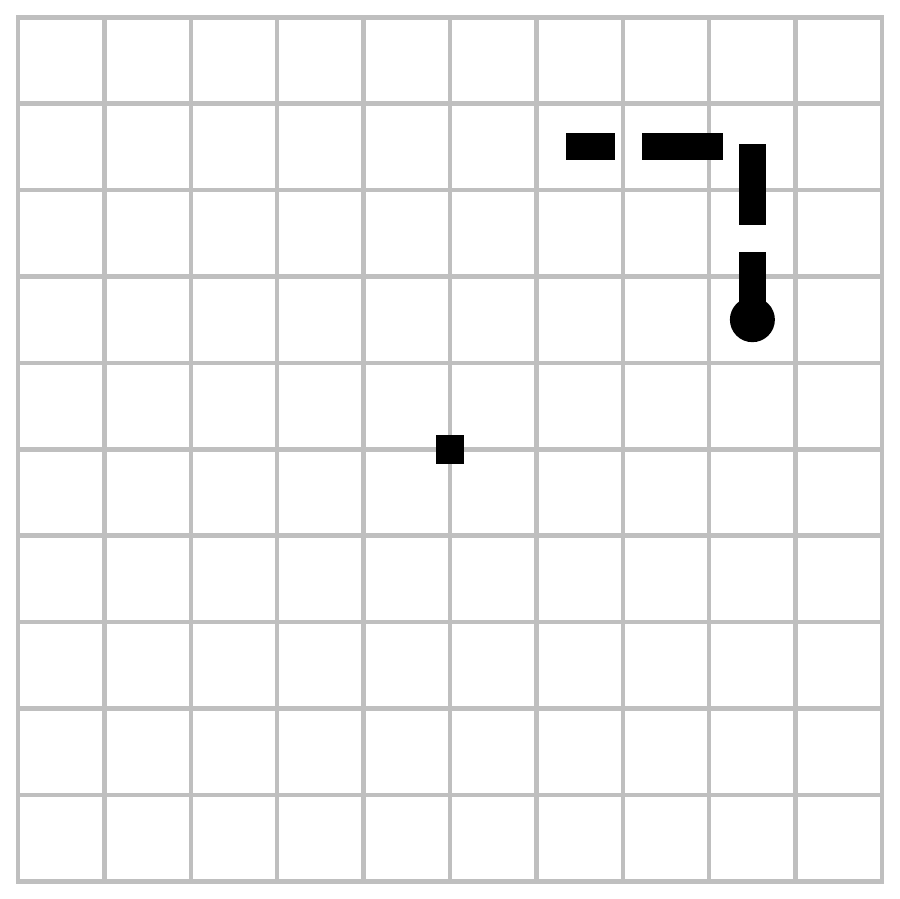} \label{fig:pxpxm-a}}
\subfigure[]{\includegraphics[width=0.3\columnwidth]{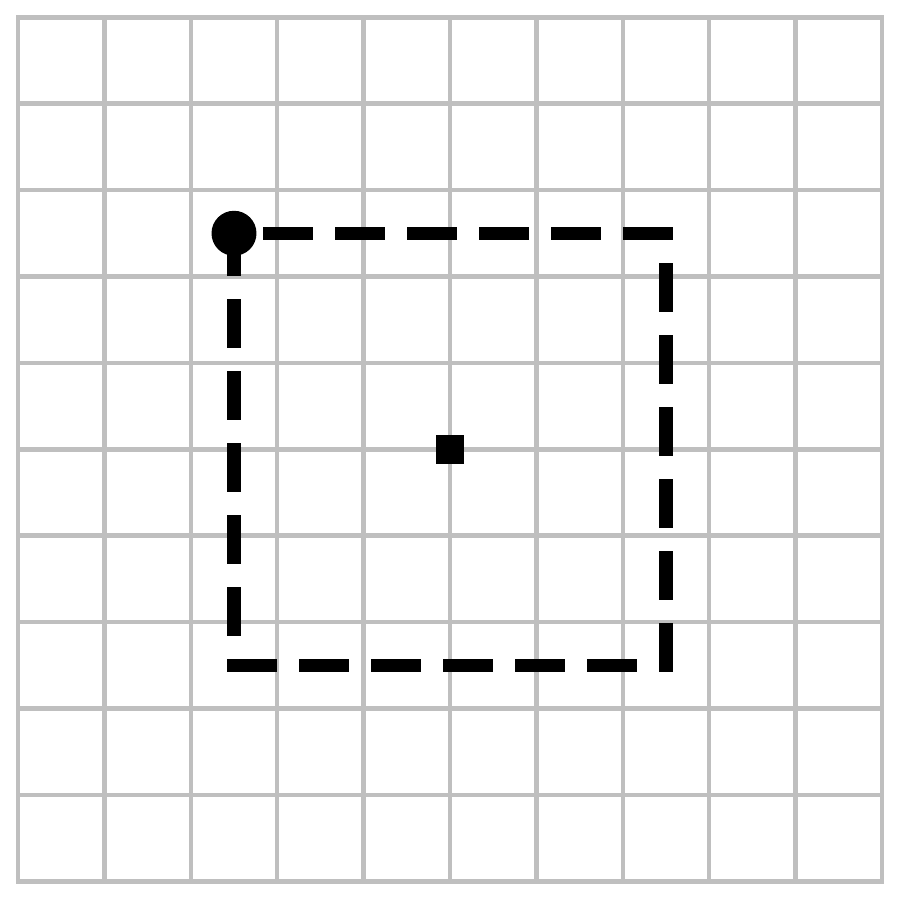} \label{fig:pxpxm-b}}
\caption{The strings for the group relation $P^{m2}_{xy}$. In (a), recall that the thicker strings are doubled; that is, they represent two strings acting in the same position.}
\end{figure}

Let us work out some of the symmetry relations.  Consider the relation $P_{xy}^2 = 1$ as it acts on a single flux.  In most cases this involves putting down a pair of identical string operators, which square to 1 trivially, as in Fig.~\ref{fig:pxpxm-a}.
However, our conventions imply that sometimes the symmetry operators involved in the relation may enclose some background charge, as in Fig.~\ref{fig:pxpxm-b}.  The relation must be constant on a given superselection sector, and this example provides a test of this claim.  The flux in question is located at $\r=(-2\tfrac{1}{2},2\tfrac{1}{2})$ in units of the lattice spacing.  Then we compute
\begin{align}
P_{xy}^m(2 \tfrac{1}{2},-2 \tfrac{1}{2}) P_{xy}^m(-2 \tfrac{1}{2},2 \tfrac{1}{2}) &= s_e^{\floor{2\tfrac{1}{2}}} s_e^{\floor{2\tfrac{1}{2}}+1} \cC \notag\\
&= s_e \cC,
\end{align}
where $\cC$ is the box drawn.  This expression must be evaluated on the ground state.  Since $\cC$ contains an odd number of background charges it evaluates to $s_e$, so that we find 
\beq
\left(P_{xy}^m\right)^2 = 1
\eeq 
in all cases, as expected.  Note that the background charge appears twice in this calculation, once explicitly and once in the construction of the single particle sign factor, and both are important in order to arrive at a consistent answer.

\begin{figure}
\subfigure[]{\includegraphics[width=0.3\columnwidth]{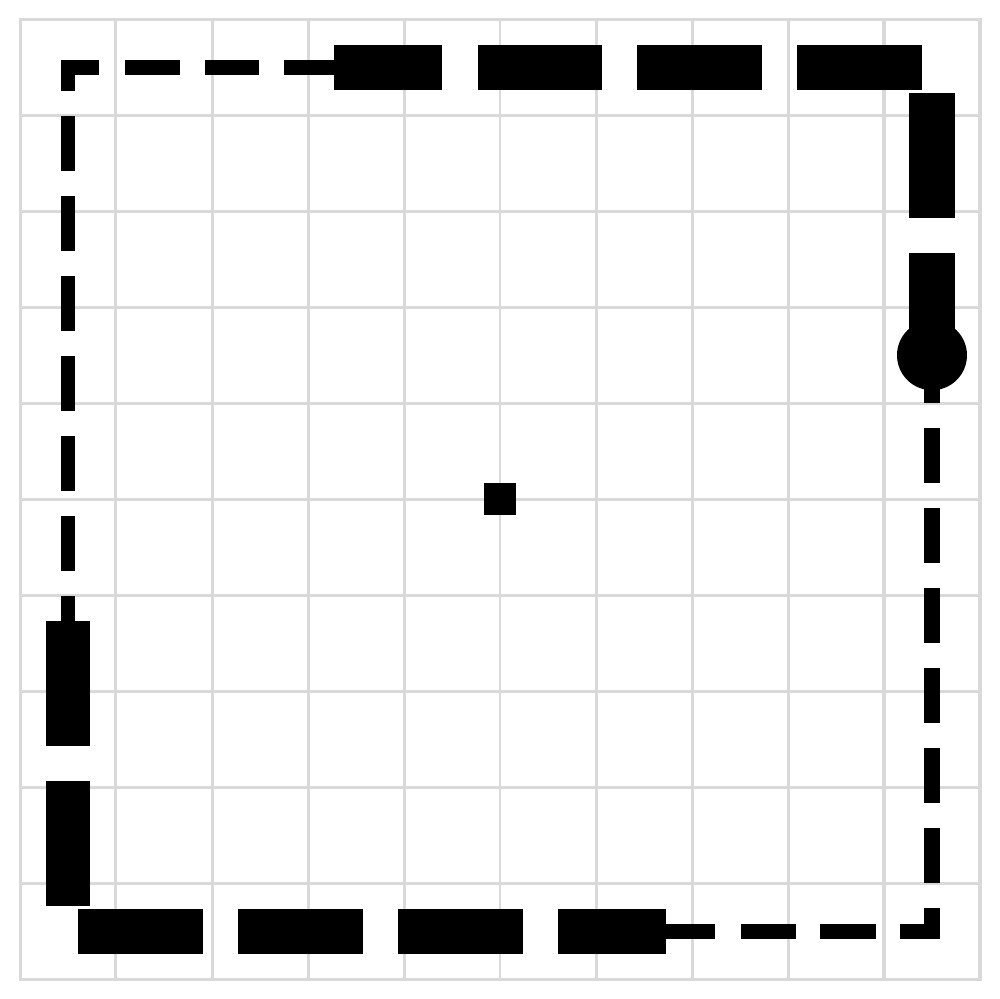} \label{fig:pxpxym}}
\subfigure[]{\includegraphics[width=0.3\columnwidth]{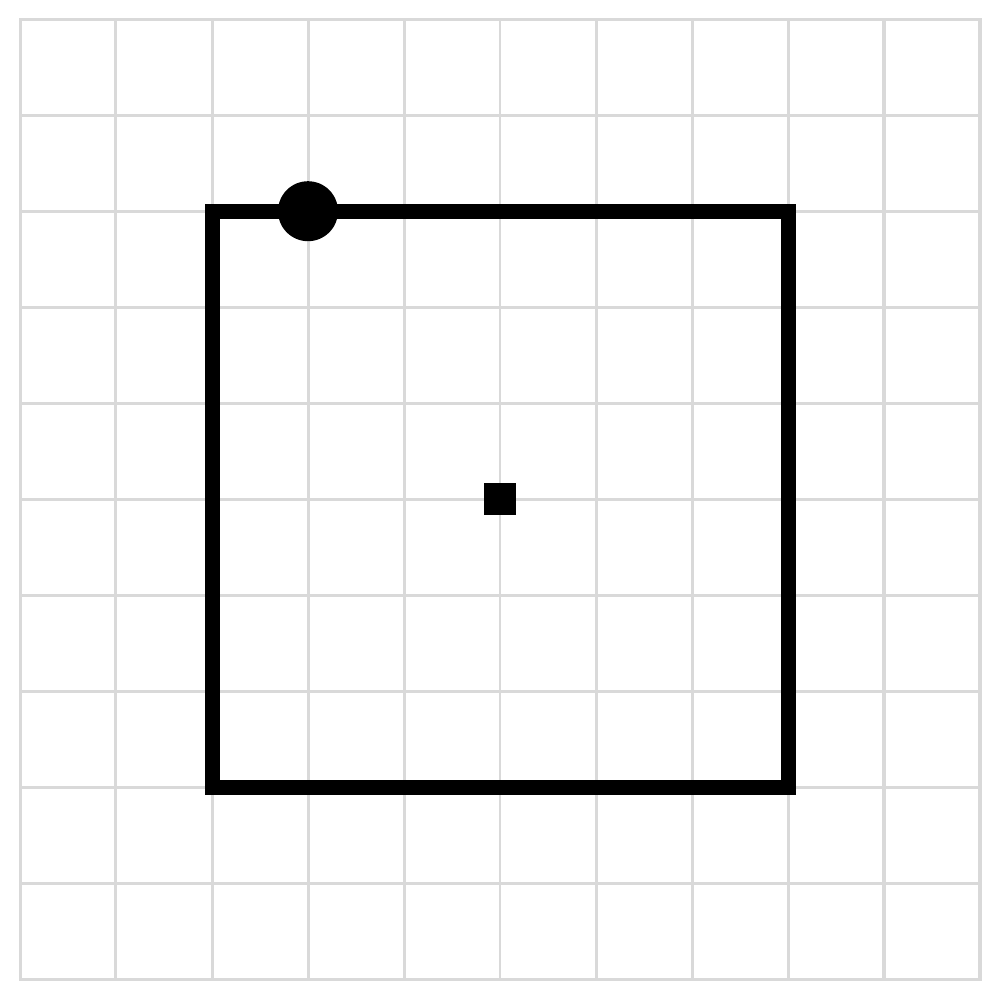} \label{fig:pxpxye}}
\caption{The strings for the group relation $(P_x P_{xy})^4$.  In (a), the darker curves are \emph{triple} strings.  In (b), all segments are drawn with the same weight. \label{fig:pxpxy}}
\end{figure}

The calculation is essentially identical for both $P_x^2$ and $P_{xy}^2$ in both the charge and flux sectors.  The other point group relation, $(P_x P_{xy})^4 = 1$, which describes fourfold rotations, is the most complicated.  We have arranged our definitions of the quasiparticle symmetry operators so as to simplify the computation of this relation.
Our definitions are such that the relation always gives a square contour that encircles the origin once, see Fig.~\ref{fig:pxpxy} (some sections of the contour are traversed three times in general).
In the case of a charge, this square encloses an even number of fluxes, while for a flux it encloses an odd number of charges.  One can work out that the single-particle signs always cancel and contribute nothing.  Therefore, one finds that 
\beq
(P_x^e P_{xy}^e)^4 = 1, \quad (P_x^m P_{xy}^m)^4 = s_e .
\eeq

The other relations involve translations.  Consider first the translation relation $T_x T_y T_x^{-1} T_y^{-1} = 1$.  On both charges and fluxes, the single-particle signs do not contribute,
\begin{multline}
T_x(x-1,y) T_y(x-1,y-1) T_x^{-1}(x,y-1) T_y^{-1}(x,y): \\
s^{\floor{y}} s^{\floor{x-1}} s^{\floor{y-1}} s^{\floor{x}} =1,
\end{multline}
so the relation just measures the background charge or flux enclosed in the elementary loop around which the group relation transports the particle in question.
That is,
\begin{equation}
T^e_x T^e_y T^{e-1}_x T^{e-1}_y = s_m , \quad
T^m_x T^m_y T^{m-1}_x T^{m-1}_y  = s_e \text{.}
\end{equation}

Next, consider the relation $T_y P_x T_y^{-1} P_x^{-1} = 1$ on the flux sector, for example.
\begin{figure}
\subfigure[]{\includegraphics[width=0.3\columnwidth]{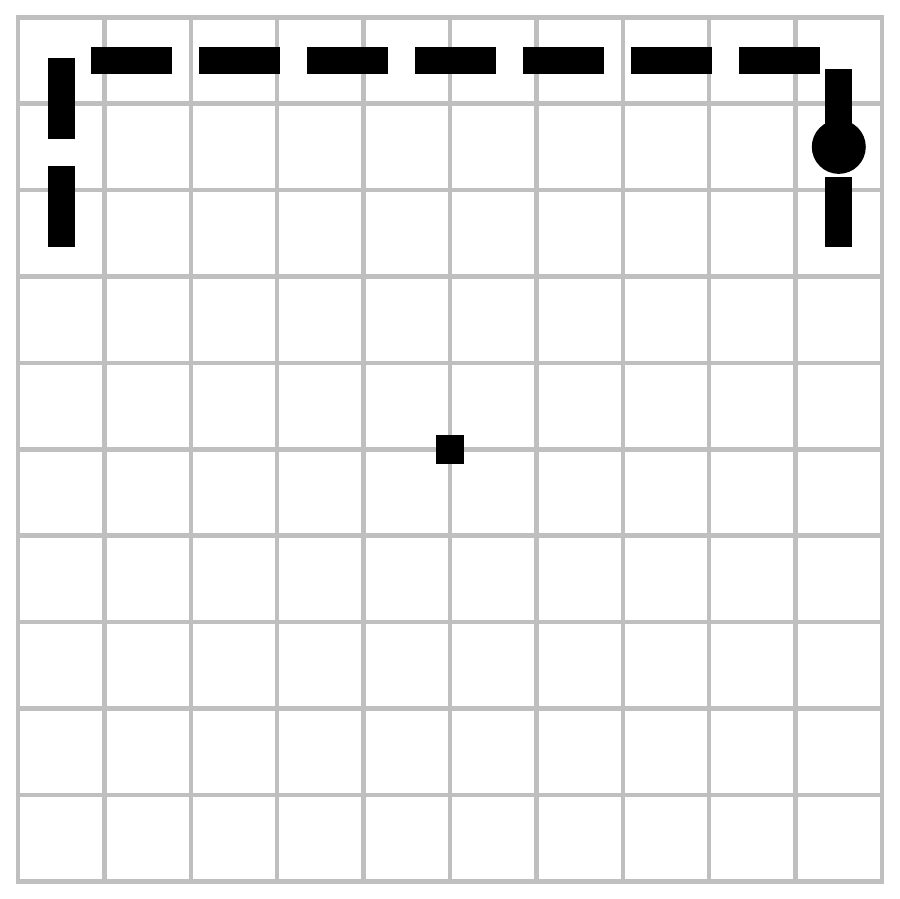} \label{fig:pxtym-a}}
\subfigure[]{\includegraphics[width=0.3\columnwidth]{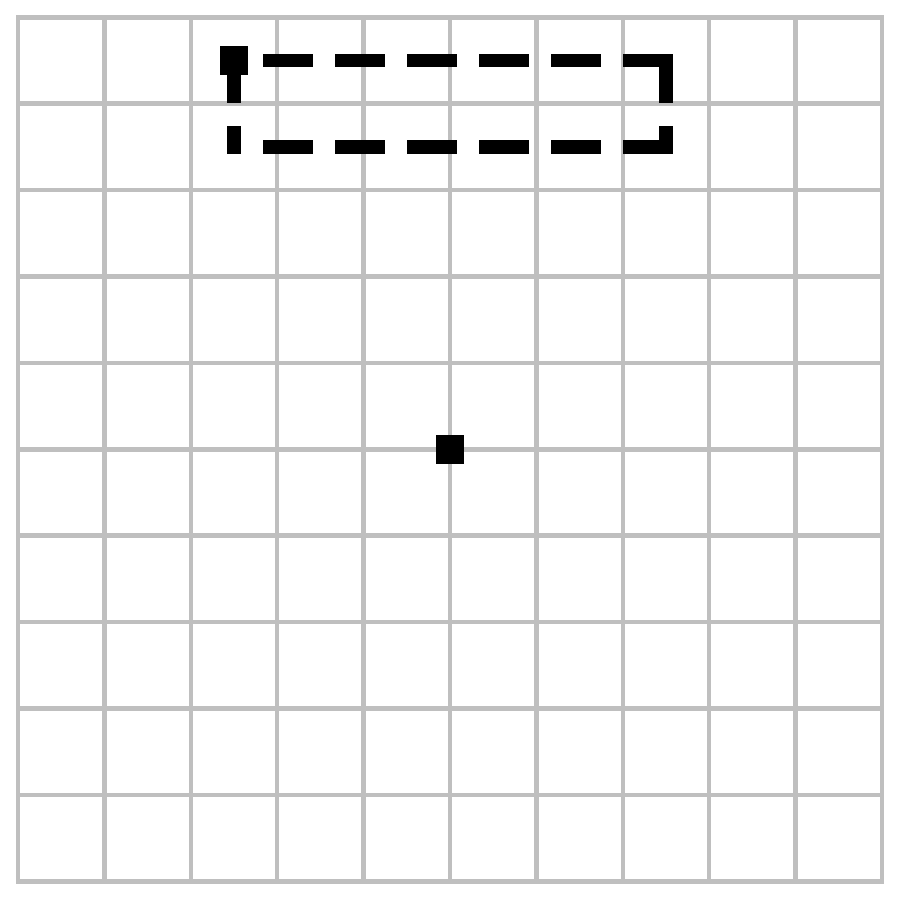} \label{fig:pxtym-b}}
\caption{The strings for the group relation $T_y^m P^m_x T_y^{m-1} P^{m-1}_x$.}
\end{figure}
There are cases where the corresponding strings do not enclose any vertices, see Fig.~\ref{fig:pxtym-a}, and others in which they do, see Fig.~\ref{fig:pxtym-b}.  For the case with no vertices enclosed, the $P_x$ segments contribute no net sign (they cancel), so one is left with the product of the $T_y$ phases, $s_e^{\floor{x} + \floor{-x}} = s_e$ since $x=n+1/2$ for a flux, for some integer $n$.  In the case where vertices are enclosed, the $P_x$ contributions do not cancel each other, but rather cancel the enclosed flux, so that the result is always
\beq
T_y^m P_x^m T_y^{m-1} P_x^{m-1} = s_e .
\eeq
In the case of a charge, the signs from $T_y$ give unity because the charges lie at integer positions, and
\beq
T_y^e P_x^e T_y^{e-1} P_x^{e-1} = 1  .
\eeq
Finally, the relation $T_x P_x T_x P^{-1}_x = 1$ is sufficiently simple to work through that we do not discuss it here.

In the end, we arrive at the result
\begin{alignat}{2}
\left(P^e_x\right)^2 &= 1,    &\quad \left(P^m_x\right)^2 &= 1,    \notag\\
\left(P^e_{xy}\right)^2 &= 1, &\quad \left(P^m_{xy}\right)^2 &= 1, \notag\\
\left(P^e_x P^e_{xy}\right)^4 &= 1,     &\quad 
\left(P^m_x P^m_{xy}\right)^4 &= s_e,    \notag\\
T^e_x T^e_y T_x^{e-1} T_y^{e-1} &= s_m, &\qquad
T^m_x T^m_y T_x^{m-1} T_y^{m-1} &= s_e,  \notag\\
T_x^e P_x^e T_x^e P_x^{e-1} &= 1,   &\qquad 
T_x^m P_x^m T_x^m P_x^{m-1} &= 1,    \notag\\
T_y^e P_x^e T_y^{e-1} P_x^{e-1} &= 1,   &\qquad 
T_y^m P_x^m T_y^{m-1} P_x^{m-1} &= s_e .
\end{alignat}
We have thus shown that, depending on $s_e$ and $s_m$, the toric code model realizes four symmetry classes.  

One interesting feature of this result is the modification of the rotation relation $\left(P_x P_{xy}\right)^4$ in the  $m$-sector.  The difference is essentially geometrical; a square with vertices (charges) at the boundary contains an even number of plaquettes (fluxes), but a square with faces at the boundary contains an odd number of vertices.

\subsubsection{Alternate approach to relations in the flux sector}

The computations above are all simpler in the charge sector, because we have chosen the point group generators to leave a vertex invariant.  We could also have centered the point group on a face of the lattice.  Indeed, there is a simple group automorphism that swaps the two: we can replace $P_x$ by $\tilde{P}_x = T_x P_x$.  This observation provides a simple way to use the $e$-sector results of the previous section to obtain the relations for the $m$-sector.  First, using the known relations for generators $P_x, P_{xy}, T_x$, we simply compute the relations for the new generators $\tilde{P}_x, P_{xy}, T_x$. We find
\begin{alignat}{2}
P_x^2 &= \sigma_{px}, &\quad 
\tilde{P}_x^2 &= \sigma_{px} \sigma_{txpx}, \notag\\
P_{xy}^2 &= \sigma_{pxy}, & 
P_{xy}^2 &= \sigma_{pxy}, \notag\\
\left(P_x P_{xy}\right)^4 &= \sigma_{pxpxy},     &\quad 
\left(\tilde{P}_x P_{xy}\right)^4 &= \sigma_{pxpxy} \sigma_{txty}, \notag\\
T_x T_y T_x^{-1} T_y^{-1} &= \sigma_{txty}, &\qquad
T_x T_y T_x^{-1} T_y^{-1} &= \sigma_{txty},  \notag\\
T_x P_x T_x P^{-1}_x  &= \sigma_{txpx},   &\qquad 
T_x \tilde{P}_x T_x \tilde{P}^{-1}_x  &= \sigma_{txpx},    \notag\\
T_y P_x T_y^{-1} P_x^{-1} &= \sigma_{typx},   &\qquad 
T_y \tilde{P}_x T_y^{-1} \tilde{P}_x^{-1} &= \sigma_{typx} \sigma_{txty} .
\end{alignat}

To use this to find the $m$-sector fractionalization class, we view $P_x$ and $P_{xy}$ as point group operations centered on a plaquette rather than a site.  Then, $\tilde{P}_x$ and $P_{xy}$ are site-centered point group generators.  In terms of the plaquette-centered generators, calculation of the $m$-sector relations is identical to the calculation of the $e$-sector generators in the previous section, and we find $\sigma_{txty} = s_e$, with all other parameters equal to unity.  Passing from plaquette to site centered generators using the relations above, we recover the results of the previous section.

\section{Quantum numbers of degenerate ground states}
\label{sec:qndgs}

In $\zz$ topologically ordered phases, it is well known that the fourfold-degenerate ground states on a torus can have different symmetry quantum numbers.  As long as the symmetry is preserved, any discrete information contained in these quantum numbers is a robust, universal property of a $\zz$ spin liquid phase.  Even though the ground states belong to the $1$-sector, their quantum numbers can be partially determined given the symmetry class.  Here, we do this for the simple case of translation symmetry alone.  The analysis here can be generalized to other symmetry groups; it may be useful to do this in future work.

Consider a finite-size system with periodic boundary conditions, and translation symmetry generated by $T_x$ and $T_y$.  We do not assume any particular Bravais lattice, and $x$ and $y$ are just labels for two primitive lattice translations.  We suppose the system has linear dimensions $(N_x, N_y)$, where for instance $N_x$ is the number of primitive cells in the $x$-direction.

The crucial observation is that the generators of the loop algebra can be viewed as translation of an anyon around a loop.  That is, we make the associations
\begin{eqnarray}
{\cal L}^e_{\mu} &=& (T^e_{\mu} )^{N_{\mu} } \\
{\cal L}^m_{\mu} &=& (T^m_{\mu} )^{N_{\mu} } \text{,} \label{eqn:m-association}
\end{eqnarray}
where $\mu = x,y$.
This suggests that, for instance,
\begin{equation}
T_x {\cal L}^e_{\mu} T^{-1}_{x}  = T^e_x (T^e_{\mu} )^{N_{\mu} }  (T^e_x)^{-1} \text{,}
\end{equation}
and so on.  

At this point, we can proceed to consider the three distinct symmetry classes.  In each case, we determine the \emph{relative} crystal momenta of the four ground states from the above relations.  We have also checked these results in the toric code model by direct calculation of the ground state quantum numbers.

In the class $\sigma^e_{txty} = \sigma^m_{txty} = 1$, we have
\begin{eqnarray}
T_{\mu} {\cal L}^e_{\nu} T^{-1}_{\mu} &=&  {\cal L}^e_{\nu} \\
T_{\mu} {\cal L}^m_{\nu} T^{-1}_{\mu} &=&  {\cal L}^m_{\nu} \text{,} 
\end{eqnarray}
independent of $(N_x, N_y)$.  Suppose $|\psi_0 \rangle$ is the ground state satisfying 
${\cal L}^e_{\mu} | \psi_0 \rangle = | \psi_0 \rangle$.  This state must have a definite crystal momentum since translations commute with ${\cal L}^e_{\mu}$.  A basis for the ground-state subspace is given by 
$\{ | \psi_0 \rangle, {\cal L}^m_x  | \psi_0 \rangle, {\cal L}^m_y  | \psi_0 \rangle, {\cal L}^m_x {\cal L}^m_y  | \psi_0 \rangle \}$, and clearly all these states have the same crystal momentum as $| \psi_0 \rangle$.  So we have determined that in this symmetry class, all four ground states have the same crystal momentum.  While this crystal momentum is not determined from the present considerations, the relative crystal momenta of the ground states are determined (and are zero).  This symmetry class is realized in the toric code for $K_e, K_m > 0$, where it is straightforward to find that all four ground states have crystal momentum $\boldsymbol{k} = 0$.
 
Next, we consider the class $\sigma^e_{txty} = 1$, $\sigma^m_{txty} = -1$.  This class is realized in the toric code when $K_e < 0$, $K_m > 0$.  If $(N_x, N_y) = (\text{even}, \text{even})$, then everything proceeds as above and the ground states all have the same crystal momentum.  In the toric code, all four ground states have crystal momentum $\boldsymbol{k} = 0$.

In the case $(N_x, N_y) = (\text{odd}, \text{even})$, we have
\begin{eqnarray}
T_x {\cal L}^m_x T^{-1}_x &=& {\cal L}^m_x \\
T_y {\cal L}^m_x T^{-1}_y &=& - {\cal L}^m_x \text{,}
\end{eqnarray}
while both $T_x$ and $T_y$ commute with ${\cal L}^m_y$ and ${\cal L}^e_{\mu}$.  We can choose $| \psi_0 \rangle$ as above, but now we see that $| \psi_0 \rangle$ and $ {\cal L}^m_y | \psi_0 \rangle$ have the same crystal momentum $\boldsymbol{k}$, while $ {\cal L}^m_x | \psi_0 \rangle$ and ${\cal L}^m_x {\cal L}^m_y | \psi_0 \rangle$ have crystal momentum $\boldsymbol{k} + (0, \pi)$.  In the toric code, we find two ground states with crystal momentum zero, and two with $(0, \pi)$.  The results are the same when $(N_x, N_y) = (\text{even}, \text{odd})$, except of course that the relative crystal momentum between pairs of ground states becomes $(\pi, 0)$.

For $(N_x, N_y) = (\text{odd}, \text{odd})$, Eq.~(\ref{eqn:m-association}) actually implies
\begin{equation}
(T^m_x)^{N_x} (T^m_y)^{N_y} (T^m_x)^{-N_x} (T^m_y)^{-N_y} = -1 ,
\end{equation}
that is, an $m$ particle translated around the ``boundary'' of the system feels a net $\pi$ flux inside.  This is not a consistent state of affairs on a periodic torus, but it can be repaired if a single $e$-particle is forced into the system.  Since this $e$-particle has no preferred spatial position, we expect the excitation spectrum to be gapless for $(N_x, N_y) = (\text{odd}, \text{odd})$.  This is precisely what occurs in the toric code model.

Finally, we consider the class $\sigma^e_{txty} = \sigma^m_{txty} = -1$.  This class is realized in the toric code when $K_e, K_m < 0$.  To analyze this case, we note that $\sigma^\epsilon_{txty} = 1$.  Therefore we can simply repeat the analysis above for the $\sigma^e_{txty} = - \sigma^m_{txty}$ class, but substituting ${\cal L}^\epsilon_{\mu}$ for ${\cal L}^e_{\mu}$.  In particular, we choose $| \psi_0 \rangle$ so that ${\cal L}^\epsilon_{\mu}  | \psi_0 \rangle = | \psi_0 \rangle$, and $|\psi_0\rangle$ has a definite crystal momentum since ${\cal L}^\epsilon_{\mu}$ commutes with translations.  Depending on $(N_x, N_y)$, the ground states thus have precisely the same relative momenta as in the $\sigma^e_{txty} = - \sigma^m_{txty}$ class.  Moreover, we also obtain the same results from the toric code model.

The above analysis points out that ground-state quantum numbers do not completely determine the symmetry class, even in the simple case of only translation symmetry.  A very interesting problem for future work is to devise a means to completely determine the symmetry class entirely from the ground-state wave functions.

\section{Comparison to projective symmetry group classification}
\label{sec:psg-comparison}

Here, we compare our symmetry classification with the projective symmetry group (PSG) classification of parton mean-field theories for spin liquids.\cite{wen02}  First, we very briefly review PSG classification in the setting where it was introduced, namely the fermionic parton approach to square lattice $S = 1/2$ Heisenberg models.  (See also Ref.~\onlinecite{gchen12} for a more extended discussion.)  We consider a system of $S = 1/2$ spins on the sites of the square lattice, and assume square lattice space group, time reversal and ${\rm SO}(3)$ spin rotation symmetries. The spin operator at site $\r$ is written as a bilinear of $S = 1/2$ fermionic partons,
\beq
\bS_{\r} = \frac{1}{2} f^\dag_{\r} \bm{\sigma} f_{\r}, \qquad
f_{\r} = \begin{pmatrix} f_{ \r \up} \\ f_{ \r \down} \end{pmatrix} \text{,}
\eeq
with the local constraint of one fermion per site.  Defining
\begin{equation}
\psi_{\r} = \begin{pmatrix} f_{ \r \up} \\ f^\dagger_{ \r \down} \end{pmatrix} \text{,}
\end{equation}
it is straightforward to show that $\bS_{\r}$ is invariant under local ${\rm SU}(2)$ gauge transformations
\begin{equation}
\psi_{\r} \to G_{\r} \psi_{\r} \text{,} \label{eqn:su2-gauge}
\end{equation}
with $G_{\r} \in {\rm SU}(2)$.  

To proceed, one writes down a mean-field Hamiltonian ${\cal H}_{{\rm MFT}}$ quadratic in the partons.  To describe a spin liquid, the mean-field theory should respect the full symmetry group. In order to leave ${\cal H}_{{\rm MFT}}$ invariant, symmetries are in general accompanied by non-trivial gauge transformations.  For instance, if
 $S: \r \to S(\r)$ is a space group operation, then in general
\begin{equation}
S : \psi_{\r} \to G^S_{\r} \psi_{S(\r) } \text{,} \label{eqn:su2-space-group}
\end{equation}
where $G^S_{\r} \in {\rm SU}(2)$.  This is permitted because the physical spin operators retain the correct transformation law $S: \bS_{\r} \to \bS_{S(\r) } $.  Such action of symmetry on the partons is a projective representation of the symmetry group, and this projective representation is referred to as a PSG.

To fully specify a PSG, it is not enough merely to specify the action of the symmetry group on the fermions.  One must also specify the subgroup of gauge transformations leaving ${\cal H}_{{\rm MFT}}$ invariant.  This subgroup is referred to as the invariant gauge group (IGG).  We restrict attention to the case of $\text{IGG} = \zz$, since in this case, one obtains a $\zz$ spin liquid upon going beyond mean-field theory (see below).  Such PSGs are referred to as $\zz$ PSGs.  The nontrivial IGG transformation is $\psi_{\r} \to - \psi_{\r}$, which clearly commutes with all symmetry operations.

PSGs can be classified up to unitary equivalence under ${\rm SU}(2)$ gauge transformations, Eq.~(\ref{eqn:su2-gauge}); this provides a symmetry classification of mean-field parton Hamiltonians with fixed IGG.  This is so because the PSG (and IGG) can be determined from any ${\cal H}_{{\rm MFT}}$ invariant under the symmetry group.  Then, keeping the IGG fixed but otherwise adding arbitrary symmetry-preserving perturbations at the mean-field level, the PSG, which is discrete, remains unchanged.  In Ref.~\onlinecite{wen02}, Wen found 272 distinct $\zz$ PSGs on the square lattice.  Actually, there are a total of 280 distinct $\zz$ PSGs.\cite{gchen12}  Wen assumed that spin rotations are not accompanied by any gauge transformations.  In Ref.~\onlinecite{gchen12}, together with G. Chen, we showed that this assumption can be relaxed, leading to eight more PSGs.

We note that many of these 280 PSGs do not lead to effective low-energy theories for gapped $\zz$ spin liquids.\cite{wen02}  Some PSGs do not actually admit a mean-field Hamiltonian with $\text{IGG} = \zz$; such PSGs are said to be only ``algebraic PSGs,'' and not ``invariant PSGs.''\cite{wen02}  For example, we found eight new algebraic PSGs with Chen in Ref.~\onlinecite{gchen12}, but only four of these are invariant PSGs admitting a mean-field Hamiltonian.  Moreover, even among invariant PSGs with $\text{IGG} = \zz$, the PSG may require the fermions to be gapless at one or more points in the Brillouin zone.  For instance, this happens for all four invariant PSGs of Ref.~\onlinecite{gchen12}.  In such cases one obtains gapless $\zz$ spin liquids, to which our classification does not apply.  

The above discussion does not address the question of symmetry classification beyond the mean-field level.  Indeed, to  connect to our classification, we need to go beyond parton mean-field theory.  For a mean-field Hamiltonian with $\text{IGG} = \zz$, this can be done by minimally coupling the fermions to a dynamical $\zz$ gauge field (see, \emph{e.g.}, Ref.~\onlinecite{gchen12} for an example of this procedure).  If ${\cal H}_{{\rm MFT}}$ endows the fermions with a gapped excitation spectrum, and if the $\zz$ gauge field is in its deconfined phase, then we obtain a low-energy effective theory for a gapped $\zz$ spin liquid.  The mean-field fermions are promoted to $\epsilon$ particles, and the $m$-particles are the gapped fluxes of the $\zz$ gauge field.

We are now in a position to compare PSG classification with our classification.  First, as already discussed in Sec.~\ref{sec:prior}, PSG classification is tied to parton formalism, and we feel that parton formalism is the wrong language with which to classify $\zz$ spin liquids.  Moreover, we know of no argument that PSG classification continues to hold beyond the mean-field level.  Next, it is already clear that PSG classification does not provide any information about the fractionalization class of the $m$-sector, and thus does not give a complete symmetry classification for that reason alone.  However, it is interesting to discuss the relationship between the two classifications for the $\epsilon$-sector.

From the discussion above, we see that a $\zz$ PSGs is a $\zz$ central extension of the symmetry group; the group $A$ is $A = \zz = \text{IGG}$.  The $\epsilon$ fractionalization class is simply the cohomology class of the PSG.  
The cohomology classification is \emph{coarser} than PSG classification, because two unitarily inequivalent PSGs may belong to the same cohomology class.  Indeed, we have found instances on the square lattice where two unitarily inequivalent PSGs belong to the same cohomology class.  However, in all cases we have found, one of the  PSGs in such a pair requires the fermions to be gapless, and our considerations do not apply.  While we have not found a case where two inequivalent PSGs for gapped $\zz$ spin liquids are equivalent under cohomology classification, we have not searched systematically for such examples, and it could be interesting to do so.

The statement that cohomology classification is coarser than PSG classification might seem somewhat puzzling, since there are $2^{11} = 2048$ distinct $\epsilon$ fractionalization classes, while there are only 280 distinct PSGs.  This occurs because $S  = 1/2$ fermionic partons are not capable of realizing every cohomology class.  For instance, restricting to point group and time reversal symmetries (\emph{i.e.}, symmetries leaving a lattice point fixed), there are cohomology classes where the smallest irreducible representation has dimension 4, but the $S = 1/2$ partons only provide a two-dimensional on-site Hilbert space.\cite{essinunpub}

In discussing PSG classification, we have focused on one particular parton representation of $S = 1/2$ spin models.  Other parton representations also exist; for instance, we could have just as well considered $S = 1/2$ \emph{bosonic} partons, and discussed the PSG classification in that case.\cite{fwang06}  For every distinct parton representation, the PSG classification needs to be redone.  Within the framework of PSG classification, it is not obvious how to compare PSGs obtained using different parton representations.  On the other hand, our classification can be applied to effective theories for gapped $\zz$ spin liquids obtained from any parton construction.  If two such effective theories belong to different symmetry classes, then they describe different $\zz$ spin liquid phases.  On the other hand, if two apparently different such effective theories belong to the same symmetry class, they \emph{may} describe the same $\zz$ spin liquid phase.

\section{Discussion}
\label{sec:discussion}

We conclude with a discussion of open issues and future directions.  Our approach to symmetry classification can likely be extended to arbitrary Abelian topological orders, including space group symmetry.  It would also be interesting to consider extension to non-Abelian topological order.  For two-dimensional $\zz$ spin liquids, the problem of full symmetry classification, where some symmetry operations may exchange $e$ and $m$ particles, is still open.  There is also, of course, the problem of full classification of symmetric $\zz$ spin liquids (\emph{i.e.}, beyond symmetry classification).  Here, $K$-matrix Chern-Simons approaches may prove useful.\cite{levin12,ymlu12}

Extending symmetry classification to three dimensional ($d = 3$)  $\zz$ spin liquids could be particularly interesting.  First, we point out a connection between our $d = 2$ classification and the classification of $d = 1$ symmetry-protected topological (SPT) phases.\cite{chen11a,turner11,fidkowski11,schuch11}  The $d=1$ SPT phases are essentially classified in terms of projective transformations of point objects bound to the ends of the $d = 1$ system (\emph{i.e.}, end states).  Quite similarly, our $d=2$ symmetry classification for $\zz$ spin liquids is based on projective transformations of point objects tied to the ends of \emph{fluctuating} one-dimensional strings, namely the anyons.  The mathematical consequence of this connection is the appearance of the second cohomology group in both classifications.  Now, in deconfined $\zz$ gauge theory in three dimensions, the topological excitations are point-like $\zz$ electric charges and extended vison loops.  A vison loop can be viewed as the boundary of a highly fluctuating two-dimensional membrane (with vanishing surface tension).  By analogy, we then speculate that there is a close connection between the ``fractionalization class'' of a vison loop---assuming it can be defined---and the classification of $d = 2$ SPT phases.

Returning to the present classification, we give an argument that some of our symmetry classes \emph{cannot} be realized strictly in two dimensions.  The argument follows Ref.~\onlinecite{vishwanath12}, which studied surface theories for three-dimensional SPT phases.  Suppose $G$ is an internal symmetry, and suppose both the $e$ and $m$ particles have a fractionalization class admitting no one-dimensional irreducible (projective) representations.  An example is time reversal symmetry with $({\cal T}^e)^2 = ( {\cal T}^m )^2 = -1$.  Strictly in two dimensions, with only internal symmetries, one expects there to be a trivial phase with no topological order or spontaneously broken symmetry (\emph{e.g.}, a dimerized phase).  To destroy the topological order, one can condense either the $e$ or $m$ particle ($\epsilon$ is a fermion and thus cannot be condensed), but in this situation this must be accompanied by spontaneous breaking of $G$-symmetry.  Therefore, there seems to be no way to leave the $\zz$ spin liquid and enter a trivial phase, so we expect this situation cannot be realized strictly in two dimensions.  On the other hand, Ref.~\onlinecite{vishwanath12} showed that such situations \emph{can} be realized on the surface of a $d=3$ SPT phase.  This discussion establishes a connection between our classification and the classification of $d=3$ SPT phases, which could be interesting to pursue in future work.

Along similar lines, there is another constraint on symmetry classes among certain (strictly two-dimensional) models.
Consider a model with translation and ${\rm SO}(3)$ spin rotation symmetries, with an odd number of $S = 1/2$ moments per unit cell.  For such a model a trivial gapped quantum paramagnet with no spontaneous symmetry breaking and no topological order is impossible.\cite{hastings04}  This implies, for instance, that both $e$ and $m$ particles must have non-trivial fractionalization classes.  If one of these particles had the trivial fractionalization class, upon condensing it, one would obtain a trivial quantum paramagnet in contradiction to the theorem of Ref.~\onlinecite{hastings04}.  All this discussion points out that it is desirable to obtain a better understanding of which symmetry classes can occur in various settings.

Eventually, we hope  our results may lead to the development of tests to distinguish different types of $\zz$ spin liquids in numerical studies.  There is, of course, the connection between symmetry classes and ground state quantum numbers discussed in Sec.~\ref{sec:qndgs}.  In cases where one has numerical access to the excitation spectrum, one could potentially obtain information about the ``coarsened'' ${\rm U}_T(1)$ fractionalization classes from multiplicities (or, more generally, decomposition into irreducible representations) of nearly degenerate energy levels, as discussed in Sec.~\ref{sec:fc-gen} for the simple example $G = \zz \times \zz$.  More ambitiously, it would be interesting and potentially useful to understand how to fully determine the symmetry class given only the ground-state wave function(s). 

While we know of no current candidate materials for a gapped $\zz$ spin liquid, it would nonetheless be interesting to devise experimental measurements of symmetry class information.  Thinking along these lines could lead to new experimental tests for fractionalization, which could potentially be applicable more broadly, for instance to gapless spin liquids.

Finally, we note that Mesaros and Ran have very recently proposed a classification of topologically ordered phases with on-site symmetry.\cite{mesaros12}  We also note very recent related results of Hung and Wen.\cite{hung12} It will be interesting to understand the relationship between our classification and these results.

\acknowledgments

We are grateful for useful and inspiring discussions with Leon Balents, Lukasz Fidkowski, Matthew Hastings, Yuan-Ming Lu, T. Senthil, Miles Stoudenmire, Ari Turner, Ashvin Vishwanath, Fa Wang,  Xiao-Gang Wen, and especially Ying Ran.  We also acknowledge useful correspondence with Alexei Kitaev, and thank Hao Song for a careful reading of the manuscript.  This work is supported by the David and Lucile Packard Foundation.  We acknowledge the hospitality and support of the Aspen Center for Physics (NSF grant No.~1066293) and the Kavli Institute for Theoretical Physics (NSF grant No.~PHY11-25915), where some of this work was carried out.

\appendix

\section{Non-singlet ground states}
\label{app:non-singlet-gs}

\begin{table*}
\begin{tabular}{|c|c|c|c|c|c|c|}
\hline
Rep. number & $P_x$ & $P_{xy}$ & $T_x$ & $U_T$ & $R_s(\theta \hat{\bf n} )$ & $\sigma$'s that are $-1$ \\
\hline
1 & $i$ & $1$ & $1$ & $1$ & $1$ & $\sigma_{px}$ \\
\hline
2 & $1$ & $i$ & $1$ & $1$ & $1$ & $\sigma_{pxy}$ \\
\hline
3 & $\mu^0$ &$\mu^0$ &$\mu^0$ & $i \mu^y$ & $\mu^0$ &  $\sigma_T$ \\
\hline
4 & $\mu^x$ & $\mu^0$ & $\mu^0$ & $\mu^z$ & $\mu^0$ & $\sigma_{Tpx}$ \\
\hline
5 & $\mu^0$ & $\mu^x$ & $\mu^0$ & $\mu^z$ & $\mu^0$ & $\sigma_{Tpxy}$\\
\hline
6 & $\mu^x$ & $(\mu^x + \mu^z)/\sqrt{2}$ & $\mu^0$ & $\mu^0$ & $\mu^0$ & $\sigma_{pxpxy}$ \\
\hline
7 & $\mu^0$&$\mu^0$ & $\mu^x$ & $\mu^z$ & $\mu^0$ & $\sigma_{Ttx}$ \\
\hline
8 & $\mu^0$& $(\mu^x + \mu^z)/\sqrt{2}$ & $\mu^x$ & $\mu^0$ & $\mu^0$ & $\sigma_{txty}$ \\
\hline
9 &$\mu^z$ & $\mu^0$ & $\mu^x$ & $\mu^0$ & $\mu^0$ & $\sigma_{txpx}, \sigma_{typx}$ \\
\hline
10 & $\mu^x$ & $(\mu^x + \mu^z)/\sqrt{2}$ & $\mu^x$ & $\mu^0$ & $\mu^0$ & $\sigma_{pxpxy}, \sigma_{txty}, \sigma_{typx}$ \\
\hline
11 & $\mu^0$ & $\mu^0$&$\mu^0$ & $i \mu^y$ & $\exp(i \theta \hat{\bf n}^i \mu^i /2 )$ & $\sigma_T, \sigma_R$  \\
\hline
\end{tabular}
\caption{Set of 11 generating representations for square lattice space group, time reversal and spin rotation symmetry.  The first column numbers the representations, 1 through 11.  The middle five columns specify generators of the group in the corresponding representation (time reversal is ${\cal T} = U_T K$, where $K$ is complex conjugation).  All representations in this table are one- or two-dimensional.  Generators of the two-dimensional representations are specified in terms of the Pauli matrices $\mu^{x,y,z}$, and the $2 \times 2$ identity matrix $\mu^0$. (We use $\mu$ rather than $\sigma$ for these matrices here, to avoid confusion with the $\zz$-valued $\sigma$ parameters.) The last column lists those $\sigma$'s that are equal to $-1$ for the corresponding representation.}
\end{table*}

When discussing symmetry localization [see, \emph{e.g.}, Eq.~(\ref{eqn:product})], we assumed we can find a ground state $| \psi_0 \rangle$ that is a singlet under all symmetry operations.  Here we describe how this assumption can be relaxed; this enables us to consider the full four-dimensional ground state subspace on the torus.

\begin{figure}
\includegraphics[width=0.5\columnwidth]{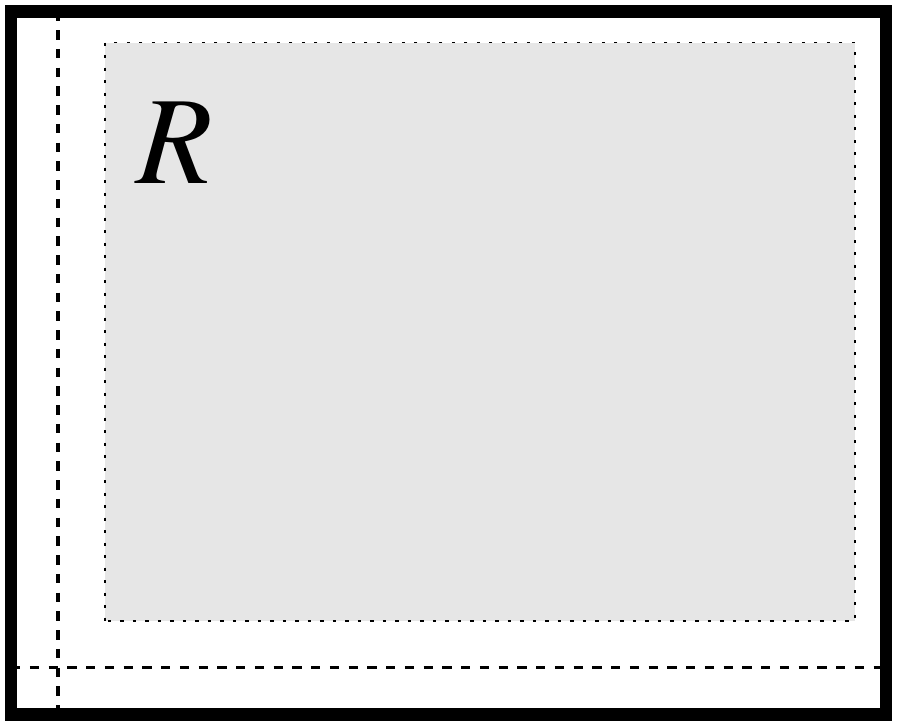}
\caption{Illustration of states used for symmetry localization to account for non-singlet ground states.  The thick solid line is the ``boundary'' of the periodic system---opposite edges are identified.  The region $R$ is shaded.  Non-contractible strings of the loop algebra act along the dashed lines, outside the region $R$.}
\label{fig:non-singlet}
\end{figure}

We consider states $| \psi_\alpha \rangle$ where all excitations, and all strings separating anyons, are confined to a large box-shaped region $R$ as shown in Fig.~\ref{fig:non-singlet}.  We take $R$ to cover almost the entire area of the system. For such states, we choose the non-contractible loops of ${\cal L}^e_x$, ${\cal L}^m_x$, and so on, to run in the space outside $R$.  Restricting to such states allows us to break the Hilbert space into ``global sectors'' associated with the four degenerate ground states in a well-defined fashion.  Formally, we decompose the Hilbert space as a tensor product ${\cal H}_R \otimes {\cal H}_G$, where ${\cal H}_R$ is the Hilbert space of excitations contained in $R$, and ${\cal H}_G$ is the four-dimensional Hilbert space of degenerate ground states.  If ${\cal O}$ is supported on $R$, then in the above Hilbert space decomposition we write ${\cal O} = {\cal O}_R \otimes 1_G$.  On the other hand, if ${\cal O}$ is a loop algebra operator, then ${\cal O} = 1_R \otimes {\cal O}_G$.  This means that acting on $| \psi_\alpha \rangle$ with any operator supported on $R$ does not affect the global sector degrees of freedom.  Conversely, acting with any loop algebra operator leaves local properties in $R$ unaffected.

This discussion motivates a generalized version of the symmetry localization assumption.  Namely, given a symmetry operation $S_a$, we assume
\begin{equation}
S_a | \psi_\alpha \rangle = S_a(R) S^G_a| \psi_\alpha \rangle \text{.}
\end{equation}
Here, $S_a(R)$ is a unitary operator supported on $R$.  The operator $S^G_a$ is a unitary linear combination of products of loop algebra generators.  Any unitary transformation on the four-dimensional ground state subspace can be written as such a linear combination, so $S^G_a$ can be thought of as a general unitary transformation among the global sectors.  Clearly we have $[ S_a(R), S^G_b ] = 0$.  

At this point we apply symmetry localization as discussed previously to the operator $S_a(R)$.  To illustrate this with a concrete example, suppose that two localized, isolated $e$-particles are contained within $R$, in regions $R^e_i$ ($i = 1,2$).  Then we write
\begin{equation}
S_a | \psi_\alpha \rangle =  S_a(R) S^G_a| \psi_\alpha = S_a(1) S_a(2) S^G_a| \psi_\alpha \rangle \text{.}
\end{equation}
Considering a group relation $S_1 \cdots S_k = 1$, we note that we must have
\begin{equation}
S^G_1 \cdots S^G_k = 1 \text{,}
\end{equation}
because $S_a = S^G_a$ on the ground-state subspace, and the ground states, of course, do not transform projectively.  Therefore the $S^G_a$ operators drop out in the group relations, and discussion of fractionalization classes proceeds exactly as in Sec.~\ref{sec:fracclasses}.

\section{Generating set of projective representations for square lattice space group plus time reversal symmetry}
\label{app:genset}

In Eqs.~(\ref{eqn:px-relation})--(\ref{eqn:spin-relation}), the $\zz$ factor sets for square lattice space group, time-reversal, and spin rotation symmetries are defined in terms of 11 $\zz$-valued 
$\sigma$ parameters.  Here, we show that all $2^{11}$ choices of the $\sigma$ parameters give consistent factor sets.  That is, it is possible to find a projective representation with any choice of the $\sigma$'s.

We proceed by constructing a ``generating set'' of 11 projective representations.  By taking tensor products of these 11 representations, one can obtain a projective representation with any of the $2^{11}$ possible choices of $\sigma$'s.  The cohomology classes of the generating set form a generating set for the group $H^2(G, \zz) = \zz^{11}$.

First, suppose we have two projective representations $A$ and $B$, each with its own set of $\sigma$'s.  That is, for representation $A$ we have $(\sigma^A_{px}, \sigma^A_{pxy}, \dots)$, and similarly for representation $B$.  It is straightforward to show that the $\sigma$'s of the tensor product representation $A \otimes B$ are given by 
\begin{equation}
(\sigma^{A \otimes B}_{px} , \sigma^{A \otimes B}_{pxy}, \dots ) 
= ( \sigma^A_{px} \sigma^B_{px} , \sigma^A_{pxy} \sigma^B_{pxy}, \dots ) \text{.}
\end{equation}
Each choice of $\sigma$'s can thus be viewed as an element of $\zz^{11}$, and if we find 11 representations whose $\sigma$'s  generate $\zz^{11}$, then a representation with any desired choice of $\sigma$'s can be obtained by taking tensor products.

Before proceeding, we note that if ${\cal T}^A = U^A_T K$ and ${\cal T}^B = U^B_T K$ are the anti-unitary time reversal operations in representations $A$ and $B$, where $K$ is the complex conjugation operator, the tensor product operation is defined as usual to be
\begin{equation}
{\cal T}^{A \otimes B} = (U^A_T \otimes U^B_T) K \text{.}
\end{equation}

At this point, we need only exhibit a generating set of 11 projective representations.  This is done in Table I.  It is straightforward to show that the corresponding 11 sets of $\sigma$'s exhibited there form a generating set for $\zz^{11}$.

\bibliography{library}

\begin{thebibliography}{59}
\bibitem [{gap()}]{gapcomment}%
  \BibitemOpen
  \href@noop {} {}\bibinfo {note} {Following common but not universal practice,
  we include the presence of an energy gap in the definition of topological
  order.}\BibitemShut {Stop}%
\bibitem [{opR()}]{opR}%
  \BibitemOpen
  \href@noop {} {}\bibinfo {note} {The operators are subject to the condition
  that they must leave the boundary of $R$ locally in the ground state. The
  reason for this requirement is that we want to avoid, for instance, moving
  the $e$-particle just barely outside of $R$. One can make the restriction on
  operators more concrete by defining a smaller region inside $R$ whose
  boundary is sufficiently far away from the boundary of $R$, and acting only
  with operators supported on this smaller region. We shall not belabor this
  point in further discussion, and simply refer to operators supported on
  $R$.}\BibitemShut {Stop}%
\end{thebibliography}

\end{document}